\documentclass[aps,prb,twocolumn,groupedaddress]{revtex4-2}
\usepackage{graphicx} 
\usepackage{dcolumn}
\usepackage{bm}
\usepackage{amsmath}
\usepackage{epsfig}
\usepackage{float}
\usepackage{epstopdf}
\usepackage{natbib}
\usepackage{subfigure}

\newcommand{\dprime}{{\prime\prime}}

\begin{document}

\title{Multi-Dimensional Coherent  Spectroscopy of Light-Driven States and their Collective Modes in Multi-Band Superconductors} 


\author{Martin~Mootz$^{1}$, Liang~Luo$^{1}$, Chuankun~Huang$^{1,2}$, Jigang~Wang$^{1,2}$, and llias~E.~Perakis$^{3*}$}

\affiliation{$^{1}$ Ames National Laboratory, U.S. Department of Energy, Ames, Iowa 50011, USA \\
$^{2}$ Department of Physics and Astronomy, Iowa State University, Ames, Iowa 50011, USA \\
 $^{3}$  Department of Physics, University of Alabama at Birmingham, Birmingham, AL 35294-1170, USA}		
	



\date{\today}
\begin{abstract}
We present a comprehensive theory of light-controlled multi-band   superconductivity, and apply it to predict distinctive signatures of light-driven  superconducting (SC) states  in terahertz multi-dimensional coherent spectroscopy (THz-MDCS) experiments. We first derive  gauge-invariant Maxwell--Bloch equations for multi-band BCS superconductors with spatial fluctuations. 
We consider  driving electromagnetic fields determined self--consistently by Maxwell's equations. By calculating the THz-MDCS spectra measured experimentally
in the clean SC limit, we  identify unique signatures of  finite-momentum Cooper-pairing  states that live longer than  the laser pulse. They are controlled by  a pair of THz laser pulses  with well-defined relative phase (pulse-pair). The pseudo-spin oscillators
that describe the properties of these SC states  are   parametrically driven by both finite-momentum Cooper pairing and  by time oscillations of the order parameter relative phase. We  show  that such strong parametric driving leads to drastic changes in the  THz-MDCS spectral shape from the predictions of  third-order nonlinear susceptibility calculations. These spectral changes  strongly depend on the interband-to-intraband interaction ratio and on the collective modes of the light-driven state.   For negligible interband interaction, the  spectra  show a transition with increasing field,  from traditional pump--probe, four-wave mixing, and third-harmonic generation 
 peaks determined by the laser frequency  to sidebands determined by the excitations of the driven system. These sidebands  emerge from difference-frequency Raman processes in the non-equilibrium SC state. For interband couplings weaker than the intraband pairing, we show that  the  Leggett phase collective mode leads to  harmonic sidebands around the traditional pump--probe peaks.  Additional Higgs collective mode peaks result from   light-induced inversion symmetry breaking in a thin film geometry.  For strong interband coupling, we find a transition from a non-equilibrium finite Cooper pair momentum  state characterized by hybrid-Higgs amplitude mode  peaks in THz-MDCS spectra to a driven state identified experimentally by  the emergence of Floquet-like  sidebands at bi-Higgs frequencies. Those dominant bi-Higgs-frequency  satellites are manifestations of  a  new order parameter relative phase collective mode that characterizes the non-equilibrium SC state.  The predicted  interaction- and field-dependent   transitions in the spectral profile  allow us to propose THz-MDCS experiments for quantum tomography of light-driven superconductivity.
\end{abstract}

  \maketitle 


\section{Introduction}
\label{sec:intro}  

Terahertz multi-dimensional coherent spectroscopy (THz-MDCS)  is developing into an
important tool for  unraveling the dynamics of electronic and vibrational excitations of
matter. In the past, such multi-dimensional spectroscopy has been  used in conventional materials to investigate  electronic excitations ~\cite{Kuehn2009,maag2016,Junginger2012,Tarekegne2020,Pal2021}, spin waves~\cite{Fuller2015,Nelson}, and vibrational modes~\cite{Johnson2019,Blank2023}, among others. In  superconducting (SC) systems, however, THz-MDCS experiments and corresponding theories have been  rare so far~\cite{Mootz2022,NatPhys,Manske2023,cheng2023evidence}. As a result, 
it is not yet clear how to obtain new information about non-equilibrium superconductivity  and other outstanding questions by analyzing THz-MDCS data. 
The SC  dipole-forbidden collective modes, which  characterize the unique properties of superconductivity, overlap in energy with the quasi-particle continuum.  In multi-band superconductors, the collective mode properties  depend on the ratio of interband over intraband interaction strength, as well as on the driving light field.  So far, 
the THz-driven dynamics of superconductors has been mostly studied with traditional single-particle  pump--probe spectroscopies~\cite{matsunaga2014,Matsunaga2017,Giorgianni2019,Chu2020,hybrid-higgs,cheng2023lowenergy}. These experiments have  been interpreted 
 by using nonlinear susceptibility expansions~\cite{Aoki2015,Aoki2017,Benfatto2019,udina2022thz}, 
Anderson  pseudo-spin models~\cite{Axt2007,krull2016,Forster2017,Schwarz2020,Mootz2020,Mootz2022} or   Green's functions~\cite{Stephen1965,Wu2017,Wu2019}, in the clean SC limit or including disorder effects~\cite{Murotani,seibold2021,Benfatto2023}. In our previous works, we have shown that THz light-wave acceleration of the Cooper-pair condensate gives access to  long-lived  SC states with finite-momentum pairing. These states are witnessed experimentally, e.~g.,  by the emergence  above critical driving field of spectral peaks centered at frequencies forbidden by the equilibrium symmetry. 
They range from quasi-particle quantum states, accessed by single-cycle THz pulses~\cite{yang2018}, to gapless-SC states with finite coherence and broken  inversion symmetry, accessed by multi-cycle THz pulses~\cite{yang2019lightwave}.
Full characterization of the different non-equilibrium states 
 requires  detection of the  collective modes~\cite{Anderson} of the underlying order parameters.
 Typical collective excitations of multi-band superconductors  include amplitude oscillations of the SC order parameter (Higgs mode) and oscillations of the relative phase of the  SC order parameters in different bands 
(Leggett mode)~\cite{krull2016,Aoki2017}.  In addition to identifying how their collective modes differ from those of the equilibrium SC state,  full characterization  of strongly-driven SC states
also requires  the resolution of high-order  correlations  that go beyond the traditional third-order nonlinear responses known to dominate close to equilibrium~\cite{Mootz2022,NatPhys}. 
Examples of such high-order correlations in light-driven SC states  include  sideband generation at bi-Higgs frequencies analogous to Floquet sidebands~\cite{NatPhys} and correlated-wave-mixing 
peaks in THz-MDCS~\cite{Mootz2022}. 

More generally, understanding the properties of quantum materials for, e.~g.,  quantum science applications, requires experiments that can measure and control correlation between different elementary excitations, like THz-MDCS. 
Quantum tomography of  THz-driven condensate states is, however,  challenging. Unlike in semiconductors, collective modes of  SC states do not couple  linearly to the electromagnetic fields without a finite Cooper-pair momentum~\cite{Klein1980,Littlewood1981,Podolsky2011,Moor2017,Shimano2019,yangPRL,Mootz2020,
Puviani2020}. In addition,  the SC energy gap is quenched coherently via Raman processes during cycles of THz light-wave oscillations ~\cite{Mootz2022}. Finally, the dynamics of the phase of the complex order parameter  must be considered. The interpretation of the experimental  spectra is further complicated by the multiple excitation pathways contributing  to the same  nonlinear  signals~\cite{Cea2016,Murotani}. Recent  simulations of THz-MDCS experiments proposed the possibility for  ultrafast
visualization and quantum control of THz-driven  SC states~\cite{Mootz2022}.  
 Resolution of high order
correlation and relative-phase collective modes has been recently realized by THz-MDCS experimental studies of an iron-based superconductor~\cite{NatPhys}. Such THz--MDCS experiments allow the identification of high--order nonlinear responses through the observation 
of new peaks centered at high frequencies. These peaks 
are distinguished  in two--dimensional frequency space from the conventional pump--probe peaks accessed through traditional one--dimensional pump--probe spectroscopies.

\begin{figure}[t!]
\begin{center}
		\includegraphics[scale=0.39]{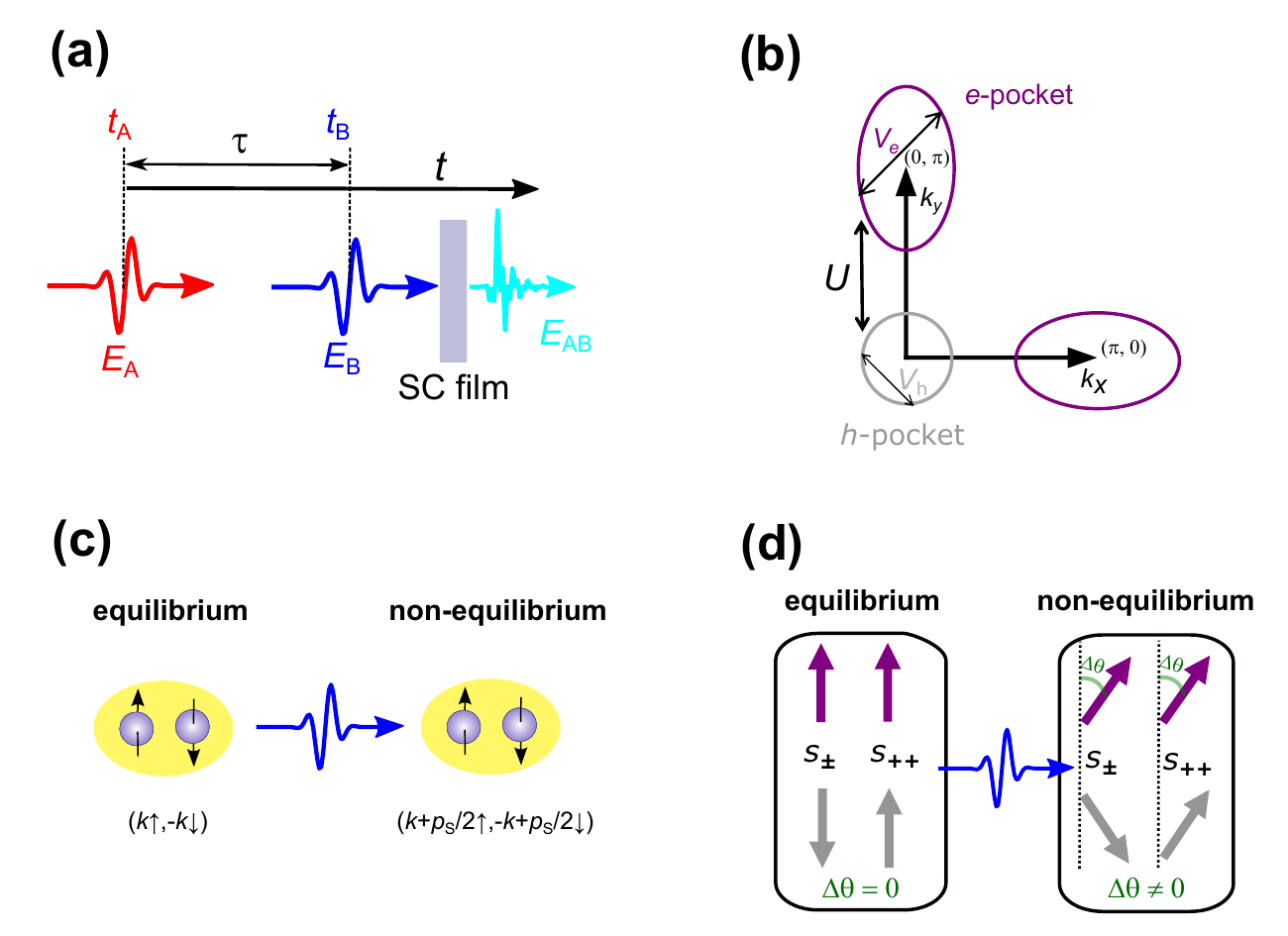}
		\caption{Multi-dimensional coherent spectroscopy of multi-band superconductors. (a) Schematic representation of the two-dimensional terahertz (THz)  configuration considered in this paper. A superconducting (SC) thin film is excited by a pulse-pair consisting of two collinear, phase-locked THz  pulses of similar strength and spectrum. Pulse $E_\mathrm{A}$ (red) is centered at $t_\mathrm{A}=0$, while pulse $E_\mathrm{B}$ (blue) is centered at $t_\mathrm{B}=\tau$. The  inter-pulse delay $\tau=t_\mathrm{B}-t_\mathrm{A}$ controls  their relative phase.  $E_\mathrm{AB}(t,\tau)$ (cyan) is the transmitted electric field after  excitation by both pulses.
The experimentally measured signal, $E_\mathrm{NL}=E_\mathrm{AB}-
E_\mathrm{A}-E_\mathrm{B}$, 
vanishes in the absence of correlations between  excitations driven
by different pulses, which distinguishes THz-MDCS from the SC nonlinear response to a single pulse and from conductivity measurements.   
(b) 3-pocket bandstructure model used in the simulations. A hole (h) pocket is centered at the $\Gamma$-point and two electron (e) pockets are located at $(\pi,0)$ and $(0,\pi)$ in ${\bf k}$--space. We consider intraband pairing interactions $V_\mathrm{e}$ and $V_\mathrm{h}$ and interband coupling $U$ between the electron and hole pockets. Two effects determine the non-equilibrium SC state: (c) The effective field inside the SC thin film accelerates the condensate and induces a finite Cooper pair  center-of-mass momentum $\mathbf{p}_\mathrm{S}$ that persists after the pulse. This THz light-wave acceleration results in a  long-lived, finite-momentum Cooper pairing state $(\mathbf{k}+\mathbf{p}_\mathrm{S} \uparrow,\mathbf{k}-\mathbf{p}_\mathrm{S} \downarrow)$; (d) Parametric driving of the superconductor by the dynamics of the order parameter relative phase. Left box:  $s_{\pm}$ ($s_{++}$) order parameter symmetry with  $\Delta \theta=0$ gives a  
pseudo-magnetic field Eq.~(\ref{B}) with  x--axis components pointing in opposite (same) 
directions in different bands. Pseudo-spins orient along this pseudo-magnetic field in equilibrium.
 Right box: Light-induced non-equilibrium state with long-lived  relative phase $\Delta\theta(t) \ne 0$ above critical pulse-pair excitation is determined by a  pseudo-magnetic field Eq.~(\ref{B}) whose $x$--$y$ components depend on  $\Delta \theta(t)$.}
		\label{fig1} 
\end{center}
\end{figure}

In this paper, we present a comprehensive theory  of  THz-MDCS experiments on multi-band BCS superconductors, and  apply it to identify unique experimental  signatures of light-driven superconductivity. By extending the 
density matrix approach of Ref.~\cite{Mootz2022} to the case of multiple coupled bands,  we derive self-consistent, gauge-invariant SC Bloch
equations (Appendix \ref{eq:be_full}). Together with Maxwell’s wave equation, these SC Bloch equations  allow us to propose distinctive experimental signatures of parametrically-driven SC states.
For this purpose, we  consider the phase-locked, collinear two-pulse geometry of Fig.~\ref{fig1}(a). Two  THz pulses with equal strength and spectral profile,
and with well-defined relative phase (phase-locked pulse-pair), 
 excite a SC thin film. These two pulses are separated by the inter-pulse time delay $\tau$ that controls their phase difference.   We study the pulse-pair excitation of a 3-pocket bandstructure model, with a hole (h) pocket centered at the $\Gamma$-point and two electron (e) pockets located at $(\pi,0)$ and $(0,\pi)$ (Fig.~\ref{fig1}(b)).  The corresponding SC order parameter components in  the different bands  are denoted as $\Delta_\lambda$, with $\lambda=\mathrm{e, h}$. 
 In this paper we consider the case of a homogeneous system and approximate  the spatial fluctuations in the full equations of motion presented in Appendix \ref{eq:be_full}, by assuming that their characteristic length  (e.g., the mean free path) exceeds the coherence length that characterizes the size of the Cooper pair. 
The SC state can then be described by introducing  Anderson pseudo-spins located at different momentum points $\mathbf{k}$~\cite{Anderson}. In this representation, up (down) pseudo-spins describe filled (empty) $(\mathbf{k} \uparrow, -\mathbf{k} \downarrow)$ Cooper-pair states. Canted pseudo-spins describe the superposition of up and down  states
on the Bloch sphere. This canting depends on the SC order parameter, which determines  the direction of the pseudo--magnetic field that describes  the pseudo--spin orientation.  For example, for a multi-band superconductor with $s_\pm$ ($s_{++}$) order parameter symmetry, a phase difference of $\pi$ (0)  leads to opposite (same) 
signs  between the order parameters of the electron and hole pockets 
(Fig.~\ref{fig1}(d), left box), which fixes  the corresponding pseudo--magnetic field and pseudo--spin components in equilibrium.
    This  order parameter relative phase  depends on the ratio between interband and intraband interaction strengths. Here we consider the effects  of   the  intraband SC pairing interactions $V_\mathrm{e,h}$ and interband Coulomb-induced coupling $U$ between  electron and hole pockets on the THz--MDCS 
    spectral profiles. We show that  the relative strength of interband vs.  intraband interactions plays a critical role in determining this spectral profile, as well as how this  profile changes with increasing driving field. In this way, the THz-MDCS spectral peaks provide direct evidence about the properties of the SC state and how the latter evolves with increasing THz field driving.  

We wish to highlight the importance of two competing effects for inducing  controllable transitions between different light-driven SC states in multi--band SCs.  First is SC driving by a finite center-of-mass Cooper pair  momentum ($\mathbf{p}_\mathrm{S}$) driven by the laser field. 
Our results   demonstrate that 
 the  effective local field  that accelerates  the SC condensate
 into a finite $\mathbf{p}_\mathrm{S}$ state 
  is  
modified from the external THz laser field by  electromagnetic wave propagation in a thin film geometry. We show that 
this modified effective driving field  
 leads to a  finite center-of-mass Cooper pair  momentum  state that can persist well after the pulse. In particular, 
 the  moving condensate momentum $\mathbf{p}_\mathrm{S}(t)$ decays slowly in time for a thin-film geometry due to radiative damping~\cite{Mootz2020}. This persistent condensate motion  
results in  $(\mathbf{k}+\mathbf{p}_\mathrm{S}/2 \uparrow,-\mathbf{k}+\mathbf{p}_\mathrm{S}/2 \downarrow)$ Cooper pairs, i.~e., to finite momentum pairing, Fig.~\ref{fig1}(c), controlled by the THz field. 
 As a result of a light-induced DC component of  $\mathbf{p}_\mathrm{S}$, the equilibrium inversion-symmetry is broken dynamically. This prediction   was verified experimentally by the observation of high-harmonic generation peaks centered at equilibrium-forbidden frequencies~\cite{vaswani2019discovery,yang2019lightwave}. Hybrid-Higgs collective modes also become  observable then. They give rise to distinct peaks in the THz-MDCS spectra,  located at the symmetry-forbidden Higgs mode frequency, even in the clean system. These Higgs spectral peaks   emerge with increasing light field strength~\cite{NatPhys}. Second, THz excitation of the SC system yields time-dependent deviations from equilibrium of the relative phase of the order parameter components between electron and hole pockets, $\Delta\theta(t)$. The latter leads to a change of the pseudo-magnetic field (Fig.~\ref{fig1}(d), right box) from its equilibrium orientation along $x$-direction (Fig.~\ref{fig1}(d), left box). For weak interband coupling $U$, the relative phase dynamics is known to describe a Leggett collective mode located within the excitation energy gap.
We show that, in this case, the THz-MDCS spectra  display strong  Leggett mode sidebands around the conventional pump--probe signals. This  Leggett mode, however, moves within the quasi-particle continuum when the interband interaction exceeds the intraband interaction, as in iron-based superconductors. Therefore, the Leggett mode becomes strongly damped with increasing $U$, which diminishes its contribution to the dynamics.  However, we will demonstrate below that, above critical THz driving, the photoexcitation leads to   undamped  oscillations of $\Delta \theta$ at the Higgs frequency, i.~e., at the same frequency as the amplitude oscillations. We will show that, with  strong interband interaction, the nonlinear coupling between pseudo-spin and relative phase oscillations at the Higgs frequency manifests itself in Floquet-like  sidebands located at twice the Higgs frequency.

The paper is organized as follows. In Sec.~\ref{sec:theory1} we summarize  the self-consistent gauge-invariant SC Bloch equations for multi-band superconductors. These equations of motion, derived in detail in Appendix~\ref{eq:be_full},  present the basis for our simulations.   In Sec.~\ref{sec:theory} we describe the excitations of the non-equilibrium SC state predicted by these equations in the homogeneous limit. In particular, we discuss the role of the competition between  finite condensate  momentum and long-lived relative phase oscillations. 
We first apply our theory in Sec.~\ref{sec:ps=0} to calculate the THz-MDCS signals for multi-band superconductors by neglecting the electromagnetic propagation effects. In this case, the Cooper pair momentum  vanishes after the pulse,  similar to previous calculations. We discuss the THz-MDCS spectral shape 
for vanishing interband interaction in Sec.~\ref{sec:U=0}.  These results extend the results of Ref.~\cite{Mootz2022}, obtained for strong narrowband pump and weak broadband probe pulses,
to the case of strong pulse-pair excitation. The   THz-MDCS spectral profiles for weak and strong  interband coupling $U$ are discussed in Sec.~\ref{sec:U>0}. 
Then we show in Sec.~\ref{sec:IS} how the above obtained results change drastically 
when the effects of electromagnetic pulse propagation in a thin film geometry are included in our calculation. 
In Sec.~\ref{sec:smallU} we present the results with such dynamical inversion symmetry breaking 
for small interband coupling strength. In Sec.~\ref{sec:largeU} we show how that THz-MDCS spectral profile  changes drastically when the interband coupling exceeds the intraband pairing strength with persisting finite-momentum pairing. We end with our summary in Sec.~\ref{sec:discussions}. In the Appendices, we present the general gauge-invariant  equations that include the effects of spatial fluctuations and show the equivalence of the equations of motion to the conventional pseudo-spin model in the case of homogeneous SC systems and excitation conditions. We also identify the different nonlinear processes that determine the overall 
spectral profile of THz-MDCS depending on the pulse-pair excitation conditions.

\section{Gauge-invariant description of THz-MDCS in multi-band superconductors}
\label{sec:theory1}

\subsection{Bogoliubov--de Gennes Hamiltonian}

We model  spatially-dependent superconductors by using the Bogoliubov--de Gennes Hamiltonian~\cite{Stephen1965,Wu2019}
	\begin{align}
	\label{eq:Ham}	H&=\sum_{\nu,\alpha}\int\mathrm{d}^3\mathbf{x}\,\psi_{\alpha,\nu}^\dagger(\mathbf{x})\left[\xi_\nu(\mathbf{p} - e\mathbf{A}(\mathbf{x},t))\right. \nonumber \\ &\qquad\qquad \left.+e\phi(\mathbf{x},t)+\mu^\nu_\mathrm{H}(\mathbf{x})+\mu^{\alpha,\nu}_\mathrm{F}(\mathbf{x})\right]\psi_{\alpha,\nu}(\mathbf{x}) \nonumber \\
		&-\sum_\nu\int\mathrm{d}^3\mathbf{x}\left[\Delta_\nu(\mathbf{x})\psi^\dagger_{\uparrow,\nu}(\mathbf{x})\psi^\dagger_{\downarrow,\nu}(\mathbf{x})+\mathrm{h.c.}\right]\,,
	\end{align}
 which is explicitly derived in Appendix~\ref{sec:gidma} by factorizing the full interacting Hamiltonian as in Refs.~\cite{Anderson,Nambu}.
	The fermionic field operators $\psi_{\alpha,\nu}^\dagger(\mathbf{x})$ and $\psi_{\alpha,\nu}(\mathbf{x})$ create and annihilate an electron in states labeled by the spin index $\alpha$ and  the hole ($\nu=\mathrm{h}$) or electron ($\nu=\mathrm{e}$) pocket index. The band dispersions give the kinetic energy contributions $\xi_\nu(\mathbf{p} -e\mathbf{A}(\mathbf{x},t))$, which depend on  the electron momentum operator $\mathbf{p}=-\mathrm{i}\nabla_\mathbf{x}$ ($\hbar =1$), the vector potential $\mathbf{A}(\mathbf{x},t)$, and the electron charge $-e$. The scalar potential is denoted by  $\phi(\mathbf{x},t)$. In the case of disordered SCs, we must add 
the impurity potential.
 The SC order parameter components at different pockets $\nu$  are given by
	\begin{align}
		\Delta_\nu(\mathbf{x})=-\sum_\lambda g_{\nu,\lambda}\langle \psi_{\downarrow,\lambda}(\mathbf{x})\psi_{\uparrow,\lambda}(\mathbf{x})\rangle=|\Delta_\nu(\mathbf{x})|\mathrm{e}^{-\mathrm{i}\theta_\nu(\mathbf{x})}\,,
		\label{eq:gap_eq}
	\end{align}
	where $\theta_\nu(\mathbf{x})$ are the corresponding phases and $g_{\lambda,\nu}$ describe  the  inter- ($\lambda\neq\nu$) and intra- ($\lambda=\nu$) band interactions.
	In Eq.~(\ref{eq:Ham}), we have added the Hartree and Fock contributions. 
	The Hartree contribution is  given by
	\begin{align}
		&\mu^\nu_\mathrm{H}(\mathbf{x})=\sum_\sigma\int\mathrm{d}^3\mathbf{x}'\,V(\mathbf{x}-\mathbf{x}')n_{\sigma,\nu}(\mathbf{x}')\,,\nonumber \\
	&  n_{\sigma,\nu}(\mathbf{x})=\langle\psi^\dagger_{\sigma,\nu}(\mathbf{x})\psi_{\sigma,\nu}(\mathbf{x})\rangle\,,
		\label{eq:mu_H}
	\end{align}
where $V(\mathbf{x})$ is the  Coulomb potential 
	with Fourier transformation  $V_\mathbf{q}=e^2/(\varepsilon_0 q^2)$.  This contribution moves the Nambu--Goldstone phase mode of the SC order parameters to the plasma energy  (Anderson--Higgs mechanism)~\cite{Anderson}. The Fock contribution is given by 
	\begin{equation}
		\mu^{\alpha,\nu}_\mathrm{F}(\mathbf{x})=-g_{\nu,\nu} n_{\alpha,\nu}(\mathbf{x}).
		\label{eq:mu_F}
	\end{equation}
	Its main role is that it ensures charge conservation.

\subsection{Gauge-Invariant SC Bloch equations}

To calculate the THz-MDCS spectra, we first model the THz-driven SC quantum dynamics by extending the density matrix approach of  Ref.~\cite{Mootz2020} to the multi-band case. In particular,  we  derive spatially-dependent gauge-invariant SC Bloch equations for the Wigner function $\tilde{\rho}^{(\nu)}(\mathbf{k},\mathbf{R})$, which, in addition to  the Cooper pair relative momentum $\mathbf{k}$, depends on  the center-of-mass coordinate $\mathbf{R}$. The full spatially-dependent Bloch equations are presented in Appendix~\ref{eq:be_full}. Here  we present the results for the THz--MDCS spectra profile for  sufficiently weak spatial ${\bf R}$ dependence, such that we can omit all orders of $\mathcal{O}(\nabla_\mathbf{k}\cdot\nabla_\mathbf{R})$ in the gradient expansion of the full spatially-dependent SC Bloch equations~(\ref{eq:eom-full1-f})--(\ref{eq:eom-full3-f}). 
We also assume homogeneous $E$-fields. Since the contribution of the Hartree potential to the nonlinear response is negligible in the case of weak spatial dependence, we can neglect  $\mu^\nu_\mathrm{H}$, but must keep the Fock energy $\mu^\nu_\mathrm{F}(t)$ to ensure charge conservation during  laser excitation.  

To make the connection with previous theoretical approaches used to describe the SC quantum dynamics, we express the gauge-invariant density matrix  in terms of Anderson pseudo-spin components at each wavevector $\mathbf{k}$:
\begin{align}
\label{eq:ps}
	\tilde{\rho}^{(\nu)}(\mathbf{k},\mathbf{R})=\sum_{n=0}^3 \tilde{\rho}^{(\nu)}_n(\mathbf{k})\sigma_n\,,
\end{align}
where $\sigma_n$, $n=1 \cdots 3$,  are the  Pauli spin matrices, $\sigma_0$ is the unit matrix,   $\tilde{\rho}^{(\nu)}_n(\mathbf{k})$, $n=1,\cdots, 3$,  are the  pseudo-spin components in band $\nu$ at momentum ${\bf k}$, 
and $\tilde{\rho}^{(\nu)}_0(\mathbf{k})$ describes the 
total charge.
By retaining only  the lowest term in the gradient expansion of the  spatially--dependent equations derived in Appendix~\ref{eq:be_full}, we obtain the following  gauge-invariant SC Bloch equations  describing a homogeneous system:  
\begin{align}
\label{eq:eoms-PS}
	&\frac{\partial}{\partial t}\tilde{\rho}^{(\nu)}_{0}(\mathbf{k})=-\,e\,\mathbf{E}(t)\cdot\nabla_\mathbf{k}\tilde{\rho}^{(\nu)}_{3}(\mathbf{k})\nonumber \\
	&-|\Delta_\nu|\left[\sin\delta\theta_\nu(\tilde{\rho}^{(\nu)}_{1}(\mathbf{k}_{-})-\tilde{\rho}^{(\nu)}_{1}(\mathbf{k}_{+}))\right. \nonumber \\
	&\qquad\quad\left.+\cos\delta\theta_\nu(\tilde{\rho}^{(\nu)}_{2}(\mathbf{k}_{-})-\tilde{\rho}^{(\nu)}_{2}(\mathbf{k}_{+}))\right]\,, \nonumber \\
	&\frac{\partial}{\partial t}\tilde{\rho}^{(\nu)}_{1}(\mathbf{k})=-E_\nu(\mathbf{k})\tilde{\rho}^{(\nu)}_{2}(\mathbf{k})-|\Delta_\nu|\sin\delta\theta_\nu N_\nu(\mathbf{k})\,,\nonumber \\
	&\frac{\partial}{\partial t}\tilde{\rho}^{(\nu)}_{2}(\mathbf{k})=E_\nu(\mathbf{k})\tilde{\rho}^{(\nu)}_{1}(\mathbf{k})-|\Delta_\nu|\cos\delta\theta_\nu N_\nu(\mathbf{k})\,, \nonumber\\
	&\frac{\partial}{\partial t}\tilde{\rho}^{(\nu)}_{3}(\mathbf{k})=-\,e\,\mathbf{E}(t)\cdot\nabla_\mathbf{k}\tilde{\rho}^{(\nu)}_{0}(\mathbf{k})\nonumber \\
	&-|\Delta_\nu|\left[\sin\delta\theta_\nu(\tilde{\rho}^{(\nu)}_{1}(\mathbf{k}_{-})+\tilde{\rho}^{(\nu)}_{1}(\mathbf{k}_{+}))\right. \nonumber \\
	&\qquad\quad\left.+\cos\delta\theta_\nu(\tilde{\rho}^{(\nu)}_{2}(\mathbf{k}_{+})+\tilde{\rho}^{(\nu)}_{2}(\mathbf{k}_{-}))\right]\,, 
\end{align}
where 
$\mathbf{k}_\pm=\mathbf{k}\pm\mathbf{p}_\mathrm{s}/2$. In the above equations, we introduced the phase-space filling contribution 
\begin{align}
N_\nu(\mathbf{k})=\tilde{\rho}^{(\nu)}_0(\mathbf{k}_{-})-\tilde{\rho}^{(\nu)}_3(\mathbf{k}_{-})-\tilde{\rho}^{(\nu)}_0(\mathbf{k}_{+})-\tilde{\rho}^{(\nu)}_3(\mathbf{k}_{+})\,,
\end{align}
and the time- and band-dependent energy
\begin{align}
E_\nu(\mathbf{k})=\xi_\nu(\mathbf{k}_{-})+\xi_\nu(\mathbf{k}_{+})+2(\mu_\mathrm{eff}+\mu^\nu_\mathrm{F})\,,
\label{Cooper}
\end{align}
where $\mu_{\rm{eff}}(t) 
=e\phi(\mathbf{x},t)+\frac{1}{2}\frac{\mathrm{d}}{\mathrm{d}t}\theta_{\nu_0}$ is the effective chemical potential with time-dependent contributions from the phase dynamics and any scalar potential. The Fock energy modifies this effective time-dependent chemical potential to ensure charge conservation at all times:  
\begin{align}
\mu^\nu_\mathrm{F}\equiv\frac{1}{2}\left(\mu^{\downarrow,\nu}_\mathrm{F}+\mu^{\uparrow,\nu}_\mathrm{F}\right)=-g_{\nu,\nu}\sum_\mathbf{k}\left[1+\tilde{\rho}^{(\nu)}_{3}(\mathbf{k})\right].
\end{align}
The order parameter amplitude 
\begin{align} 
&|\Delta_\nu| \nonumber=-\mathrm{e}^{-\mathrm{i}\,\delta\theta_\nu}\sum_{\lambda,\mathbf{k}} g_{\nu,\lambda}\left[\tilde{\rho}_{1}^{(\lambda)}(\mathbf{k})-\mathrm{i}\tilde{\rho}_{2}^{(\lambda)}(\mathbf{k})\right] \nonumber \\
&=-\sum_{\lambda,\mathbf{k}} g_{\nu,\lambda}\left[\cos\delta\theta_\nu\tilde{\rho}_{1}^{(\lambda)}(\mathbf{k})-\sin\delta\theta_\nu\tilde{\rho}_{2}^{(\lambda)}(\mathbf{k})\right]
\end{align}
remains real-valued in the numerical calculation at all times and fields. 
The above equations in the homogeneous limit provide a starting point to add weak disorder effects through a Born approximation treatment of the impurity potential. Disorder effects will enhance the collective mode contributions predicted here to  result from a persisting inversion-symmetry breaking controlled by light. 
In the case of strong spatial fluctuations (e.g., dirty limit SC), a more appropriate starting point 
are the equations of motion for $\tilde{\rho}^{(\nu)}(\mathbf{r},\mathbf{R})$
obtained from Eqs.~(\ref{eq:eom-full1b})--(\ref{eq:eom-full3b})  by Fourier transform.

The SC Bloch equations~(\ref{eq:eoms-PS}) include four  different dynamical sources that drive
the Anderson pseudo-spins:  (i) The light-induced condensate center-of-mass momentum $\mathbf{p}_\mathrm{S}$  differs from the laser vector potential ${\bf A}(t)$ considered in previous works due to electromagnetic propagation effects.   In addition to its  oscillatory contribution during photoexcitation, it also exhibits a static component, which is generated by difference-frequency Raman processes~\cite{Mootz2022}. This DC momentum  
 remains finite after the pulse and decays slowly in time,  leading to a non-equilibrium condensate state with finite momentum Cooper  pairing  (Fig.~\ref{fig1}(c)); (ii) The effective  chemical potential $\mu_\mathrm{eff}(t)$, whose time dependence is determined by the  non-equilibrium order parameter phase in reference band $\nu_0$:
\begin{align}
&\frac{d \theta_{\nu_0}}{d t}= \nonumber \\
&-2\,e\,\phi+\frac{1}{|\Delta_{\nu_0}|}\sum_{\nu,\mathbf{k}} g_{\nu_0,\nu}\left[\xi_\nu(\mathbf{k}_{-})+\xi_\nu(\mathbf{k}_{+})+2\mu^\nu_\mathrm{F}\right]\tilde{\rho}_{1}^{(\nu)}(\mathbf{k}) \nonumber \\ &+\frac{1}{|\Delta_{\nu_0}|}\sum_{\nu,\mathbf{k}} g_{\nu_0,\nu}|\Delta_\nu|N_{\nu}(\mathbf{k})\cos(\delta\theta_\nu)\,;
\end{align}
(iii) The time-dependent phase difference of the order parameters in different bands $\nu$, $\delta\theta_\nu=\theta_{\nu_0}-\theta_\nu$, is given by the equation of motion
\begin{align}
&\frac{d\delta\theta_\nu}{d t}= \frac{1}{|\Delta_\nu|}\sum_{\lambda,\mathbf{k}}g_{\nu,\lambda}\left[\left(\xi_\nu(\mathbf{k}_{-})+\xi_\nu(\mathbf{k}_{+})+2\mu^\nu_\mathrm{F}\right)\right.\nonumber \\ &\quad\qquad\left.\times\left(\cos\delta\theta_\nu\tilde{\rho}^{(\lambda)}_1(\mathbf{k})-\sin\delta\theta_\nu\tilde{\rho}^{(\lambda)}_2(\mathbf{k})\right)\right. \nonumber \\
&\qquad\qquad\left. -|\Delta_\lambda|N_\nu(\mathbf{k})\cos(\delta\theta_\nu-\delta\theta_\lambda)\right]\,.
\label{eq:dtheta}
\end{align}
This relative phase dynamics follows from the constraint imposed by  the gauge invariance,   
\begin{align}
\sum_{\lambda,\mathbf{k}} g_{\nu,\lambda}\left[\sin\delta\theta_\nu\tilde{\rho}_{1}^{(\lambda)}(\mathbf{k})+\cos\delta\theta_\nu \tilde{\rho}_{2}^{(\lambda)}(\mathbf{k})\right]=0\,. \label{gauge-inv}
\end{align}
This constraint is satisfied  exactly at all times and for any strong driving field;  (iv) The 
order parameter amplitudes $|\Delta_\nu(t)|$ in each band, which are coherently quenched during cycles of THz light-field oscillations via difference-frequency Raman processes~\cite{Mootz2022}.

To model the THz-MDCS spectra measured in the experiments, we express  the gauge-invariant supercurrent density in terms of $\tilde{\rho}_{0}^{(\lambda)}(\mathbf{k})$ as 
\begin{align}
J(t)=\frac{2e}{V}\sum_{\mathbf{k},\lambda}\nabla_\mathbf{k}\xi_\lambda(\mathbf{k})\tilde{\rho}_{0}^{(\lambda)}(\mathbf{k})\,,
\label{eq:current2}
\end{align}
with normalization volume $V$. The measured nonlinear differential transmission  follows  from the transmitted electric field  obtained by solving Maxwell's equations as described in Ref.~\cite{Mootz2020}.
For a thin film geometry, we can neglect the spatial dependence, so the effective field becomes
\begin{align}
\label{eq:trans}
E(t) = E_\mathrm{THz}(t)-\frac{\mu_0\,c\,d}{2n}J(t)\,. 
\end{align}
Here, $E_\mathrm{THz}(t)$ corresponds to the applied THz laser electric field, while $n$ denotes the refractive index of the SC system, $d$ is the thickness of the SC thin film, and $c$ is the speed of light. The dynamics of the current density~(\ref{eq:current2}) 
in linear response can be described by the London equation, $\partial J(t)/\partial t=n_\mathrm{s}e^2/m E(t)$, where $n_\mathrm{s}$ is the superfluid density. Using Eq.~(\ref{eq:trans}), one obtains
\begin{align}
   \frac{\partial J(t)}{\partial t}=\frac{n_\mathrm{s}e^2}{m}E_\mathrm{THz}(t)-\frac{J(t)}{\tau}\,.
\label{eq:dJt}
\end{align}
The current density decays due to radiative damping with lifetime
\begin{align}
    \tau=\frac{2nm}{\mu_0 c n_\mathrm{s}e^2 d}\,,
    \label{eq:tau}
\end{align}
where 
\begin{align}
    n_\mathrm{s}=\frac{4m}{\hbar^2 V}\sum_{\nu,\mathbf{k}}\nabla_\mathbf{k}\xi_\nu(\mathbf{k})\nabla_\mathbf{k}\tilde{\rho}_3^{(\nu),0}(\mathbf{k})\,
\end{align}
is superfluid density and $\tilde{\rho}_3^{(\nu),0}(\mathbf{k})$ is the equilibrium $z$-component of the pseudo-spin.

Following Refs.~\cite{Mootz2020,Mootz2022},
the correlated signal measured in THz-MDCS experiments is given by 
\begin{align}
E_\mathrm{NL}(t,\tau)=E_\mathrm{AB}(t,\tau)-E_\mathrm{A}(t)-E_\mathrm{B}(t,\tau)\,.
\label{eq:Enl}
\end{align}
This expression applies to the case of the collinear 2-pulse geometry considered in this paper (Fig.~\ref{fig1}(a)). Here, $E_\mathrm{AB}(t,\tau)$ is the transmitted $E$-field induced by {\em both} pulses A and B, which depends on both the real time $t$ and the delay time between the two pulses, $\tau$.  $E_\mathrm{A}(t)$ ($E_\mathrm{B}(t,\tau)$) is the transmitted electric field induced by pulse A (B). The THz-MDCS spectra are obtained by Fourier transform of  $E_\mathrm{NL}(t,\tau)$ with respect to both $t$ (frequency $\omega_t$) and $\tau$ (frequency $\omega_\tau$). Equation~(\ref{eq:trans}) also provides the driving field 
of   the SC Bloch equations~(\ref{eq:eoms-PS}), whose solution determines 
the current density according to Eq.~(\ref{eq:current2}). The dependence of the driving  field on the current density leads to a {\em self-consistent calculation},  whose results  differ from the calculation with driving field given by the two laser pulses.

In this paper, we solve the full Bloch equations~(\ref{eq:eoms-PS}) for a 3-pocket  model bandstructure,  which includes a hole (h) pocket centered at the $\Gamma$-point and two electron (e) pockets located at $(\pi,0)$ and $(0,\pi)$ (Fig.~\ref{fig1}(b)). We include  inter-pocket $e$-$h$ couplings between the hole and the two electron pockets ($g_\mathrm{e,h}= g_\mathrm{h,e}$),  as well as intra-pocket pairing interactions  ($V_\lambda=g_{\lambda,\lambda}$, $\lambda =e, h$). We neglect the inter-electron pocket  interactions for simplicity and calculate the THz-MDCS spectra  as a function of the interband-to-intraband interaction ratio  $U=g_\mathrm{e,h}/V_\mathrm{h}$.
The electron and hole band energies are described by using the square lattice nearest-neighbor tight-binding dispersion $\xi_\nu(\mathbf{k})=-2\,[J_{\nu,x}\mathrm{cos}(k_x a)+J_{\nu,y}\mathrm{cos}(k_y a)]+\mu_\nu$, with hopping parameters $J_{\nu,i}$, band-offset $\mu_\nu$, and lattice constant $a$. We choose a circular hole pocket with $J_{h,x}=J_{h,y}=25.0$\,~meV and $\mu_\mathrm{h}=-15.0$\,~meV. We  introduce a particle-hole asymmetry between electron and hole pockets~\cite{Liu2010,Fernandes2010,yangPRL} by considering elliptical electron pockets with $J_{e,x}=-25.0$\,~meV, $J_{e,y}=-50.0$\,~meV, and $\mu_e=15.0$\,~meV. Such asymmetry strongly suppresses the higher band Higgs mode in our calculated spectra,  as discussed in Ref.~\cite{hybrid-higgs}. As lattice constant we choose $a=4.0$~\AA~while  we consider a sample thickness of $d=20$~nm which is a typical thickness of superconductor thin films used in THz spectroscopy experiments~\cite{yang2019lightwave}. We assume 
equilibrium SC order parameters $\Delta_h=4.1$\,meV for the hole pocket and $\Delta_e=8.2$\,meV for the electron pockets. We consider an excitation protocol where the multi-band SC system is excited by a pulse-pair consisting of two equal few-cycle broadband pulses with well-defined relative phase controlled  by the time delay $\tau$ and central laser frequency $\omega_0=1$~THz.    The  two  electric fields used in the calculation  are  $\mathbf{E}_\mathrm{A}(t^\prime) \sin(\omega_\mathrm{A} t^\prime)$ and $\mathbf{E}_\mathrm{B}(t^\prime) \sin(\omega_\mathrm{B} t^\prime)$ with Gaussian envelope functions $\mathbf{E}_\mathrm{A,B}(t^\prime)$.
Similar to previous studies in  semiconductors~\cite{Kuehn2009,Kuehn2011,Junginger2012,maag2016}, we introduced ``time vectors", $t^\prime=(t,\tau)$ and ``frequency vectors", $(\omega_t,\omega_\tau)$ . 
The  frequency vectors of the two pulses A and B are then  $\omega_\mathrm{A}=(\omega_0,0)$ and $\omega_\mathrm{B}=(\omega_0,-\omega_0)$.

\section{Excitations of the Driven Superconductor }

\label{sec:theory}

Prior to presenting our numerical results, we first identify the main drivers of nonlinearity that  determine the most striking  features of the  THz-MDCS spectra of SCs. For this purpose, we  extract from the full SC Bloch equations {\em pseudo-spin nonlinear coupled oscillator} equations of motion, obtained by 
extending Anderson's random phase approximation  treatment of  pseudo-spin flips and Higgs collective modes~\cite{Anderson} to multi-band SCs without  linearization.  We characterize the non-equilibrium state by the  density matrix $\tilde{\rho}^{(\nu)}({\bf k})$, which we  decompose  as 
\begin{align}
\label{eq:rho-dev}
	\quad \tilde{\rho}^{(\nu)}(\mathbf{k}) =  \tilde{\rho}^{(\nu),\mathrm{0}}(\mathbf{k}) +\Delta\tilde{\rho}^{(\nu)}(\mathbf{k})\,,
\end{align} 
where $\tilde{\rho}^{(\nu),\mathrm{0}}(\mathbf{k})$ denotes the density matrix of  the  thermal state, where $\partial_t \tilde{\rho}^{(\nu),\mathrm{0}}(\mathbf{k})=0$, and 
$\Delta\tilde{\rho}^{(\nu)}(\mathbf{k})$ describes the  full non-thermal change induced by the driving field. 
We then derive nonlinear pseudo-spin oscillator equations for the $x$ and $y$ pseudo-spin component deviations $\Delta\tilde{\rho}^{(\nu)}_{1,2}(\mathbf{k})$ by taking the second time derivatives of Eqs.~(\ref{eq:eoms-PS}). 
Substituting 
the decomposition Eq.~(\ref{eq:rho-dev})
in the  resulting coupled nonlinear equations, we obtain 
driven coupled oscillator equations of motion that 
describe the non-thermal deviations $\Delta \tilde{\rho}^{(\nu)}_{1,2}(\mathbf{k})$ of the transverse pseudo-spin
 components along the $x$ and $y$ axes:
\begin{align}
&	\partial_t^2 \,
	\Delta \tilde{\rho}^{(\nu)}_1(\mathbf{k}) + \left[ E^2_\nu(\mathbf{k}) 
	+4 |\Delta_\nu|^2 \, \sin^2 \Delta \theta_\nu 
	 \right]
	\, 
	\Delta \tilde{\rho}^{(\nu)}_1(\mathbf{k})
	\nonumber \\
	&+ \left[
	\partial_t E_\nu(\mathbf{k})
	+
	2 |\Delta_\nu|^2 \, \sin 2 \Delta \theta_\nu \right] \,
\Delta\tilde{\rho}_2^{(\nu)}(\mathbf{k})
	 \nonumber \\
	&=S^{(1)}_\nu(\mathbf{k})-\left[\partial_t \delta \Delta^\dprime_\nu - 
	\delta  \Delta^\prime_\nu  E_\nu(\mathbf{k})  \right] 
	 N_\nu(\mathbf{k})   \,, \nonumber \\
&	\partial_t^2 \,
	\Delta \tilde{\rho}^{(\nu)}_2(\mathbf{k}) + \left[ E^2_\nu(\mathbf{k}) +
	4 |\Delta_\nu|^2 \, \cos^2 \Delta \theta_\nu 
	\right] \Delta \tilde{\rho}^{(\nu)}_2(\mathbf{k}) 
	\nonumber \\
	&+
\left[ -\partial_t E_\nu(\mathbf{k}) +
	2 |\Delta_\nu|^2 \, \sin 2 \Delta \theta_\nu \right] \, 
\Delta	 \tilde{\rho}_1^{(\nu)}(\mathbf{k})
	\nonumber \\
	&=S^{(2)}_\nu(\mathbf{k})-
	\left[\partial_t \delta \Delta^\prime_\nu + 
	\delta  \Delta^\dprime_\nu  E_\nu(\mathbf{k})  \right] 
	 N_\nu(\mathbf{k})\,,
	\label{eq:eom_t2}
\end{align} 
where $\Delta\theta_\nu(t)
 =\delta\theta_\nu-\delta\theta^0_\nu$ describes the canting of the pseudo-spins away from their  equilibrium directions. The latter directions are determined  by $\delta\theta^0_\nu=0, \pi$ ($\delta\theta^0_\nu=0$) for $s_{\pm}$ ($s_{++}$) order parameter symmetry. In Eq.~(\ref{eq:eom_t2}) we defined  real and imaginary parts of the SC order parameters, $\Delta^\prime_\nu$ and $\Delta^\dprime_\nu$ respectively, as 
\begin{align} 
&\Delta^\prime_\nu=
|\Delta_{\nu}| \cos \Delta \theta_\nu
=-\sum_{\lambda,\mathbf{k}} g_{\nu,\lambda} \ \tilde{\rho}_{1}^{(\lambda)}(\mathbf{k})\,,
\nonumber \\
&\Delta^\dprime_\nu=
|\Delta_{\nu}| \sin \Delta \theta_\nu 
=\sum_{\lambda,\mathbf{k}} g_{\nu,\lambda}\tilde{\rho}_{2}^{(\lambda)}(\mathbf{k})\,.
\label{D12}
\end{align} 
Equation~(\ref{eq:eom_t2}) describes  two coupled driven oscillators. The two terms on the right-hand side (rhs) of Eqs.~(\ref{eq:eom_t2}) are the  driving forces, 
which drive the oscillators at frequencies $\sim 2 \omega_0$  as well as  at  the SC  elementary excitation frequencies. 
The  first terms  on the rhs, $S^{(1,2)}_\nu(\mathbf{k})$, come from sum- and difference-frequency Raman as well as from quantum transport processes~\cite{Mootz2022}:
\begin{align}
S^{(1)}_\nu(\mathbf{k})&= -
	\left[ \Delta E^2_\nu(\mathbf{k}) 
	+4 |\Delta_\nu|^2 \sin^2 \Delta \theta_\nu  \right]
	\, \tilde{\rho}_1^{(\nu), 0}(\mathbf{k})
	\nonumber \\
	&+\Delta \left[
	 N_\nu(\mathbf{k})
	E_\nu(\mathbf{k})) \right] |\Delta^{0}_\nu| 
	\nonumber \\
	& 
	- |\Delta_\nu| \sin \Delta \theta_\nu  \left[e\,\mathbf{E}\cdot\nabla_{\mathbf{k}_{+}}(\tilde{\rho}_0^{(\nu)}(\mathbf{k}_{+})+\tilde{\rho}_3^{(\nu)}(\mathbf{k}_{+}))\right. \nonumber \\
	&\left.+e\,\mathbf{E}\cdot\nabla_{\mathbf{k}_{-}}(\tilde{\rho}_0^{(\nu)}(\mathbf{k}_{-})-\tilde{\rho}_3^{(\nu)}(\mathbf{k}_{-}))\right]\,,
	\nonumber \\
 S^{(2)}_\nu(\mathbf{k})&=\left[ \partial_t E_\nu(\mathbf{k})
-2 |\Delta_\nu|^2 \, \sin 2 \Delta \theta_\nu  \right]
\, \tilde{\rho}_1^{(\nu), 0}(\mathbf{k})
\nonumber \\
& -|\Delta_\nu| \cos \Delta \theta_\nu  \left[e\,\mathbf{E}\cdot\nabla_{\mathbf{k}_{+}}(\tilde{\rho}_0^{(\nu)}(\mathbf{k}_{+})+\tilde{\rho}_3^{(\nu)}(\mathbf{k}_{+}))\right. \nonumber \\
&\left.+e\,\mathbf{E}\cdot\nabla_{\mathbf{k}_{-}}(\tilde{\rho}_0^{(\nu)}(\mathbf{k}_{-})-\tilde{\rho}_3^{(\nu)}(\mathbf{k}_{-}))\right]\,.
	\label{eq:source}
\end{align} 
The  second terms on the rhs of Eq.~(\ref{eq:eom_t2}),  proportional to the light-induced deviations $ \delta \Delta_1^{(\nu)}$ and $\delta  \Delta_2^{(\nu)}$  of the 
order parameters from their equilibrium values,  describe the collective  effects, including the Higgs and Leggett collective modes  generated by the coupling of all $\mathbf{k}$–point pseudo-spins. 
The excitation energies  are given by the left-hand side of Eq.~(\ref{eq:eom_t2}), 
which determines  the frequencies of the coupled $x$ and $y$ pseudo-spin oscillators.  These frequencies describe pseudo-spin precession around the time-dependent magnetic field 
\begin{equation} 
\mathbf{B}_{\mathbf{k},\nu}(t) =
\begin{bmatrix}
          -2 |\Delta_{\nu}| \cos \Delta \theta_\nu \\
           2 |\Delta_{\nu}| \sin \Delta \theta_\nu\\
E_\nu(\mathbf{k})       
         \end{bmatrix}\,.
         \label{B}
\end{equation}
The transverse $x$ and $y$ components of this magnetic field are given by the real and imaginary part of the order parameters, 
Eq.~(\ref{D12}). In particular, the $y$-component is generated by  light-induced  relative phase oscillations,
$\Delta \theta_\nu (t) \ne 0$. On the other hand, 
 the time-dependence of the longitudinal magnetic field $z$-component is determined by the condensate momentum, with additional contributions from the time-dependent effective chemical potential. The latter, however, does not play an important role for our results here. 
The oscillator frequencies  are determined by $\sqrt{[B^z_{\mathbf{k},\nu}(t)]^2 + [B^y_{\mathbf{k},\nu}(t)]^2}$ ($\Delta \tilde{\rho}^{(\nu)}_1$) 
and  by $\sqrt{[B^z_{\mathbf{k},\nu}(t)]^2 + [B^x_{\mathbf{k},\nu}(t)]^2}$ ($\Delta \tilde{\rho}^{(\nu)}_2$). Importantly, they depend on the time-dependent coupling between  the two transverse components, given by 
$\pm \partial_t B^z_{\mathbf{k},\nu}(t) + B^x_{\mathbf{k},\nu}(t) B^y_{\mathbf{k},\nu}(t)$.  The latter transverse  light-induced  coupling does not contribute to the third-order response. It becomes important  in multi-band SCs when enhanced by 
 the dynamics of the phase difference between electron and hole pockets, $\Delta\theta_\nu(t)$. As we show below, in the case of  strong interband coupling exceeding the intraband interaction, the product  $B^x_{\mathbf{k},\nu}(t) B^y_{\mathbf{k},\nu}(t)$ is enhanced  by a light-induced relative-phase collective mode of the  non-equilibrium SC state. Unlike for the strongly damped Leggett mode, this phase mode at the Higgs frequency  is underdamped. The resulting long-lived  superconductivity state is characterized by second harmonic (Floquet) sidebands at bi-Higgs frequencies in the THz-MDCS spectra, discussed in Sec.~\ref{sec:largeU}. 

To elaborate further, the  above non-equilibrium state is driven by the time-dependent coupling term 
\begin{align}
&\left[ \partial_t E_\nu(\mathbf{k}) -
2|\Delta_\nu|^2  \sin 2\Delta\theta_\nu(t) \right] \Delta \mathrm{\rho}_1^{(\nu)}\nonumber \\
&=\left[ e\,({\bf E}(t) \cdot \nabla_{{\bf k}}) ({\bf p_{\rm{S}}} \cdot \nabla_{{\bf k}}) \, \xi_\nu({\bf k}) \right. \nonumber \\ &\left. -
2|\Delta_\nu|^2  \sin 2\Delta\theta_\nu(t) \right] \Delta \mathrm{\rho}_1^{(\nu)}+\mathcal{O}(\mathbf{p}_\mathrm{S}^2)\,,
\label{coupling}
\end{align}
where we expanded the band dispersions in powers of the center-of-mass momentum $\mathbf{p}_\mathrm{S}(t)$ in the last step. The time-dependence of the first term of Eq.~(\ref{coupling}) 
drives pseudo-spins via difference-frequency Raman  processes $\omega_\mathrm{A,B}-\omega_\mathrm{A,B} \sim 0 $ and sum-frequency  Raman  processes $\omega_\mathrm{A,B}+\omega_\mathrm{A,B} \sim 2 \omega_0$. Here, $\omega_0$ is the center frequency of the applied two laser pulses, which are denoted by A and B. This term contributes beyond  the previously studied third-order nonlinear response and    
leads to the formation of sidebands at Leggett- and Higgs-mode energies in the THz-MDCS spectra, discussed in Secs.~\ref{sec:U=0} and \ref{sec:U>0}. The second term of Eq.~(\ref{coupling}) describes parametric driving of pseudo-spins by the amplitude and phase collective mode oscillations  of  the driven system. As we show in Sec.~\ref{sec:largeU},  second harmonic sidebands at bi-Higgs frequency $\sim 2 \omega_\mathrm{H,h}$ thus emerge in the THz-MDCS spectra above critical driving, but only for multi-band SC systems with strong interband coupling exceeding intraband interaction. In general, the two  different processes described by  Eq.~(\ref{coupling}) are competing.  The result of this competition depends on the electric field strength of the applied laser pulses, as well as on  the DC current  induced by electromagnetic pulse propagation.  In Sec.~\ref{sec:largeU} we discuss the experimental  signatures of the two different processes.

To  illustrate further the light-induced  $\Delta\tilde{\rho}^{(\nu)}(\mathbf{k})$
 driven by $\Delta \theta_\nu(t)$, we may transform the equations of motion~(\ref{eq:eoms-PS}) to the rotated frame
 defined by the angle $\Delta\theta_\nu(t)$:
\begin{align}
&\mathrm{P}_1^{(\nu)}(\mathbf{k})=\tilde{\rho}^{(\nu)}_1(\mathbf{k})\cos\Delta\theta_\nu-\tilde{\rho}^{(\nu)}_2(\mathbf{k})\sin\Delta\theta_\nu\,,\nonumber \\
&\mathrm{P}_2^{(\nu)}(\mathbf{k})=\tilde{\rho}^{(\nu)}_1(\mathbf{k})\sin\Delta\theta_\nu+\tilde{\rho}^{(\nu)}_2(\mathbf{k})\cos\Delta\theta_\nu'\,.
\end{align}
In this rotating frame, the SC Bloch equations  take the form
\begin{align}
	&\frac{\partial}{\partial t}\tilde{\rho}^{(\nu)}_{0}(\mathbf{k})=-\,e\,\mathbf{E}(t)\cdot\nabla_\mathbf{k}\tilde{\rho}^{(\nu)}_{3}(\mathbf{k}) \nonumber \\
	&\qquad\qquad-|\Delta_\nu|\left[\mathrm{P}^{(\nu)}_{2}(\mathbf{k}_{-})-\mathrm{P}^{(\nu)}_{2}(\mathbf{k}_{+})\right]\,, \nonumber \\
	&\frac{\partial}{\partial t}\mathrm{P}^{(\nu)}_{1}(\mathbf{k})=-(E_\nu(\mathbf{k})+\partial_t\Delta\theta_\nu)\mathrm{P}^{(\nu)}_{2}(\mathbf{k})\,,\nonumber \\
	&\frac{\partial}{\partial t}\mathrm{P}^{(\nu)}_{2}(\mathbf{k})=(E_\nu(\mathbf{k})+\partial_t\Delta\theta_\nu)\mathrm{P}^{(\nu)}_{1}(\mathbf{k})-|\Delta_\nu|N_\nu(\mathbf{k})\,, \nonumber\\
	&\frac{\partial}{\partial t}\tilde{\rho}^{(\nu)}_{3}(\mathbf{k})=-\,e\,\mathbf{E}(t)\cdot\nabla_\mathbf{k}\tilde{\rho}^{(\nu)}_{0}(\mathbf{k})\nonumber \\
	&\qquad\qquad-|\Delta_\nu|\left[\mathrm{P}_2^{(\nu)}(\mathbf{k}_{-})+\mathrm{P}^{(\nu)}_2(\mathbf{k}_{+})\right]\,.
	\label{eq:eom_trans}
\end{align}
Analogously to the discussion above, we take the second derivative of Eq.~(\ref{eq:eom_trans}) and obtain
\begin{align}
& \partial_t^2\ \Delta \mathrm{P}_2^{(\nu)}(\mathbf{k})+\left[ \left(E_\nu(\mathbf{k})+\partial_t\Delta\theta_\nu\right)^2+4|\Delta_\nu|^2 \right] \Delta \mathrm{P}_2^{(\nu)}(\mathbf{k}) \nonumber \\
&-\left[\partial_t E_\nu(\mathbf{k})+\partial^2_t\Delta\theta_\nu\right]
\ \Delta \mathrm{P}_1^{(\nu)}(\mathbf{k}) 
=\nonumber \
S_\nu(\mathbf{k})+ \partial_t | \Delta_\nu| N_\nu(\mathbf{k}), \nonumber \\
& \partial_t \Delta \mathrm{P}^{(\nu)}_{1}(\mathbf{k})=-(E_\nu(\mathbf{k})+\partial_t\Delta\theta_\nu) \, \Delta \mathrm{P}^{(\nu)}_{2}(\mathbf{k})\,,
\label{eq:eom2_trans}
\end{align}
with driving source term 
\begin{align}
S_\nu(\mathbf{k})& = \ \left[\partial_t E_\nu(\mathbf{k})+\partial^2_t\Delta\theta_\nu\right]
\ \mathrm{P}_1^{(\nu),0}(\mathbf{k}) \nonumber \\
&-|\Delta_\nu|\left[e\,\mathbf{E}\cdot\nabla_{\mathbf{k}_{+}}(\tilde{\rho}_0^{(\nu)}(\mathbf{k}_{+})+\tilde{\rho}_3^{(\nu)}(\mathbf{k}_{+}))\right. \nonumber \\
&\left.+e\,\mathbf{E}\cdot\nabla_{\mathbf{k}_{-}}(\tilde{\rho}_0^{(\nu)}(\mathbf{k}_{-})-\tilde{\rho}_3^{(\nu)}(\mathbf{k}_{-}))\right]\,.
	\label{eq:source2}
\end{align} 
The nonlinear oscillator equations (\ref{eq:eom2_trans}) are formally equivalent to the parametric oscillator equations of an one-band superconductor derived in Ref.~\cite{Mootz2022}. In multi-band superconductors with strong interband coupling, the pseudo-spin
deviations from equilibrium, determined by the time-dependent rotating frame angle $\Delta\theta(t)$,  generates high-order sidebands in the THz-MDCS spectra at high driving fields. These sidebands  are analogous to Floquet sidebands, as demonstrated in Sec.~\ref{sec:U>0}. 

According to Eqs.~(\ref{eq:current2}) and (\ref{eq:Enl}), the THz-MDCS nonlinear signal $E_\mathrm{NL}$ follows from the dynamics of 
$\tilde{\rho}^{(\nu),\mathrm{NL}}_{0}(\mathbf{k})=\tilde{\rho}^\mathrm{(\nu),AB}_0(\mathbf{k})-\tilde{\rho}^\mathrm{(\nu),A}_{0}(\mathbf{k})-\tilde{\rho}^\mathrm{(\nu),B}_{0}(\mathbf{k})$.  Here, $\tilde{\rho}^\mathrm{(\nu),AB}(\mathbf{k})$ denotes the gauge-invariant  density matrix of the non-equilibrium state driven by both pulses.  $\tilde{\rho}^\mathrm{(\nu),A}(\mathbf{k})$ ($\tilde{\rho}^\mathrm{(\nu),B}(\mathbf{k})$) is the gauge-invariant density matrix of the non-equilibrium state driven by pulse A (B). For interpreting the results of the full numerical calculation, it is useful to  distinguish between nonlinear  processes due to the excitations by a single pulse that are sensed by the other pulse and the excitations generated by {\em both} pulses simultaneously. These two different processes lead to different experimental features. For this, we decompose the density matrix $\tilde{\rho}^\mathrm{(\nu),AB}(\mathbf{k})$ as
\begin{align}
\label{eq:rho-dev3}
	\tilde{\rho}^\mathrm{(\nu),AB}(\mathbf{k}) =  \tilde{\rho}^\mathrm{(\nu),A}(\mathbf{k}) +\tilde{\rho}^\mathrm{(\nu),B}(\mathbf{k})+\Delta\tilde{\rho}^\mathrm{(\nu),AB}(\mathbf{k})\,,
\end{align} 
where $\Delta\tilde{\rho}^\mathrm{(\nu),AB}(\mathbf{k})$ is generated by interference between SC excitations of both pulses A and B. 
Inserting Eq.~(\ref{eq:rho-dev3}) into the equation of motion for $\tilde{\rho}^{(\nu),\mathrm{NL}}_{0}(\mathbf{k})$ leads to
\begin{align}
\label{eq:delta_rho0}
&\frac{\partial}{\partial t}\tilde{\rho}^{(\nu),\mathrm{NL}}_{0}(\mathbf{k})=-\,e\,\mathbf{E}_\mathrm{B}(t)\cdot\nabla_\mathbf{k}\tilde{\rho}^{(\nu),\mathrm{A}}_{3}(\mathbf{k}) \nonumber \\
&-|\Delta^\mathrm{A}_\nu|\nabla_\mathbf{k}\left(\tilde{\rho}^{(\nu),\mathrm{A}}_{2}(\mathbf{k})+\Delta\theta_\mathrm{A}\tilde{\rho}^{(\nu),\mathrm{A}}_{1}(\mathbf{k})\right)\,\mathbf{p}_\mathrm{S}^\mathrm{B} \nonumber \\
&-|\Delta^\mathrm{A}_\nu|\nabla_\mathbf{k}\left(\tilde{\rho}^{(\nu),\mathrm{B}}_{2}(\mathbf{k})+\Delta\theta_\mathrm{A}\tilde{\rho}^{(\nu),\mathrm{B}}_{1}(\mathbf{k})\right)\,\mathbf{p}_\mathrm{S}^\mathrm{B} \nonumber \\
&-|\Delta^\mathrm{A}_\nu|\nabla_\mathbf{k}\left(\tilde{\rho}^{(\nu),\mathrm{B}}_{2}(\mathbf{k})+\Delta\theta_\mathrm{A}\tilde{\rho}^{(\nu),\mathrm{B}}_{1}(\mathbf{k})\right)\,\mathbf{p}_\mathrm{S}^\mathrm{A}
\nonumber \\
&-|\Delta^\mathrm{A}_\nu|\Delta\theta_\mathrm{B}\nabla_\mathbf{k}\left(\tilde{\rho}^{(\nu),\mathrm{A}}_{1}(\mathbf{k})\,\mathbf{p}_\mathrm{S}^\mathrm{B}+\tilde{\rho}^{(\nu),\mathrm{B}}_{1}(\mathbf{k})\,\mathbf{p}_\mathrm{S}^\mathrm{A}\right. \nonumber \\
&\left.\qquad\qquad\qquad +\tilde{\rho}^{(\nu),\mathrm{A}}_{1}(\mathbf{k})\,\mathbf{p}_\mathrm{S}^\mathrm{A}+\tilde{\rho}^{(\nu),\mathrm{B}}_{1}(\mathbf{k})\,\mathbf{p}_\mathrm{S}^\mathrm{B}\right)
\nonumber \\ &+\Delta\tilde{\rho}^\mathrm{(\nu),AB}+\mathcal{O}(\mathrm{p}_\mathrm{S}^2)+\mathcal{O}((\Delta\theta)^2)+\mathrm{A}\leftrightarrow\mathrm{B}\,,
\end{align} 
where we expanded in terms of $\mathbf{p}_\mathrm{S}$ and $\Delta\theta$ and neglected all contributions of order $\mathcal{O}(\mathbf{p}_\mathrm{S}^2)$ and $\mathcal{O}((\Delta\theta)^2)$. 
The terms in the first two lines of Eq.~(\ref{eq:delta_rho0})  contribute to the nonlinear response when  pulse B arrives after  pulse A to sense the SC excitations by pulse A, as in conventional pump--probe spectroscopy.
This signal exists even if the two pulses do not overlap in time. The corresponding contribution to the THz-MDCS spectra shows up in our results  as harmonic sidebands ($\pm \omega_0$) around the quasi-particle, Higgs and Legget mode frequencies.  
The THz-MDCS spectral peaks at high driving fields are mainly generated by the terms in the third line on the rhs of Eq.~(\ref{eq:delta_rho0}), which dominate over the other terms in the non-perturbative excitation regime. Here,  the observed signals only appear along $(\omega_t,0)$, i.~e.,  along the frequency vector of pulse A, $\omega_\mathrm{A}=(\omega_0,0)$, and along $(\omega_t,-\omega_t)$, i.~e., along the frequency vector of pulse B, $\omega_\mathrm{B}=(\omega_0,-\omega_0)$.  The  $\Delta\tilde{\rho}^\mathrm{(\nu),AB}$ contributions result from the interference between SC excitations of pulses A and B  and lead to  the correlated wave-mixing peaks  discussed in Ref.~\cite{Mootz2022}.
The latter new signals  arise from parametric driving of superconductivity by the pump--probe coherent modulation of the order parameter by {\em both} pulses A and B. 
In this paper, we consider a different pulse excitation scheme from Ref.~\cite{Mootz2022}, with two strong few-cycle pulses of similar {\em broad} spectral  shape. For this excitation protocol, the $\Delta\tilde{\rho}^\mathrm{(\nu),AB}(\mathbf{k})$ interference terms only contribute to the THz-MDCS spectra at low fields. These $\Delta\tilde{\rho}^\mathrm{(\nu),AB}$ contributions are discussed in more detail in Appendix~\ref{sec:pp}.

\section{THz-MDCS  without persistent symmetry breaking}

\label{sec:ps=0}

To demonstrate the importance of the electromagnetic propagation effects in determining the effective driving field,   we first consider in this section a driving field given by  the laser pulse $E_\mathrm{THz}(t)$, instead of   
 the effective  local  pulse Eq.~(\ref{eq:trans}).  The Cooper pair momentum 
 $\mathbf{p}_\mathrm{S}(t)$ then vanishes after the pulse, as in previous   theories. 
  First we present the spectra as a function of the driving pulse-pair strength with interband interaction set to zero. Then, we show how these spectra  change for finite interband interaction.

\subsection{Zero  interband interaction}
\label{sec:U=0}

\begin{figure}[t!]
\begin{center}
		\includegraphics[scale=0.45]{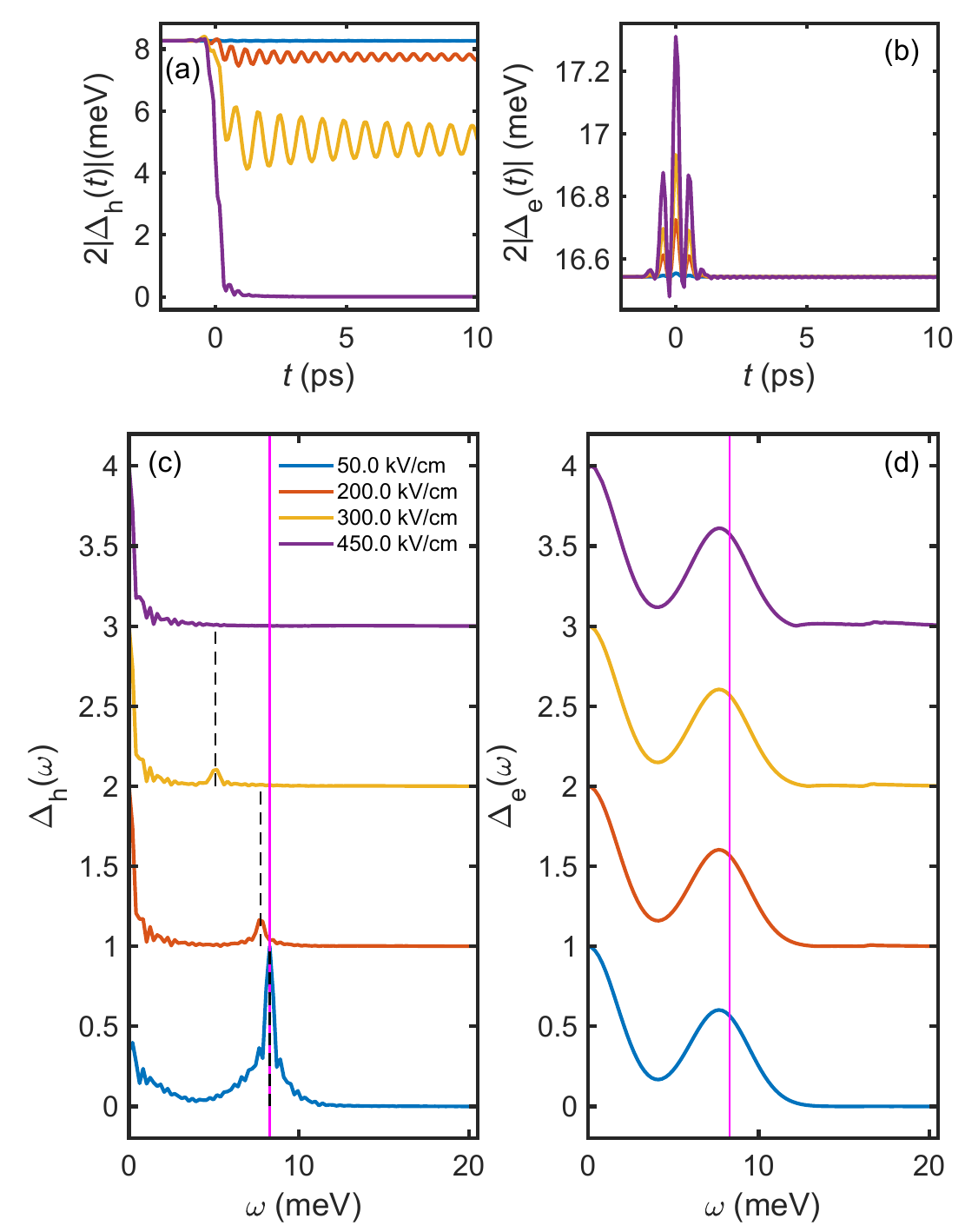}
		\caption{THz driven non-equilibrium order parameter dynamics. Time evolution of (a) the order parameter amplitude of the hole pocket, $2|\Delta_\mathrm{h}|$, and (b) the order parameter amplitude of the electron pocket, $2|\Delta_\mathrm{e}|$ for four different driving fields. The corresponding spectra are shown in (c) and (d). Traces are offset for clarity. The vertical magenta line indicates the second harmonic resonance at $\omega=2\omega_0$, while the Higgs mode frequency of the hole pocket, $\omega_\mathrm{H,h}$, is marked by vertical dashed black lines.}
		\label{fig2} 
\end{center}
\end{figure}


We start with the calculation of the THz-MDCS spectrum  by setting the  interband interaction to zero, $g_\mathrm{e,h}=0$. The results of this calculation with uncoupled bands extend the corresponding one-band results  of Ref.~\cite{Mootz2022}, obtained   for excitation by a strong narrowband pump and a weak broadband probe, to  the case of excitation by a pair of few-cycle  pulses with identical broad spectral shape and strength. The choice of excitation protocol is important for controlling  the  different nonlinear process that can  dominate the THz-MDCS.

We  first make a connection with previous works by calculating  the order-parameter dynamics driven by a single pump pulse with  $g_\mathrm{e,h}=0$. The upper panel of 
Fig.~\ref{fig2} shows the time-dependence  of  (a) the hole pocket order parameter amplitude, $2|\Delta_\mathrm{h}|$, and (b) the electron pocket order parameter amplitude, $2|\Delta_\mathrm{e}|$. It compares between four different electric field strengths in the case of  resonant excitation of the hole pocket SC gap. The corresponding order parameter spectra 
are presented in Figs.~\ref{fig2}(c) and \ref{fig2}(d).
For weak  excitation (blue line), both electron and hole order parameters show second-harmonic oscillations, with frequency $2\omega_0=2$~THz, which occur during the short pulse  (vertical magenta lines in Figs.~\ref{fig2}(c) and \ref{fig2}(d)). These harmonic oscillations are  followed  by damped Higgs mode oscillations with frequencies $\omega_\mathrm{H,\nu}=2\Delta_{\infty,\nu}$, where $\Delta_{\infty,\nu}$ are the quenched order parameters of the steady state  after the pulse. For the weakest driving field studied here (blue line), $\omega_\mathrm{H,h} \sim 2\omega_0$ due to near-resonant excitation across the equilibrium state energy gap.  As a result,  we only observe a single resonance, since  
the Higgs frequency  $\omega_\mathrm{H,h}$ overlaps  with the second-harmonic resonance at 2$\omega_0$. With increasing driving field, however, the SC order parameter of the hole pocket is quenched in the non-equilibrium steady state as compared to the initial state. This quantum quench of the order parameter occurs via difference-frequency Raman processes~\cite{Mootz2022} during cycles of field oscillations, which results in a low-frequency coherent enhancement of $2|\Delta_\mathrm{h}|$. It also results in a red-shift of the  Higgs mode frequency, $\omega_\mathrm{H,h}=2\Delta_{\infty,\mathrm{h}}< 2\Delta_{0,\mathrm{h}}=2\omega_0$ (vertical dashed black line in Fig.~\ref{fig2}(c)). As seen in Fig.~\ref{fig2}(c), with increasing field, the  low-frequency enhancement of the order parameter spectrum  dominates over the Higgs mode resonance. The latter peak vanishes completely for the highest studied driving field (purple line).
Compared to the hole pocket, the spectrum of $2|\Delta_\mathrm{e}|$ in the electron pocket always  shows a broad second harmonic peak (vertical magenta line), which increases with growing field strength. There is no Higgs mode peak there.
The difference in the dynamics between hole and electron pockets is  due to the off-resonant excitation condition $2\omega_0\ll\omega_\mathrm{H,e}$ for the electron band.  

\begin{figure}[t!]
\begin{center}
		\includegraphics[scale=0.55]{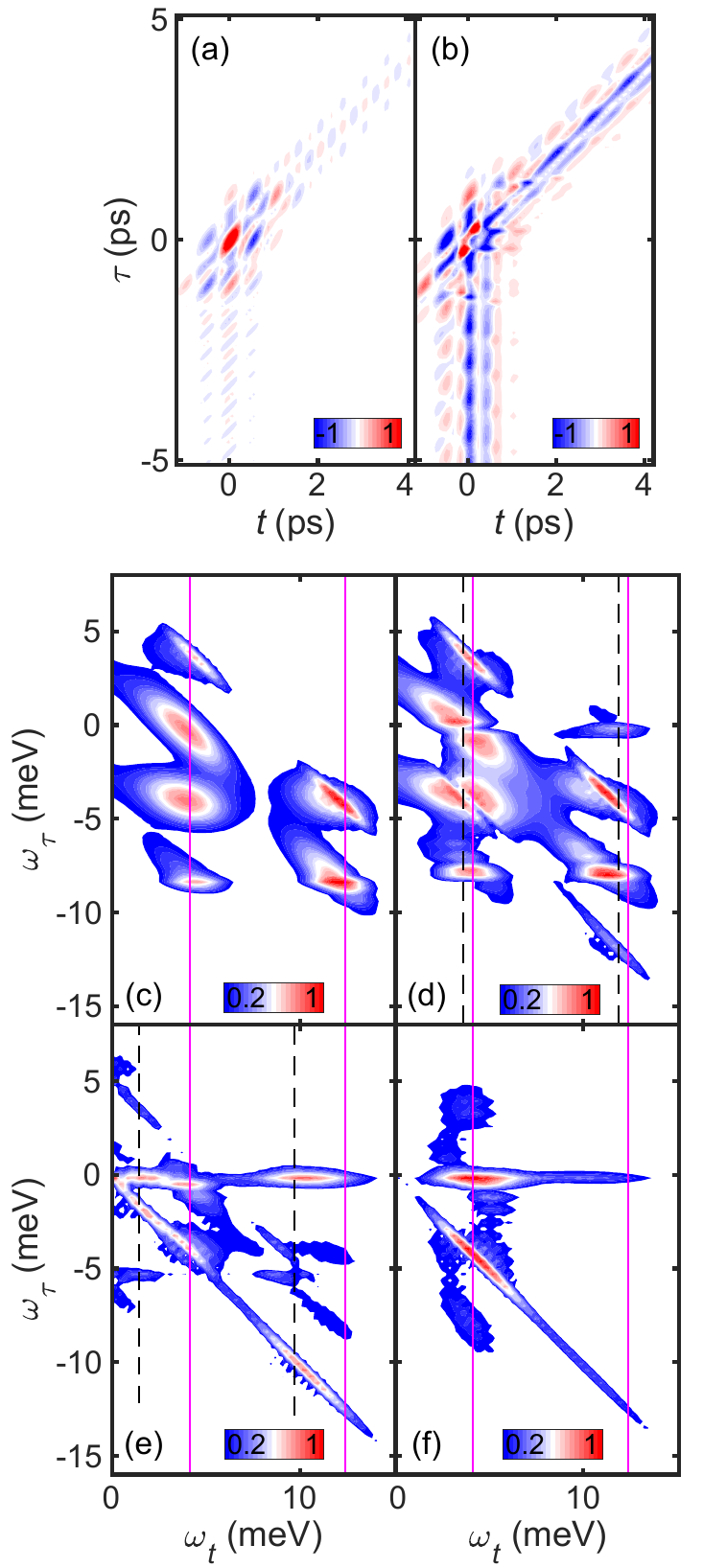}
		\caption{THz-MDCS of a multi-band SC system with negligible interband interaction. Normalized $E_\mathrm{NL}(t,\tau)$ as a function of real time $t$ and inter-pulse delay $\tau$ for a electric field strength of (a) 50~kV/cm and (b) 300~kV/cm. (c)--(f) THz-MDCS spectra $E_\mathrm{NL}(\omega_t,\omega_\tau)$  for an electric field strength of (c) 50~kV/cm, (d) 200~kV/cm, (e) 300~kV/cm, and (f) 450~kV/cm. Vertical magenta lines indicate $\omega_t=\omega_0$ and $\omega_t=3\omega_0$ while $\omega_t=\omega_\mathrm{H,h}\pm\omega_0$ are marked by vertical dashed black lines.}
		\label{fig3} 
\end{center}
\end{figure}

To fully characterize the non-equilibrium condensate state,  we must turn to the THz-MDCS spectra.  Figures~\ref{fig3}(a) and \ref{fig3}(b) show two examples of the calculated nonlinear differential field $E_\mathrm{NL}(t,\tau)$ as a function of real time $t$ and inter-pulse time delay $\tau$.  By comparing the results obtained with electric  fields of $50$~kV/cm and $300$~kV/cm,  we observe   a change in the overall  time dependence with increasing field. For low driving (Fig.~\ref{fig3}(a)), the oscillations along the $t$- and $\tau$-axes are determined by the laser frequency  $\omega_0$. As seen by comparing with Fig.~\ref{fig3}(b), the oscillation frequencies change  for the higher field. To identify the nonlinear processes leading to this change, we study the THz-MDCS spectra obtained by Fourier transform of  $E_\mathrm{NL}(t,\tau)$ with respect to both $t$ (frequency $\omega_t$) and $\tau$ (frequency $\omega_\tau$). Figures~\ref{fig3}(c)--(e) show the $E_\mathrm{NL}(\omega_t,\omega_\tau)$ two-dimensional spectra for the four driving fields studied in Fig.~\ref{fig2}. We compare the positions of the THz-MDCS peaks with  respect to 
 the laser pulse  frequency vectors  $\omega_\mathrm{A}=(\omega_0,0)$  and $\omega_\mathrm{B}=(\omega_0,-\omega_0)$, which determine the positions of the peaks in semiconductors and other conventional materials. 
For the weakest studied field (Fig.~\ref{fig3}(c)), the order parameters remain close to their equilibrium values. Similar to  semiconductors,   the quasi-particle excitation energy gap  then does not change  from equilibrium. 
As  a result, the THz-MDCS spectrum in Fig.~\ref{fig3}(c)  shows peaks at multiples of the laser frequency $\omega_0$. This behavior  is similar to  semiconductors~\cite{Kuehn2011,maag2016} and recovers  the results of  a perturbative  susceptibility analysis~\cite{Mootz2022}. The peaks observed in Fig.~\ref{fig3}(c) 
can be attributed to the  third-order nonlinear responses of the equilibrium SC state. These can be described  by using the third-order susceptibility, which in superconductors measures the elementary excitations of the equilibrium  state that do not contribute to the linear response. 
Figure~\ref{fig3}(c) displays  the conventional pump--probe (PP)  peaks  at $(\omega_t,\omega_\tau)=(\omega_0,0)$ and  $(\omega_0,-\omega_0)$, which are generated by the nonlinear processes $(\omega_\mathrm{B}-\omega_\mathrm{B}) + \omega_\mathrm{A}$  and $(\omega_\mathrm{A}- \omega_\mathrm{A}) + \omega_\mathrm{B} $, respectively. 
The peaks at  $(\omega_0,-2 \omega_0)$ and $(\omega_0,\omega_0)$ correspond to four-wave mixing signals arising from the processes $(\omega_\mathrm{B}-\omega_\mathrm{A})+\omega_\mathrm{B}$ and $( \omega_\mathrm{A} - \omega_\mathrm{B})+\omega_\mathrm{A}$.
 Finally, third-harmonic generation peaks are visible at $(3 \omega_0,-2\omega_0)$  and $(3 \omega_0,-\omega_0)$. These conventional third-order nonlinear peaks characterize the dynamics of a SC elementary excitation, quasi-particle or collective mode. As known from  previous works,  quasi-particle excitations dominate over the Higgs collective mode contributions in this case. 

With increasing driving field, the  THz quantum quench of the SC order parameter  separates the excitation energy  of the non-equilibrium state, $\omega_\mathrm{H,h}=2 \Delta_{\infty,\mathrm{h}}$,   from the second harmonic frequency $2 \omega_0 \sim 2 \Delta_{0,\mathrm{h}}$
that excites the equilibrium state energy gap. Figure~\ref{fig3}(d) then shows multiple new THz-MDCS peaks. In particular,  Fig.~\ref{fig3}(d) displays  strong peaks   at frequencies $\omega_t=\omega_\mathrm{H,h}\pm\omega_0$ (dashed black lines),  which split from the pump--probe and third-harmonic generation peaks of Fig~\ref{fig3}(c) 
(solid magenta lines). These new peaks  are mainly generated by contributions to the current
Eq.~(\ref{eq:current2}) arising from the  source terms of the 
form $\sim \tilde{\rho}^{(\nu),\mathrm{A}}_{2}(\mathbf{k})
\,\mathbf{p}_\mathrm{S}^\mathrm{B}$
in the 
second term on the rhs of Eq.~(\ref{eq:delta_rho0}). 
Such nonlinear contributions reflect the 
time dependence of the excitations of the non-equilibrium SC state, which give oscillations of the  pseudo-spin 
component $\tilde{\rho}^{(\nu),\mathrm{A,B}}_{2}(\mathbf{k})$ at frequencies $\sim \omega_\mathrm{H,h}$ that are sensed by the other pulse ($\mathbf{p}_\mathrm{S}^\mathrm{B,A}$). 
The  contribution of these excitations to the THz-MDCS
mainly arises   
 when pulse B (A) arrives after the  pulse A (B)  driving the 
 $\omega_\mathrm{H,h}$ excitation.  
 This nonlinear signal  
  does not require any interference or interaction between 
 excitations by  different pulses.  For high fields, it measures the 
  dynamics of a single non-equilibrium SC state  excitation 
  by one pulse (Fig.~\ref{fig2}), which is probed by the other pulse.
 
 The THz-MDCS signal and additional peaks observed  in Fig~\ref{fig3}(d) cannot be attributed  to the  Higgs collective mode. 
  In particular,  our calculated  $\tilde{\rho}^{(\nu)}_2(\mathbf{k})$ spectra are dominated by $\mathbf{k}$-dependent peaks centered at the quasi-particle continuum  energies $E^{\mathrm{qp, (h)}}_\mathbf{k}=2\sqrt{E_\mathrm{h}^2(\mathbf{k})+|\Delta_{\infty,h}|^2}$. This is the case for the few-cycle  broadband pulses used in this paper~\cite{Mootz2022}. The largest contribution of $\tilde{\rho}^{(\nu)}_2(\mathbf{k})$ stems from the quasi-particle excitations in proximity to the excitation energy minimum (energy gap $\sim \omega_\mathrm{H,h}$). The second term on the rhs of Eq.~(\ref{eq:delta_rho0}) then  generates the third-order nonlinear processes $\omega_\mathrm{H,h;A}\pm\omega_\mathrm{B}$, where $\omega_\mathrm{H,i;A}=(\omega_\mathrm{H,i},0)$ ($i=$~e, h)
describes a single SC excitation by pulse A.  
  These processes lead to THz-MDCS 
  peaks at $(\omega_\mathrm{H,h}-\omega_0,\omega_0)$ and $(\omega_\mathrm{H,h}+\omega_0,-\omega_0)$
  (dashed black lines in Fig~\ref{fig3}(d)). These peaks  split from the conventional PP and HHG peaks of Fig~\ref{fig3}(c), which are located at  
 $(\omega_0,\omega_0)$ and $(3 \omega_0,-\omega_0)$,  as $\omega_\mathrm{H,h}< 2 \omega_0$ with increasing field. 
     Exchanging labels A and B gives the third-order nonlinear processes $\omega_\mathrm{H,h;B}\pm\omega_\mathrm{A}$ where  $\omega_\mathrm{H,i;B}=(\omega_\mathrm{H,i},-\omega_\mathrm{H,i})$ ($i=$~e, h), which generate  peaks at $(\omega_\mathrm{H,h}-\omega_0,-\omega_\mathrm{H,h})$ and $(\omega_\mathrm{H,h}+\omega_0,-\omega_\mathrm{H,h})$.
 We conclude that the above  THz-MDCS peaks  mostly characterize the dynamics of the quasi-particle  excitations of the SC steady state with quenched energy gap.

The THz-MDCS spectrum shows four additional competing peaks, two along direction $(\omega_t, 0)$ and two along $(\omega_t,-\omega_t)$.
By increasing the driving field further, Fig.~\ref{fig3}(e) shows that the latter signals, with peaks at   2D frequencies $\sim (\omega_\mathrm{H,h}\pm\omega_0,-\omega_\mathrm{H,h}\mp\omega_0)$ and  $\sim (\omega_\mathrm{H,h}\pm\omega_0,0)$, dominate over the third-order signals at $\omega_t=\omega_\mathrm{H,h}\pm\omega_0$  in
Fig.~\ref{fig3}(d).
We  next discuss their origin.  For the field strength of 300~kV/cm used in
Fig.~\ref{fig3}(e), the SC order parameter of the hole pocket is  strongly quenched   by  the difference-frequency coherent Raman  process $2\omega_\mathrm{A}-2\omega_\mathrm{A}$.  As seen in Fig.~\ref{fig2}(c) (yellow curve, black dashed line), the resulting peak of  the $|\Delta^\mathrm{A}_\mathrm{h}|$ spectrum at $\omega \sim 0$ 
dominates over the peak at $\omega_\mathrm{H,h}$ (black dashed line). At the same time, $\Delta\tilde{\rho}^{(\nu)}_{2}$ in
Eq.~(\ref{eq:eom_t2})  
 is driven by the source term  Eq.~(\ref{coupling}).  Without interband interaction, the latter time-dependent 
transverse canting is driven by the condensate acceleration leading to $\partial_t E_{\nu}({\bf k}) \ne 0$. 
 Equation~(\ref{eq:eom_t2}) describing the coupling between 
 the light-induced $x$ and $y$ pseudo-spin component deviations from equilibrium then 
 reduces to $ \partial_t E_{\nu}({\bf k}) \Delta \mathrm{\rho}_1^{(\nu)}\approx e\,({\bf E}(t) \cdot \nabla_{{\bf k}}) ({\bf p_{\rm{S}}} \cdot \nabla_{{\bf k}}) \, \xi_\nu({\bf k})\Delta \mathrm{\rho}_1^{(\nu)}$. Noting that the   $\Delta\tilde{\rho}^{(\nu)}_1(\mathbf{k})$ dynamics  is determined by the 
 frequency $\omega_\mathrm{H,h}$,  the source term Eq.~(\ref{coupling}) resonantly drives $\Delta \tilde{\rho}_2^{(\nu)}(\mathbf{k})$ with frequency 
 $\omega_\mathrm{H,h;B}+(\omega_\mathrm{B}-\omega_\mathrm{B})$. 
 Together with the  order parameter quantum quench through the frequency-difference 
 coherent process  $2\omega_\mathrm{A}-2\omega_\mathrm{A}$,  we obtain  a signal, determined by the 
 third term on the rhs of Eq.~(\ref{eq:delta_rho0}),  that is generated by the high-order 
  nonlinear process $(2\omega_\mathrm{A}-2\omega_\mathrm{A})+\omega_\mathrm{H,h;B}+(\omega_\mathrm{B}-\omega_\mathrm{B})\pm\omega_\mathrm{B}$.  
  This  higher Raman  process  is of ninth order with respect 
  to the equilibrium state.  It 
generates THz-MDCS peaks at 2D frequencies $(\omega_\mathrm{H,h}\pm\omega_0,-\omega_\mathrm{H,h}\mp\omega_0)$ through a combination of transverse pseudo-spin canting and order parameter coherent modulation. 
 In the same way, by exchanging $\textrm{A}\leftrightarrow\textrm{B}$,  we obtain peaks at $(\omega_\mathrm{H,h}\pm\omega_0,0)$. Figure~\ref{fig3}(e) demonstrates that  the above  high-order Raman processes  dominate over the conventional third-order processes in the THz-MDCS nonlinear spectra  for sufficiently strong pulse-pair driving of a non-equilibrium SC steady state. 
 Eventually, in  the extreme nonlinear excitation regime (Fig.~\ref{fig3}(f)) where the SC order parameter amplitude $|\Delta_\mathrm{h}|$ and corresponding excitation energy gap is completely quenched  (purple line in Fig.~\ref{fig2}(c)),   the THz-MDCS spectral lineshape displays  sharp pump--probe peaks at $(\omega_0,0)$ and  $(\omega_0,-\omega_0)$ that dominate over the third-harmonic peaks of Fig.~\ref{fig3}(c).

\begin{figure}[t!]
\begin{center}
		\includegraphics[scale=0.38]{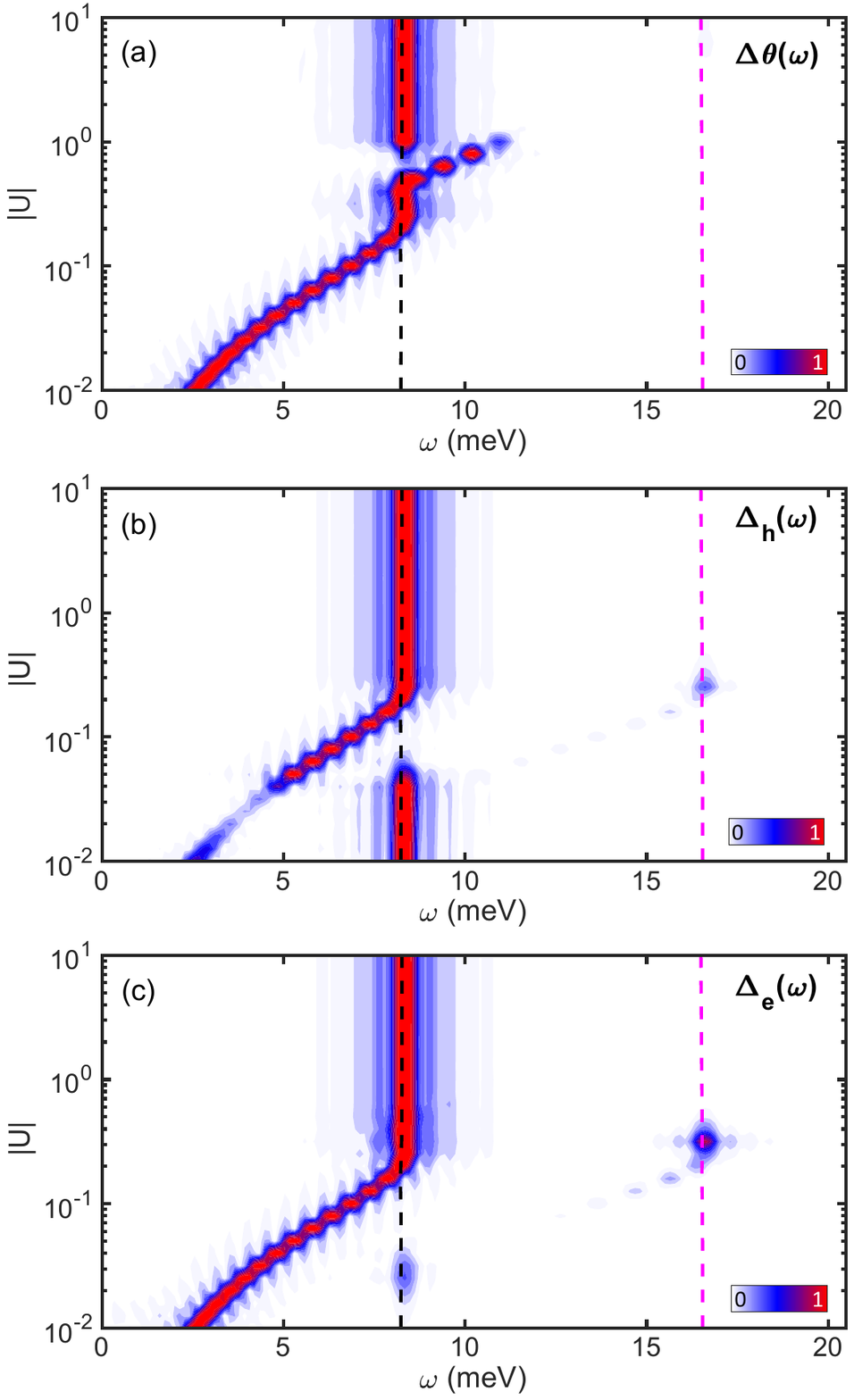}
		\caption{Dependence of amplitude and relative phase collective modes on the interband interaction strength. (a) $\Delta\theta$ spectrum, (b) $|\Delta_\mathrm{h}|$ spectrum, and (c) $|\Delta_\mathrm{e}|$ spectrum as a function of interband interaction strength $|U|$ for weak $E_0=100.0$~kV/cm excitation. Spectra are normalized to one for a given $|U|$. Dashed black (magenta) line indicates $\omega_\mathrm{H,h}$ ($\omega_\mathrm{H,e}$).}
		\label{fig4} 
\end{center}
\end{figure}

\begin{figure*}[t!]
\begin{center}
		\includegraphics[scale=0.48]{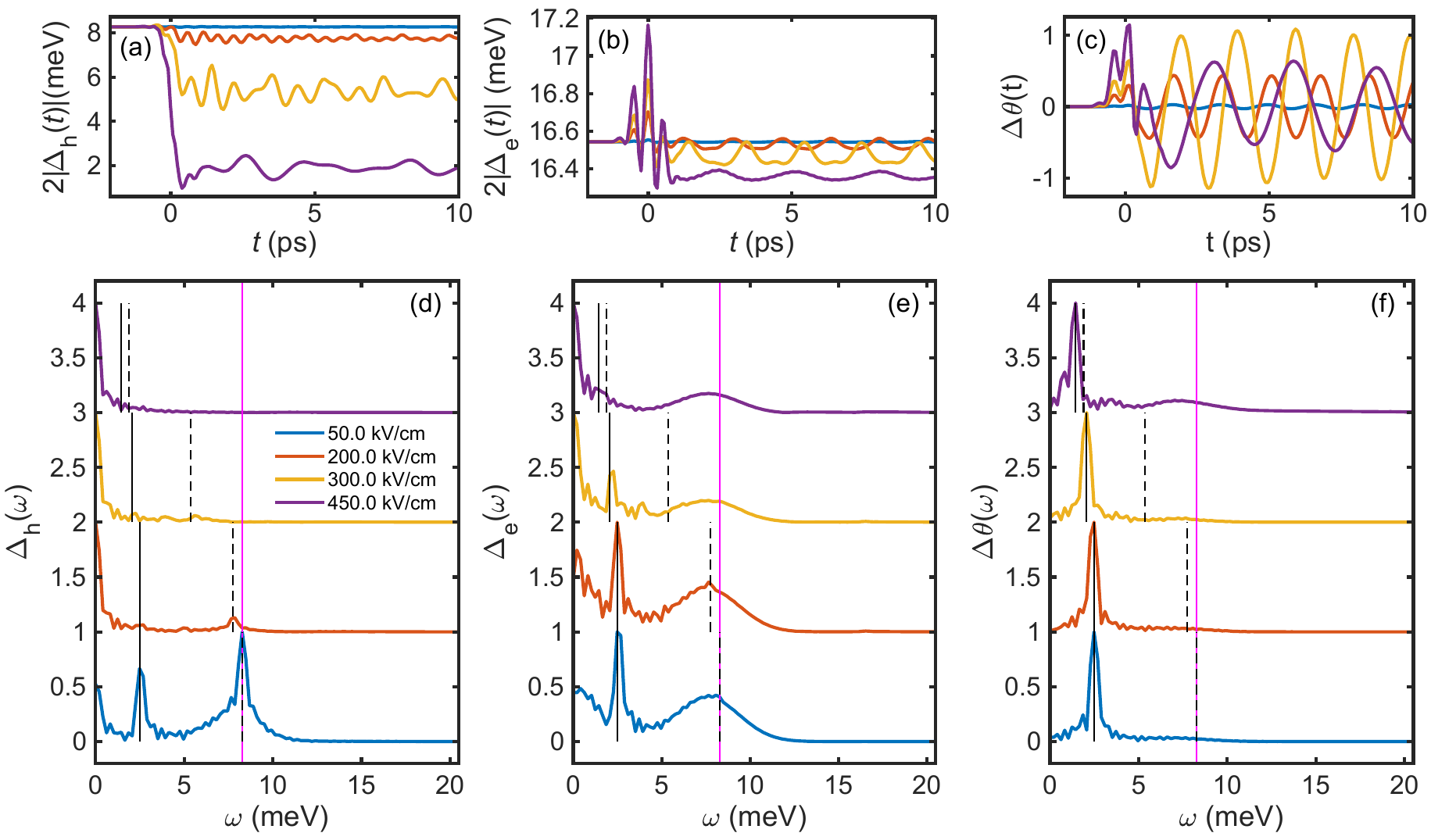}
		\caption{THz-driven dynamics of amplitude and relative phase modes for weak interband interaction $U$=0.01 and four different driving fields. Time evolution of (a) the order parameter amplitude of the hole pocket, $2|\Delta_\mathrm{h}|$, (b) the order parameter amplitude of the electron pocket, $2|\Delta_\mathrm{e}|$, and (c) the relative phase, $\Delta\theta$. The corresponding spectra are shown in (d)--(f). Vertical magenta lines indicate the second harmonic resonance at $\omega=2\omega_0$ while the Higgs mode frequency of the hole pocket, $\omega_\mathrm{H,h}$, (Leggett mode $\omega_\mathrm{L}$) is marked by vertical dashed (solid) black lines.}
		\label{fig5} 
\end{center}
\end{figure*}

We end this section by noting that a comparison of  the $U$=0 few-cycle  pulse-pair excitation results with the 
narrowband strong pump-broadband weak probe  results of Ref.~\cite{Mootz2022}
 suggests that the excitation protocol can be designed to highlight different nonlinear processes in the THz-MDCS spectra (see also Appendix~\ref{sec:pp}). In all cases, the main  qualitative differences in the spectral profile of superconductors as compared to  other systems originate  from the coherent modulation of the elementary excitation energy. In particular,  THz-MDCS of superconductors measures excitations of the light-induced  non-thermal  steady state, which differs from the equilibrium states. Additional differences arise from nonlinear pseudo-spin processes driven by the condensate acceleration by the applied THz field, which determine the properties of the time-dependent SC state.

\subsection{Non-zero interband interaction}

\label{sec:U>0}

We now consider how interband interactions can change  the above $U=0$ picture.
Before presenting the THz-MDCS results, 
we  first study the multi-band order parameter amplitude and relative  phase spectra  as a function of the interband interaction. We consider  excitation by a single, relatively weak, few-cycle THz electric field, so that we can map the excitations of the equilibrium state for different  $U$. 
Figure~\ref{fig4} shows the dependence on  $|U|=|g_{\lambda,\nu}|/V_\mathrm{h}$, $\nu\neq\lambda$, of  (a) relative phase spectrum, $\Delta\theta(\omega)$, (b)  hole order parameter amplitude spectrum, $|\Delta_\mathrm{h}|$, and (c) electron order parameter amplitude spectrum, $|\Delta_\mathrm{e}|$. In the simulations, we fixed the ratio $U=g_\mathrm{e,h}/V_\mathrm{h}$ and adjusted $V_\mathrm{e}$ and $V_\mathrm{h}$ to keep $\omega_\mathrm{H,h}$ and $\omega_\mathrm{H,e}$ unchanged. A resonance in 
$\Delta\theta(\omega)$ is seen in Fig.~\ref{fig4}(a), which 
results from a phase collective mode. 
For small $|U|$, we obtain such a  Leggett mode with energy $\omega_\mathrm{L}\sim 2.5$~meV. This phase mode  lies well below the lower hole band excitation energy gap  $\omega_\mathrm{H,h}$,  which is marked  in Fig.~\ref{fig4} by the dashed black line.
The position of the Leggett mode within this energy gap  agrees with earlier studies on multi-band superconductors with dominating intraband interaction~\cite{krull2016}. As $|U|$ increases, however, the  phase mode blueshifts towards  $\omega_\mathrm{H,h}$ and crosses $\omega_\mathrm{H,h}$ around $|U|\sim 0.5$. It moves towards $\omega_\mathrm{H,e}$ (magenta dashed line) as $|U|$ further increases. In this regime of strong $|U|$,  the Leggett mode  becomes strongly damped, as it is located within the hole band quasi-particle continuum. As a result, for weak photoexcitation, the effects of the relative phase dynamics  on the THz-MDCS  are small for  large $U$, unlike for small $U$.   Next, we study  how this order parameter dynamics changes by increasing the driving field, and compare the field  dependence between weak and strong interband interactions. 

\subsubsection{Dominant intraband interaction}

We first study the behavior of the order parameter for small but finite $U$, $0<|U|\ll 1$. This is the case, e.~g., in MgB$_2$ superconductors~\cite{Blumberg,Klein2010,Aoki2017,Murotani,Cea2016b,Ortolani,krull2016,Giorgianni2019}. The main difference introduced by the  small $|U|$ is the emergence of a Leggett phase mode whose frequency lies within the energy gap. Figure~\ref{fig5} presents the time dependence  of (a) the order parameter amplitude in the hole pocket, $2|\Delta_\mathrm{h}|$, (b)  the order parameter amplitude in the electron pocket, $2|\Delta_\mathrm{e}|$, and (c) the phase difference between the electron and hole pocket order parameters, $\Delta\theta$. These results were obtained  for four different electric field strengths with $U=0.01$.  Figure~\ref{fig4} shows that, for such parameters,  the Leggett mode is not damped.
  The corresponding order parameter spectra are plotted in Figs.~\ref{fig5}(d)--(f). 
The main difference from the uncoupled band case, Fig.~\ref{fig2}, is the emergence of an additional peak in $\Delta \theta(\omega)$, 
Fig.~\ref{fig5}(f). This resonance
also appears in the 
 order parameter amplitude spectra
 (solid black line in Figs.~\ref{fig5}(d) and (e)).  

In  the weak excitation regime (blue line), the $\Delta\theta$ spectrum shows a sharp peak at the Leggett mode energy $\omega_\mathrm{L}\sim$2.5~meV (solid black line). As discussed above, this peak is located well below the low energy excitation energy gap  $\omega_\mathrm{H,h}\sim$~8.2~meV (dashed black line). This Leggett mode (solid black line) also shows up in the $|\Delta_\mathrm{h}|$ ($|\Delta_\mathrm{e}|$) amplitude spectrum, in addition to  the Higgs mode (dashed black line, $\omega_\mathrm{H,h}$) and  second harmonic generation ($2 \omega_0$, solid magenta line) peaks. This coexistence of 
phase and amplitude mode peaks demonstrates that the amplitude  mode and the phase mode are coupled despite their different frequencies. With increasing field, 
the Leggett mode  slightly redshifts, while  the order parameter becomes quenched. The latter coherent quench of the hole 
and electron order parameter amplitudes is seen as  a low-frequency ($\omega=0$)  enhancement of $|\Delta_{\mathrm{h}}|$ and $|\Delta_\mathrm{e}|$ spectra  in Figs.~\ref{fig5}(d) and \ref{fig5}(e). The order parameter amplitude quench dominates over the collective mode resonances in these 
amplitude spectra  with increasing field,  analogously to the case without interband interaction in  
Fig.~\ref{fig2}.

\begin{figure}[t!]
\begin{center}
		\includegraphics[scale=0.48]{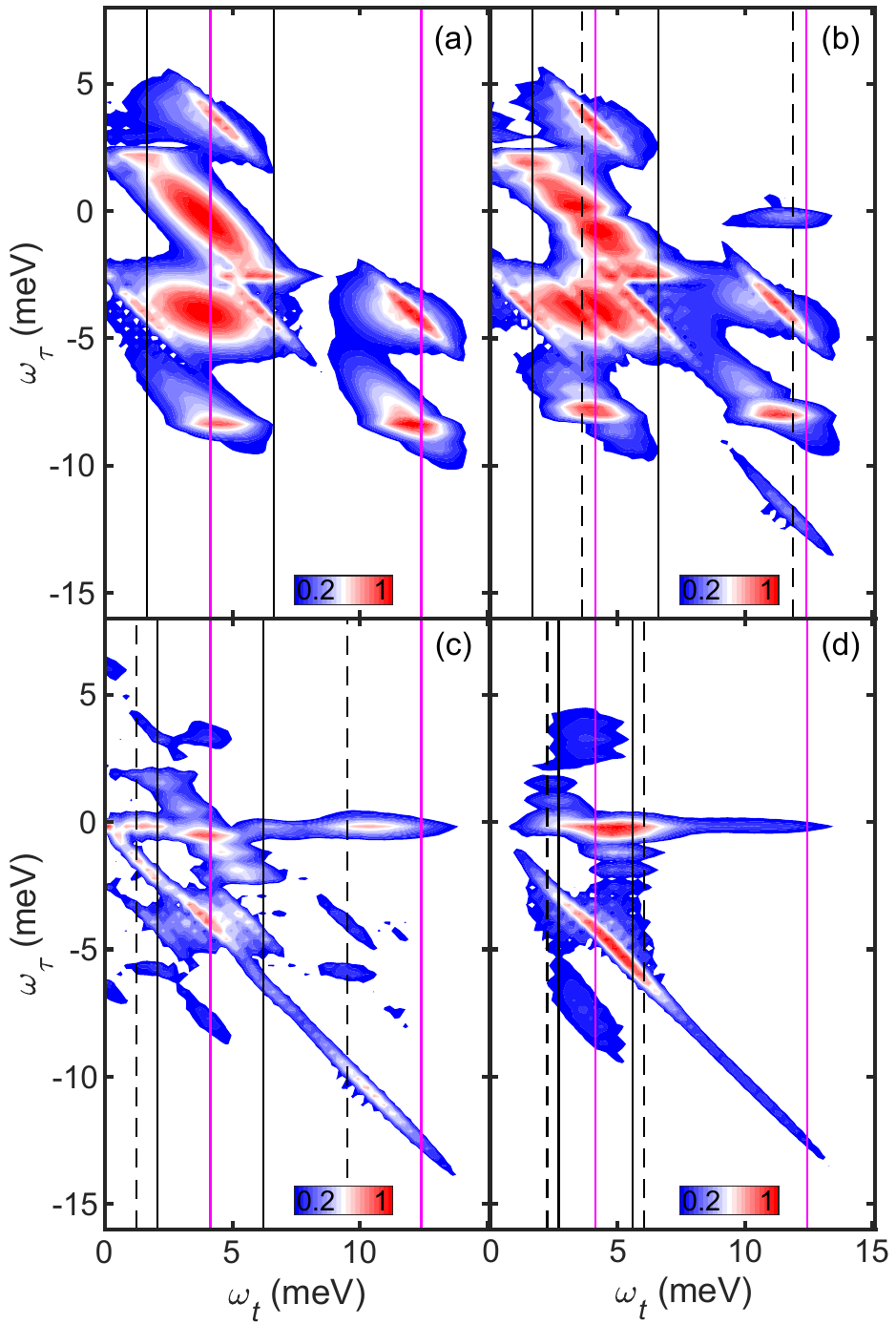}
		\caption{THz-MDCS of a multi-band superconductor with dominant intraband interaction ($\mathbf{p}_\mathrm{S}=0$ after the pulse).
	 Normalized THz-MDCS spectra $E_\mathrm{NL}(\omega_t,\omega_\tau)$  for an electric field strength of (a) 50~kV/cm, (b) 200~kV/cm, (c) 300~kV/cm, and (d) 450~kV/cm. Vertical magenta lines indicate $\omega_t=\omega_0$ and $\omega_t=3\omega_0$, while $\omega_t=\omega_\mathrm{H,h}\pm\omega_0$ ($\omega_t=\omega_0\pm\omega_\mathrm{L}$) are marked by vertical black dashed (solid) lines.}
		\label{fig6} 
\end{center}
\end{figure}

Figure~\ref{fig6} presents the THz-MDCS spectra for the four electric field strengths studied in Fig.~\ref{fig5} with finite $U$. For low driving field, Fig.~\ref{fig6}(a), the THz-MDCS spectrum is dominated by the pump--probe, four-wave mixing, and third harmonic peaks discussed in Sec.~\ref{sec:U=0}. The order parameter for this  low field (blue curves in Fig.~\ref{fig5}) displays sharp Higgs and Leggett modes. By comparing 
with the corresponding spectra for $U=0$, Fig.~\ref{fig3}(c), 
the presence of the Leggett phase mode for $U \ne 0$  leads to the formation of THz-MDCS satellites (vertical solid black lines) around the  $(\omega_0,0)$ and $(\omega_0,-\omega_0)$ pump--probe peaks.  These Leggett mode peaks originate from the fifth term on the rhs of Eq.~(\ref{eq:delta_rho0}), $\propto \Delta \theta(t)
\mathbf{p}_\mathrm{s}(t)$. 
In particular, the relative phase oscillations at frequency 
$\omega_L$ (Leggett collective mode) 
lead to  $(\omega_0\pm\omega_\mathrm{L},-\omega_0)$ peaks (black solid lines in Fig.~\ref{fig6}(a)) via the 
 third-order nonlinear processes
 $\omega_\mathrm{B}\pm\omega_\mathrm{L;A}$, where $\omega_\mathrm{L;A}=(\omega_\mathrm{L},0)$. Exchanging pulses A and B produces similar Leggett mode  peaks at $(\omega_0\pm\omega_\mathrm{L},\mp\omega_\mathrm{L})$, generated by the 
 third-order nonlinear processes
 $\omega_\mathrm{A}\pm\omega_\mathrm{L;B}$,  where $\omega_\mathrm{L;B}=(\omega_\mathrm{L},-\omega_\mathrm{L})$.

With increasing field,   Fig.~\ref{fig5}(a) (red line), the  order parameter coherent quench,
$\omega_\mathrm{H,h} < 2 \omega_0$,
 results in the formation of  THz-MDCS peaks at $\omega_t=\omega_\mathrm{H,h}\pm\omega_0$, similar to  Section~\ref{sec:U=0}. These  peaks, marked by the vertical dashed lines in Fig.~\ref{fig6}(b), are additional to the relative phase mode  peaks (solid black lines) and the conventional third order response peaks (solid magenta lines). As the field increases further, Fig.~\ref{fig6}(c) shows that, similar to the $U=0$ case,  the THz-MDCS spectrum is dominated by sharp peaks along $(\omega_t, 0)$ and $(\omega_t,-\omega_t)$ directions. The difference from  $U = 0$  are  peaks  at $\omega_t=\omega_0\pm\omega_\mathrm{L}$ marked by  vertical solid black lines.  These peaks are  generated by the nonlinear coupling of the $x$ and $y$ transverse pseudo-spin components described by Eq.~(\ref{coupling}). 
 The pseudo-spin are then driven via sum-frequency nonlinear processes $\omega_0+\omega_0+\omega_\mathrm{L}$,  instead of  $\omega_0-\omega_0+\omega_\mathrm{H,h}$ processes.    
We thus obtain  peaks at $\omega_t=\omega_0\pm\omega_\mathrm{L}$ 
 via the ninth-order nonlinear processes $(2\omega_\mathrm{A}-2\omega_\mathrm{A})\pm\omega_\mathrm{L,B}+(\omega_\mathrm{B}+\omega_\mathrm{B})-\omega_\mathrm{B}$.
 These peaks at  two-dimensional frequencies $(\pm\omega_\mathrm{L}+\omega_0,\mp\omega_\mathrm{L}-\omega_0)$ are marked by the solid black lines.  Exchanging pulses A and B produces similar peaks at $(\pm\omega_\mathrm{L}+\omega_0,0)$. Finally, for the highest studied field strength of 450~kV/cm, the Leggett mode is close to  $\omega_\mathrm{H,h}$ in Fig.~\ref{fig5} (purple line). As a result, the collective mode and quasi-particle peaks cannot be resolved anymore in the THz-MDCS spectrum of Fig.~\ref{fig6}(d). We then obtain broad THz-MDCS peaks around $\omega_t=\omega_0$ (magenta solid line).

\begin{figure}[t!]
\begin{center}
		\includegraphics[scale=0.48]{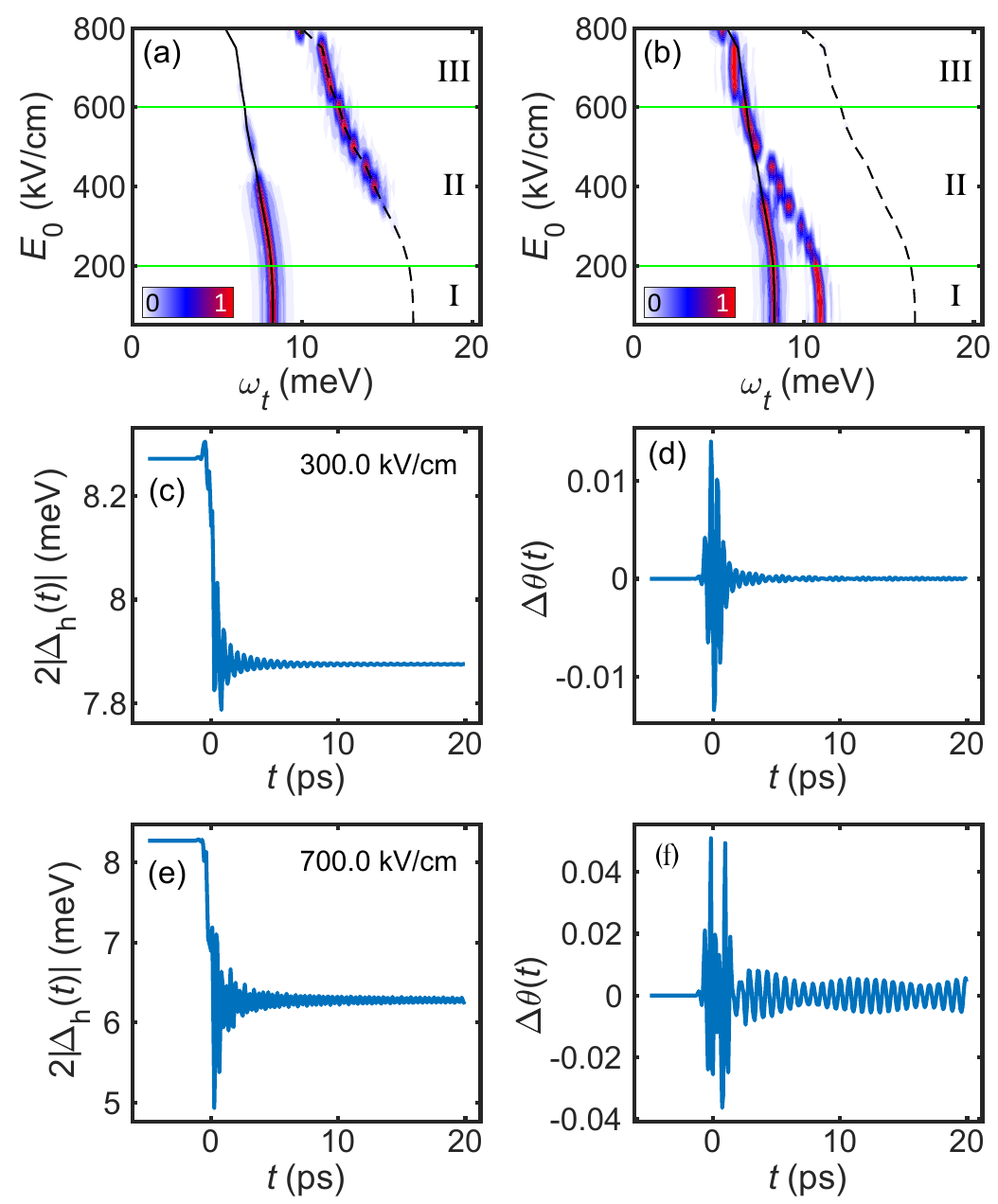}
		\caption{Dependence of amplitude and relative phase collective modes on the driving field strength for interband interaction dominating over intraband interaction. Electric field strength dependence of (a) the SC order parameter spectrum $|\Delta_\mathrm{h}|$ and (b) the relative phase spectrum $\Delta\theta$. Spectra are normalized to one for a given $E_0$. The Higgs mode $\omega_\mathrm{H,h}$ ($\omega_\mathrm{H,e}$) is marked with a solid (dashed) black line. Horizontal lines indicate the excitation regimes I--III. (c),(e) $|\Delta_\mathrm{h}|$ dynamics for driving fields of 300~kV/cm (c) and 700~kV/cm (e). The corresponding relative phase dynamics is plotted in (d) and (f).}
		\label{fig7} 
\end{center}
\end{figure}

\begin{figure}[t!]
\begin{center}
		\includegraphics[scale=0.55]{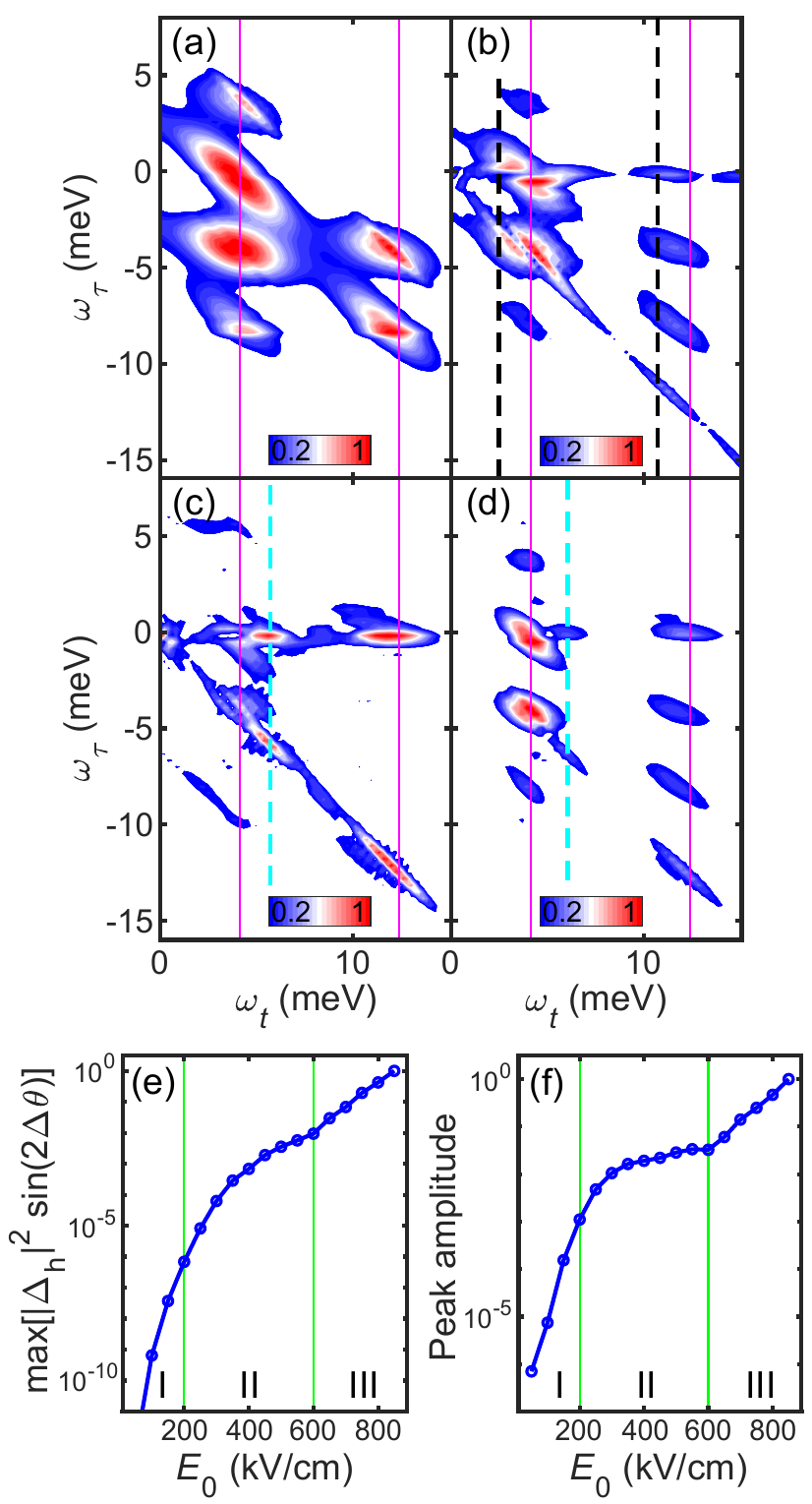}
\end{center}		
		\caption{Formation of bi-Higgs energy sideband in the THz-MDCS spectra of a multi-band SC with strong interband coupling in the absence of persisting  dynamical inversion symmetry breaking ($\mathbf{p}_\mathrm{S}=0$ after the pulse) . (a)--(c) Normalized THz-MDCS spectra $E_\mathrm{NL}(\omega_t,\omega_\tau)$  for electric field strengths of (a) 200~kV/cm, (b) 400~kV/cm, and (c) 800~kV/cm. (d) Normalized THz-MDCS spectrum resulting from a calculation without phase--amplitude coupling, Eq.~(\ref{eq:biHiggs}), for electric field strength of 800~kV/cm. Vertical dashed black lines mark $\omega_t=\omega_\mathrm{H,h}\pm\omega_0$.  Vertical solid magenta lines indicate $\omega_t=\omega_0$ and $\omega_t=3\omega_0$ while dashed cyan lines denote $\omega_t=2\omega_\mathrm{H,h}-\omega_0$.  
		(e),(f) Correlation in the field strength dependence of the maximum of the $2|\Delta_\mathrm{h}|^2  \sin 2\Delta\theta$ spectrum (e) and the $\omega_t = 2\omega_\mathrm{H,h} - \omega_0$ sideband peaks in the THz-MDCS spectra (f). Vertical lines indicate the excitation regimes I--III.}
		\label{fig8} 
\end{figure}

\subsubsection{Dominant interband interaction}


We now compare the above THz-MDCS spectral profiles for small $U$ with the case where the interband interaction exceeds the intraband interaction, $|U|>1$. This strong interband coupling  condition is realized in several iron-based superconductors~\cite{Fernandes2010,Patz2014,Patz2017,yangPRL}.  Here we
present results for  $U=2$ as in such systems. We first characterize the  SC order-parameter  dynamics induced by a single pulse. Figure~\ref{fig7} shows the field dependence of (a) the order parameter amplitude spectrum $|\Delta_\mathrm{h}(\omega)|$ and (b) the relative phase spectrum $\Delta\theta(\omega)$. In the weak photoexcitation regime, $|\Delta_\mathrm{h}(\omega)|$ exhibits a peak at the Higgs mode frequency  $\omega_\mathrm{H,h}$ (solid black line). In addition to this Higgs mode peak, the spectrum of the relative phase $\Delta\theta(\omega)$ in Fig.~\ref{fig7}(b) shows a Leggett mode peak.  The difference for   large  $U$ is that the Leggett mode lies within the quasi-particle continuum, $\omega_\mathrm{L}\sim 11~\textrm{meV} > \omega_\mathrm{H,h}$, and is thus strongly damped for low fields.  With increasing driving field, the $\omega_\mathrm{H,h}$-mode  red shifts slightly. In addition, a new peak emerges in the hole order parameter spectrum $\Delta_\mathrm{h}(\omega)$, at the electron pocket  frequency $\omega_\mathrm{H,e}$ (dashed black line). This peak is absent for small $U$ and  arises from the coherent coupling of the Higgs modes in the electron and hole  pockets (hybrid-Higgs mode~\cite{hybrid-higgs}).
 Light-induced nonlinear  couplings of collective modes in different bands coupled via Coulomb interaction have been proposed before  to affect the  pump--probe spectra via the formation of hybrid modes~\cite{shah}. 
 Here, a  {\em hybrid-Higgs} amplitude collective mode  forms due to  the strong coupling between the electron and hole pockets for large $U$~\cite{hybrid-higgs}. 
 At the same time, the Leggett mode peak redshifts towards
$\omega_\mathrm{H,h}$. As a result, a new relative phase collective mode forms with frequency around  $\omega_\mathrm{H,h}$. This new mode characterizes the light-driven non-equilibrium SC state and is not observed close to equilibrium. We show below that the relative phase mode  parametrically drives the coupled nonlinear harmonic oscillator equations of motion~(\ref{eq:eom_t2})  at the frequency $\omega_\mathrm{H,h}$, 
via  Eq.~(\ref{coupling}), which
 results in the formation of dominant bi-Higgs-frequency sidebands in the THz-MDCS spectra, discussed below.

Next, we compare  the THz-MDCS spectra in the  three excitation regimes indicated by the horizontal lines in Figs.~\ref{fig7}(a) and \ref{fig7}(b). In  Regime~I (lowest fields), the order parameters are close to their equilibrium values, such that the response is describable by a susceptibility perturbative expansion around the equilibrium state. Regime~II (intermediate fields) is characterized by the formation of the hybrid Higgs collective mode.  Regime~III (highest fields) is governed  by the relative-phase collective mode. Figures~\ref{fig7}(c) and \ref{fig7}(e) show examples of the order-parameter dynamics in Regime~II and Regime~III, respectively. The corresponding relative phase dynamics is presented in Figs.~\ref{fig7}(d) and \ref{fig7}(f). The Higgs mode oscillations observable in $|\Delta_\mathrm{h}(t)|$  are damped in both excitation regimes. The relative phase oscillations are also weak and damped in Regime~II. However, they are undamped and amplified in Regime~III, showing a beating oscillation pattern. The observed beating oscillations of $\Delta\theta(t)$ in Regime~III signify  the coupling between the amplitude and relative phase,  now  both oscillating close to the same Higgs frequency  $\sim\omega_\mathrm{H,h}$.

We now turn to the THz-MDCS spectra. Figures~\ref{fig8}(a)--(c) present example spectra in Regime~I (Fig.~\ref{fig8}(a)), Regime~II (Figs.~\ref{fig8}(b)), and Regime~III (Fig.~\ref{fig8}(c)). In Regime~I, the THz-MDCS spectrum shows pump--probe, four-wave mixing, and third harmonic generation peaks similar to the weak-excitation result without interband coupling (Fig.~\ref{fig3}(c)). 
With increasing pulse-pair excitation in Regime~II, Fig.~\ref{fig8}(b), difference-frequency Raman peaks form at frequencies $\omega_t=\omega_\mathrm{H,h}\pm\omega_0$ (vertical dashed black lines). Similar to the $U=0$ result presented in Fig.~\ref{fig3}(d),  these peaks are generated by ninth-order nonlinear processes that take into account the light-induced  modulation of the SC energy gap. They split from the conventional peaks located at $\omega_t=\omega_0, 3\,\omega_0$ (vertical solid magenta lines). Unlike in the previous two excitation regimes, however, in Regime~III (Fig.~\ref{fig8}(c)), new dominant peaks emerge at  a {\em different frequency} of  $\omega_t=2\omega_\mathrm{H,h}-\omega_0$ (dashed cyan line).  These  peaks at {\em bi-Higgs frequencies} are not observable for  small $U$ (Fig.~\ref{fig6}) or without interband interaction (Fig.~\ref{fig3}).

We next demonstrate that the bi-Higgs-frequency  sidebands   observed  for strong driving fields and large $U$ are generated by parametrical driving of pseudo-spins by relative phase dynamics at frequency $\omega_\mathrm{H,h}$ arising from  the relative-phase collective mode (see  Sec.~\ref{sec:theory}).
 The parametrical driving of pseudo-spins by Eq.~(\ref{coupling})
 is enhanced by the light-induced formation of the relative phase mode seen in Fig.~\ref{fig7}(b). 
Several observations associate the emergence of bi-Higgs-frequency sidebands in the THz-MDCS spectra with the formation of the relative phase mode at $\omega_\mathrm{H,h}$   of a  light-driven non-equilibrium state absent for small $U$.
 First, we note that the bi-Higgs sideband is clearly observed  in Regime~III, i.~e., above the  critical driving field threshold where the relative phase mode has formed. This result indicates that the relative phase dynamics is crucial for the formation process of the bi-Higgs-frequency sideband. Second, as demonstrated in Fig.~\ref{fig8}(d), the THz-MDCS  peaks at $\omega_t=2\omega_\mathrm{H,h}-\omega_0$ vanish when the term $|\Delta_\mathrm{h}|^2 \sin(2 \Delta \theta)\Delta\tilde{\rho}_1^{(\mathrm{h})}$ in Eq.~(\ref{coupling}) is switched off. This term causes the time-dependent transverse coupling of the pseudo-spin components driven by $\Delta \theta(t)$. This coupling is enhanced by the strong modulation of the superfluid
density, which is characterized by the light–induced changes in $|\Delta_\mathrm{h}|^2$. Expansion of this  coupling in terms of $\Delta\theta$ gives  
\begin{align}
	2|\Delta_\mathrm{h}|^2  \sin 2\Delta\theta(t)\Delta\mathrm{\rho}_1^{(\mathrm{h})}=4|\Delta_\mathrm{h}|^2 \Delta \theta\Delta\tilde{\rho}_1^{(\mathrm{h})}+\mathcal{O}((\Delta \theta)^2)\,,
	\label{eq:biHiggs}
\end{align}
to lowest order in $\Delta\theta$. In Regime~III,  long-lived  relative phase oscillations   at  frequency $\omega_\mathrm{H,h}$ arise from the formation of the under-damped relative phase mode. As discussed in Sec.~\ref{sec:U=0}, the $\Delta\tilde{\rho}_1^{(\nu)}$ dynamics is dominated by quasi-particle excitations also at frequency $\sim \omega_\mathrm{H,h}$. Therefore,  the transverse coupling term~(\ref{eq:biHiggs}) drives the nonlinear coupled parametric oscillator equations~(\ref{eq:eom_t2}) with the bi-Higgs frequency $\sim 2\omega_\mathrm{H,h}$. Based on Eq.~(\ref{eq:delta_rho0}), we then find that the the transverse coupling of pseudo-spin components manifests itself via bi-Higgs-frequency  sideband peaks located at $(2\omega_\mathrm{H,h}-\omega_0,0)$ and $(2\omega_\mathrm{H,h}-\omega_0,-2\omega_\mathrm{H,h}+\omega_0)$. These Floquet-like sidebands  are clearly  observed  as distinct  peaks in our numerical results.  They are generated  by  ninth-order nonlinear processes $2\omega_\mathrm{A,B}-2\omega_\mathrm{A,B}+2\omega^\mathrm{B,A}_\mathrm{H,h}-\omega_\mathrm{B,A}$. Third, to provide further evidence for the proposed physical picture, we show in Fig.~\ref{fig8}(e) the field dependence of the maximum of the $2|\Delta_\mathrm{h}|^2  \sin 2\Delta\theta$ spectrum (amplitude of the driving term~(\ref{eq:biHiggs})) while the field dependence of  the bi-Higgs-frequency sideband peak in the THz-MDCS spectra at $\omega_t=2\omega_\mathrm{H,h}-\omega_0$ is plotted in Fig.~\ref{fig8}(f).   Both quantities are smaller than $10^{-3}$ in Regime~I. In regime II, where the relative phase mode red shifts towards the Higgs mode energy, both quantities increase up to an amplitude of about $10^{-2}$. Specifically, this increase flattens close to the transition from Regime II to Regime III. In Regime III however, where the relative phase mode is located close to the Higgs mode energy, both quantities grow strong compared to the intensity range of $[400, 600]$~kV/cm in Regime II. This strong increase of both quantities indicates that the bi-Higgs frequency signals results from the long-lived dynamics of the order parameter relative phase.

In summary, a comparison of Fig.~\ref{fig3} ($U=0$), Fig.~\ref{fig6} ($|U| <1$) and 
Fig.~\ref{fig8} ($|U|>1$ demonstrates how the changes in the spectral profile of THz-MDCS with increasing  pulse-pair driving reflects the properties of the SC non-equilibrium states for different interband interactions in the case of zero Cooper pair momentum after the pulse. In the next section, we show how the dynamical breaking of inversion symmetry achieved via electromagnetic propagation effects changes the above spectral profiles in significant ways that allow the direct observation of the non-equilibrium state collective modes without applying any symmetry-breaking static fields.

\section{THz-MDCS  with light-induced  dynamical inversion symmetry breaking}
\label{sec:IS}

\begin{figure}[t!]
\begin{center}
		\includegraphics[scale=0.55]{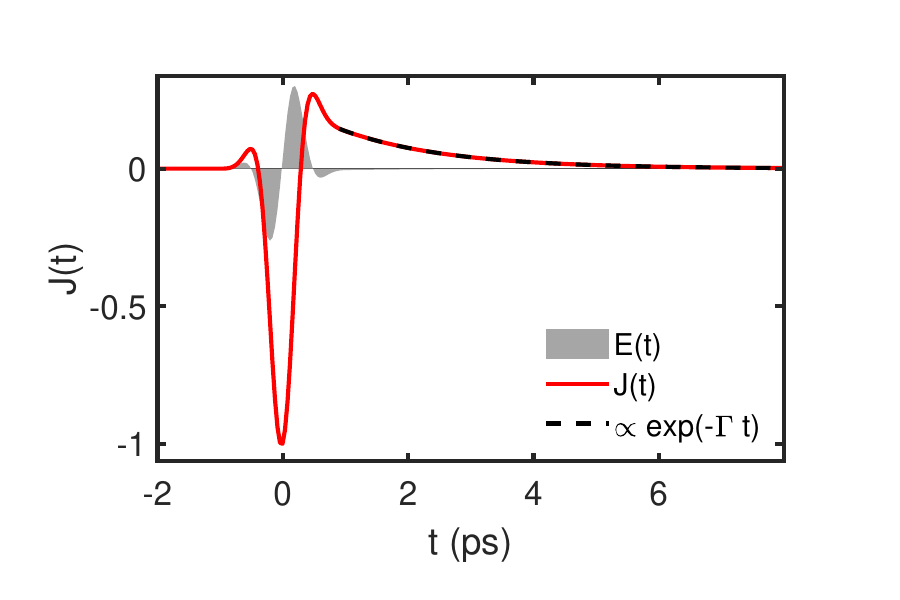}
		\caption{Time dependence of the current density $J(t)$ (red curve)  when electromagnetic propagation effects are taken into account in a thin film geometry with film thickness $d=20$~nm.  By comparing the time-dependence of $J(t)$ with that of the THz laser pulse (shaded area), it is clear that the current lasts longer than the pulse duration.  It decays exponentially after the pulse, with the dashed line showing the radiative damping with decay time $\sim 1.7$~ps.}
		\label{rad_ps} 
\end{center}
\end{figure}

So far, we have presented results without including the electromagnetic propagation effects. The latter effects change the effective  driving field from the laser field considered in the previous sections, which results in 
a current that lasts much longer than the laser pulse (Fig.~\ref{rad_ps}). This current leads to persistent light-induced inversion symmetry breaking, which  has significant effects on the THz-MDCS spectral peaks that can be directly  observed experimentally. To identify the corresponding spectral features, we consider the full self-consistent solution  of the SC Bloch equations, which  takes into account light-induced finite momentum Cooper pairing  persisting well after the pulse.  

\begin{figure*}[t!]
\begin{center}
		\includegraphics[scale=0.55]{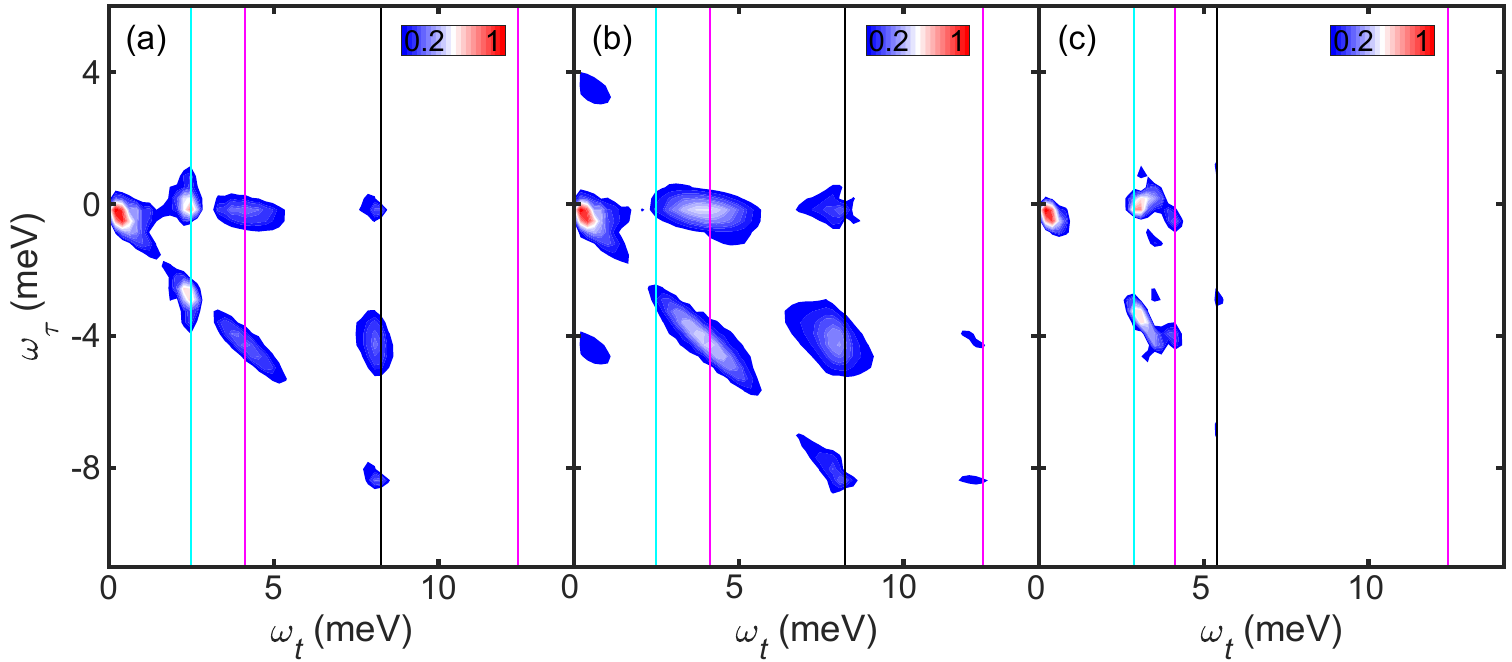}
		\caption{Sensing of the Higgs and Leggett collective modes and THz dynamical inversion symmetry breaking in the THz-MDCS spectra for weak interband interaction  $U=0.01$ as in Fig.~\ref{fig6}. (a) Normalized THz-MDCS spectrum $E_\mathrm{NL}(\omega_t,\omega_\tau)$ resulting from a simulation with light-wave electromagnetic propagation effects  for an electric field strength of 200~kV/cm. (b),(c) Normalized $E_\mathrm{NL}(\omega_t,\omega_\tau)$ for a calculation with propagation effects but without Leggett collective mode effects and a calculation with propagation effects but without Higgs collective mode effects. Vertical magenta lines denote $\omega_t=\omega_0$ and $\omega_t=3\omega_0$. The Leggett (Higgs) mode signals at $\omega_t=\omega_\mathrm{L}$ ($\omega_t=\omega_\mathrm{H,h}$) are indicated with a solid cyan (black) line.}
		\label{fig9} 
\end{center}
\end{figure*}

\subsection{Dominant intraband interaction}

\label{sec:smallU}

For one-band SCs, Ref.~\cite{Mootz2022}
 showed  that photo-generation of a DC current  that persists well after the pulse 
leads to 
Higgs collective mode distinct peaks in the THz-MDCS spectra. The latter peaks  are generated by high-order  difference-frequency Raman processes~\cite{Mootz2020}, with inversion  symmetry-breaking breaking achieved  dynamically via  light-wave electromagnetic propagation.
We have already shown that 
such  photogenerated DC supercurrent yields high-harmonic generation at equilibrium-symmetry forbidden frequencies, and also leads to gapless superconductivity~\cite{Mootz2020,yang2019lightwave,vaswani2019discovery}. Here, we extend these results to 
multi-band SCs and 
 demonstrate direct sensing with  THz-MDCS of Leggett and Higgs collective excitations of strongly driven states.  While the Higgs mode can  also be observed for weak fields if a DC external current is applied to break the equilibrium inversion  symmetry, with increasing field,  it is  hidden by the quasi-particle continua in conventional one-dimensional spectra ~\cite{NatPhys}.    As a result,  we cannot discern the collective modes of the strongly driven non-equilibrium state in this way, unlike for  THz-MDCS.
 
 To take into account the  light-wave electromagnetic propagation effects, we  solved self-consistently the Bloch equations~(\ref{eq:eoms-PS}) together with Eq.~(\ref{eq:trans}) in the homogeneous system with penetration depth larger than the film thickness. Figure~\ref{fig9}(a) presents an example of the THz-MDCS spectrum obtained with this calculation for weak interband interaction $U$=0.01 similar to Fig.~\ref{fig6}.  While we used the same field strength and interband interaction strength as in Fig.~\ref{fig6}(b), the THz-MDCS of Fig.~\ref{fig9}(a) displays a different spectral profile. In particular, the 
finite-momentum Cooper pairing results in 
  a THz-MDCS spectrum showing distinct peaks at the Leggett mode energy ($\omega_t=\omega_\mathrm{L}$, solid cyan line) and at the Higgs mode energy ($\omega_t=\omega_\mathrm{H,h}$, solid black line). 
  These sharp peaks replace the broad lineshape of  Fig.~\ref{fig6}(b). To verify that they  are indeed generated by Leggett and Higgs collective effects, rather than by quasi-particle excitations, 
   we show in Figs.~\ref{fig9}(b) and \ref{fig9}(c) the corresponding results of our simulations without the  Leggett  and Higgs collective  contributions to Eq.~(\ref{eq:eom_t2}), respectively. In particular, the peaks at $\omega_t=\omega_\mathrm{L}$ vanish without the phase mode collective effects in Eq.~(\ref{eq:eom_t2}), while the signals at $\omega_t=\omega_\mathrm{H,h}$ are suppressed without the Higgs collective  contribution. These peaks at the collective mode frequencies become observable due  to light-induced symmetry breaking 
   as in Fig.~(\ref{rad_ps})   and replace the sidebands at 
$\omega_{H,h} \pm \omega_0$ and    $\omega_\mathrm{L} \pm \omega_0$ observed without persistent  Cooper pair momentum in Fig.~\ref{fig6}(b). 
   This result demonstrates that the light-induced dynamical inversion-symmetry breaking allows THz-MDCS signals to  be used for  sensing the collective modes of multi-band superconductor non-equilibrium states under 
  strong THz excitation. 
   
It is important to note that the Leggett- and Higgs-mode peaks in the THz-MDCS spectra are generated by high-order coherent nonlinear processes. These high-order nonlinearities
include difference-frequency coherent Raman processes leading to coherent photogeneration of a DC supercurrent and quantum quench of the energy gap during light field oscillation cycles. 
 In particular, 
the observed collective mode peaks characterize a superfluid state with a finite Cooper pair momentum $\mathbf{p}_\mathrm{S}$ that includes both oscillating and static $\omega=0$  components. The finite-momentum Cooper pairing is in addition to the $\omega_0$ oscillatory  component determined by the laser frequency. In its presence, 
 the  sum- and difference-frequency Raman processes  that  led to  the peaks at $\omega_t=\omega_\mathrm{H,h}\pm\omega_0$ and $\omega_t=\omega_0\pm\omega_\mathrm{L}$ in the THz-MDCS spectra of Fig.~\ref{fig6}  now generate new  peaks at $\omega_t=\omega_\mathrm{L}$ and $\omega_t=\omega_\mathrm{H,h}$. 
These new peaks reflect the light-induced inversion symmetry breaking that persists  after the pulse in a moving condensate state. 
 To explain the THz-MDCS  experiments~\cite{NatPhys},  an important difference from semiconductors in the strong excitation regime is that  we must consider nonlinear processes around a light-driven non-equilibrium SC state   that differs from the equilibrium state.  As a result, THz-MDCS can be used to observe  the excitations of  non-equilibrium superconductor states controlled by a pulse-pair. 
More generally, unlike for the rigid properties of  conventional materials determined by bandstructure,  the soft properties of quantum materials can be coherently modified  by light, which must be considered when interpreting their nonlinear response~\cite{Lingos2017,Lingos2021}

\subsection{Dominant interband interaction}

\label{sec:largeU}

\begin{figure}[t!]
\begin{center}
		\includegraphics[scale=0.55]{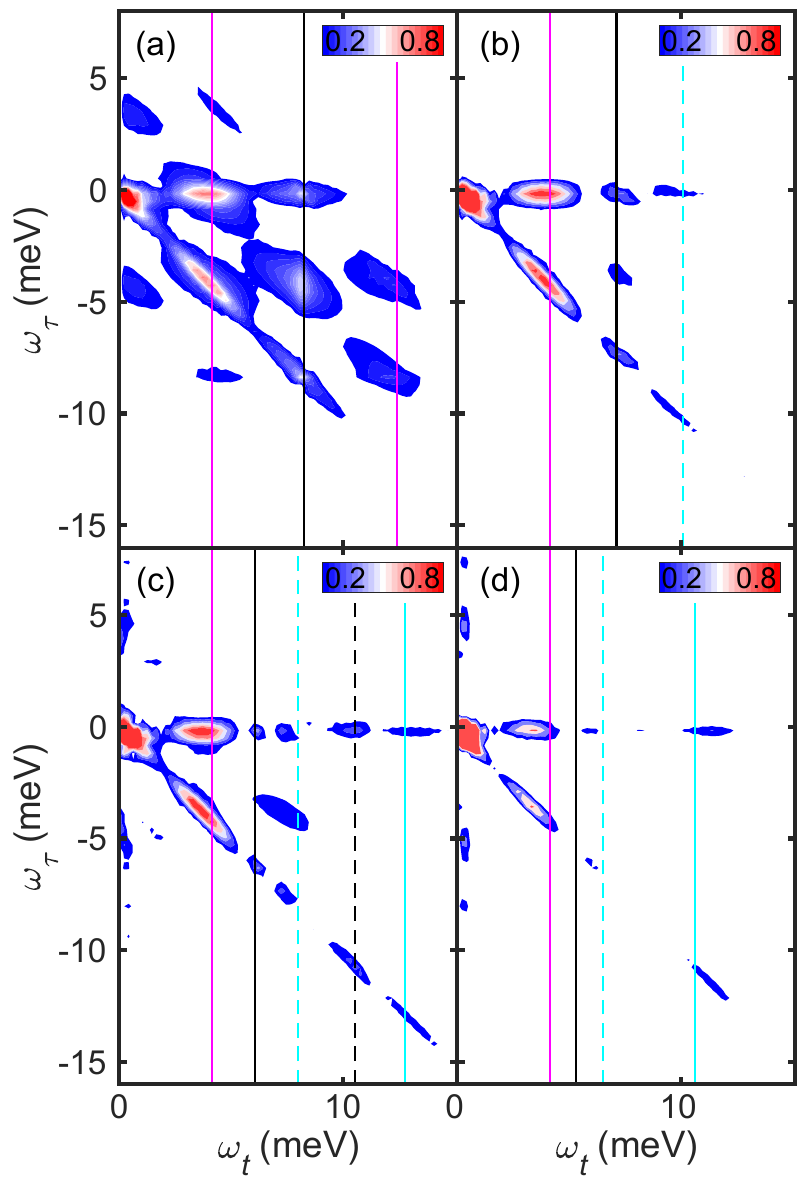}
		\caption{Transition from a non-equilibrium state characterized by a hybrid-Higgs amplitude collective mode to a non-equilibrium state determined by a relative phase mode.
				Normalized THz-MDCS spectra $E_\mathrm{NL}(\omega_t,\omega_\tau)$ resulting from a simulation with light-wave electromagnetic propagation effects for electric field strength of (a) 150~kV/cm, (b) 400~kV/cm, (c) 650~kV/cm, and (d) 900~kV/cm. Vertical magenta lines indicate $\omega_t=\omega_0$ and $\omega_t=3\omega_0$. The signals at $\omega_t=\omega_\mathrm{H,h}$ and $\omega_t=2\omega_\mathrm{H,h}$ are denoted by vertical solid black and cyan lines, respectively. The signals at $\omega_t=\omega_\mathrm{H,h}+\omega_0$ and $\omega_t=2\omega_\mathrm{H,h}-\omega_0$ are marked by vertical dashed black and cyan lines, respectively.}
		\label{fig10} 
\end{center}
\end{figure}

The light-induced transition for strong  $U$=2 from a non-equilibrium state characterized by the Higgs mode to a non-equilibrium state determined by the relative phase of Sec.~\ref{sec:U>0} is directly observable in the THz-MDCS spectra when dynamical inversion symmetry breaking is induced by the electromagnetic propagation.  Figure~\ref{fig10} shows examples of the calculated THz-MDCS spectra in Regime~I (Fig.~\ref{fig10}(a)), Regime~II (Fig.~\ref{fig10}(b)), at the transition from Regime~II to Regime~III (Figs.~\ref{fig10}(c)), and in Regime~III (Fig.~\ref{fig10}(d)). In Regime~I (lowest fields), the THz-MDCS spectrum shows strong peaks at the Higgs mode frequency $\omega_t=\omega_\mathrm{H,h}$. These peaks measure directly  the Higgs amplitude collective mode (black solid line) as discussed in Sec.~\ref{sec:smallU}. Unlike for conventional one-dimensional spectroscopies, the Higgs frequency peaks  are well  separated from the pump--probe, four-wave mixing, and third harmonic generation peaks (vertical magenta lines) in 2D frequency space. By providing a  high-resolution two-dimensional visualization of the non-equilibrium SC state,  THz-MDCS allows   to directly detect the collective modes of the non-equilibrium driven states. 
In Regime~II (intermediate fields, Fig.~\ref{fig10}(b)), a relative phase starts to form from the Higgs and hybrid-Higgs amplitude modes. The bi-Higgs-frequency sidebands at $\omega_t=2\omega_\mathrm{H,h}-\omega_0$ (dashed cyan line) then start to emerge slightly above the Higgs mode signals at $\omega_t=\omega_\mathrm{H,h}$. 
At the transition from Regime~II to Regime~III, Fig.~\ref{fig10}(c), the THz-MDCS spectrum shows both Higgs (dashed black line) and bi-Higgs-frequency satellites (dashed cyan line), at $\omega_t=\omega_\mathrm{H,h}+\omega_0$ and $\omega_t=2\omega_\mathrm{H,h}-\omega_0$, in addition to the Higgs mode signals at $\omega_t=\omega_\mathrm{H,h}$ (vertical solid black line). Most importantly, 
new signals now emerge at the bi-Higgs  frequency $\omega_t=2\omega_\mathrm{H,h}$ (vertical solid cyan line), which represent a direct observation of the relative phase mode. 
 These bi-Higgs-frequency peaks become observable due to persisting inversion-symmetry breaking by  photo-generated DC supercurrent and finite-momentum Cooper pairing. The bi-Higgs-frequency signals dominate over the Higgs peaks in Regime~III (higher fields, Fig.~\ref{fig10}(d)). 
In this excitation regime, the strong nonlinear increase of the $2|\Delta_\mathrm{h}|^2  \sin 2\Delta\theta$ spectral
peak in Fig.~\ref{fig8}(e)  
 exceeds the corresponding light-induced DC current contribution. As a result,  the
THz–MDCS signals generated by the  relative phase collective mode at $\omega_t=2\omega_\mathrm{H,h}$ dominate
over the Higgs collective mode signals at $\omega_t=\omega_\mathrm{H,h}$. 
The observed change in THz-MDCS spectral profile reflects the transition from a non-equilibrium state characterized by the hybrid Higgs collective mode to a non-equilibrium state determined by the relative phase mode, observed experimentally  in Ref.~\cite{NatPhys}. The agreement between experiment and theory indicates that long-lived, light-accelerated
moving condensate driven states can be sensed and controlled with super-resolution by THz MDCS since it provides the necessary resolution to observe higher-order correlations and corresponding nonlinear responses in contrast to one-dimensional coherent spectroscopy where multiple different nonlinear processes contribute to the same frequencies.

\section{Summary}
\label{sec:discussions}

In this paper, we have provided a detailed microscopic theory  allowing us to propose  that THz-MDCS experiments on multi-band superconductors with varying interband couplings can be used  to both drive different   non-equilibrium SC states  and characterize them uniquely by measuring their collective modes. For this purpose,  we first extended the gauge-invariant density matrix approach of Refs.~\cite{Mootz2020,Mootz2022} to the multi-band case.  The resulting superconductor Maxwell--Bloch equations describe the nonlinear dynamics of light-wave acceleration of different finite-momentum Cooper-pair condensate states that live after the pulse. We have shown that these non-equilibrium states  are characterized by quasi-particle excitations and  collective amplitude and phase modes that differ from those of the equilibrium SC state. We have  pointed  out smoking-gun signals in the THz-MDCS spectra
that signify the transition to such light-driven SC states. For this,  we have derived nonlinearly-coupled  Anderson pseudo-spin oscillator equations of motion. These coupled oscillators  are   driven by the competition between the time-dependent superfluid momentum and the order parameter relative phase. This light-controlled competition depends critically on the interband interaction strength. The superfluid-momentum  is controlled by both the applied THz laser field  and  by light-wave electromagnetic propagation effects. As a result, the effective field pulse that drives the superconductor differs from the applied laser pulse. The spectrum of this effective driving pulse can include an inversion  symmetry-breaking DC component, which can lead to formation of a DC supercurrent and long-lived  finite-momentum Cooper pairing.  On the other hand, a  transverse driving of the Anderson pseudo-spin oscillators by the relative phase mode oscillations  can dominate over analogous effects driven by the Cooper pair momentum for multi-band superconductors with strong interband coupling. Here, unlike for  the strongly damped Leggett collective mode of the equilibrium state, 
which lies within the quasi-particle continuum, 
the relative phase collective mode has frequency close to the SC energy gap in the case of  strong interband interaction. 

 We have applied the presented theory in the homogeneous limit  to calculate the  THz-MDCS spectra for different interband-to-intraband interaction ratios and for weak to strong pulse-pair excitation. The presented results demonstrate a direct ultrafast visualization of THz-driven moving condensate states by using THz-MDCS. In particular, in the case of small  interband interactions, the THz-MDCS spectra show a transition from traditional third-order nonlinear signals, i.~e. pump--probe, four-wave mixing, and third-harmonic generation signals, to harmonic  sidebands that  split  from the conventional signals due to the quantum quench of the superconductor excitation energy gap as a result of Raman difference-frequency nonlinear processes  during cycles of laser field oscillations.   With increasing driving field, ninth-order nonlinear signals (with respect to the equilibrium state)  emerge  as a result of  quasi-particle excitations of the non-equilibrium state assisted by difference-frequency Raman processes. For intraband interaction exceeding the interband coupling, the Leggett phase mode lies within the energy gap and leads to the formation of harmonic sidebands around the pump--probe peaks. We have shown that the photo-excitation of a  long-lived DC supercurrent via light-wave propagation inside a thin superconducting film, neglected in previous works,  allows the Leggett and Higgs collective modes to be  simultaneously detected as distinct peaks in the THz-MDCS spectra. 

We also studied the case of multi-band superconductors with interband coupling exceeding the intraband pairing  interaction. The drastic change of the THz-MDCS spectral profile in this case reveals  a transition from a non-equilibrium state characterized by a hybrid-Higgs amplitude collective mode to a state with a relative phase collective mode. The latter strongly-driven state manifests itself directly in the THz-MDCS spectra via the formation of Floquet-like sidebands at bi-Higgs frequencies.  We have demonstrated that such new bi-Higgs-frequency signals are generated by ninth-order nonlinear processes (with respect to the equilibrium state) and arise from the interaction of the  relative phase collective mode with quasi-particle excitations, both at the Higgs mode energy. The relative phase  collective mode forms above critical driving when the relative phase Leggett mode, which close to equilibrium is located within the quasi-particle continuum, has shifted towards the low-energy Higgs mode. The formed relative phase collective  mode parametrically drives pseudo-spin oscillators. This light-induced effect is enhanced by the strong modulation of the superfluid density, such that the relative phase mode driving of the pseudo-spin oscillators exceeds their superfluid momentum driving at elevated  effective fields. As a result, bi-Higgs-frequency  signals exceed the harmonic Higgs peaks in the THz-MDCS spectra, but only  in the presence of photo-generated DC supercurrent and persistent finite-momentum Cooper pairing.

The theoretical approach presented here can be extended to study the THz-driven non-equilibrium dynamics in  systems with SC order parameter coupled to spin or charge order, as in iron-based and topological superconductors. There, THz-MDCS could be used to detect new non-equilibrium states and to identify the SC order-parameter equilibrium symmetry. Specifically, we expect that superconductors with $d$-wave order parameter symmetry will show a very different THz-MDCS spectrum than $s$-wave superconductors due to the presence of gap nodes. In this context, it will be also interesting to see how the real and imaginary parts of the THz-MDCS spectra change as a function of interband-interaction strength and field strength for different SC order-parameter equilibrium symmetries. Our theoretical approach can also be applied to study more complex amplitude modes of the SC order parameter, such as  e.~g.~, soliton-like states~\cite{Yuzbashyan:2006,Yuzbashyan2008}. The latter non-equilibrium states are characterized by persisting oscillations of the SC order parameter with unchanged amplitude and frequency~\cite{Mootz2020,Balseiro}. These states  emerge above critical THz-field strengths and are driven by Rabi--Higgs flopping, to be discussed elsewhere. Such non-equilibrium state modes are expected to show up as unique  distinct peaks in the THz-MDCS spectra. 
Similarly, THz-MDCS can be used to identify the role of coherent interactions or correlations between many collective modes of the non-equilibrium states. Moreover, the THz-MDCS approach and understanding presented here can be readily applied to explore the intriguing realm of quantum materials featuring topological~\cite{Luo,Luo2019,Yang2020,cheng2023chirality,vasw2020}, magnetic~\cite{Patz2015,femtomag}, and charge density wave properties~\cite{Song2023}. 

In summary, by using our developed theory to analyze THz-MDCS experiments and drive quantum nonlinear dynamics with strong phase-locked pulse-pairs, it is possible to reconstruct non-equilibrium
driven quantum states with super resolution in a broad range of interesting quantum systems, thus achieving quantum tomography of non-equilibrium light-driven quantum states.

\appendix 

\section{Gauge-invariant density matrix approach}	
\label{sec:gidma}

In this appendix, we extend the gauge-invariant density matrix approach of Ref.~\cite{Mootz2020} to the multi-band case. We first present the derivation of Hamiltonian~(\ref{eq:Ham}). We start from the general Hamiltonian introduced by Nambu in Ref.~\cite{Nambu} extended to the multi-band case:
\begin{align}
\label{eq:Ham_full}
&H=\sum_{\nu,\alpha}\int\mathrm{d}^3\mathbf{x}\,\psi_{\alpha,\nu}^\dagger(\mathbf{x})\left[\xi_\nu(\mathbf{p}-e\mathbf{A}(\mathbf{x}))+e\phi(\mathbf{x})\right]\psi_{\alpha,\nu}(\mathbf{x}) \nonumber \\
		&-\frac{1}{2}\sum_{\alpha\beta \nu\lambda}\int\mathrm{d}^3\mathbf{x}\mathrm{d}^3\mathbf{x}'\psi_{\alpha,\nu}^\dagger(\mathbf{x})\psi_{\beta,\lambda}^\dagger(\mathbf{x}')\nonumber \\ &\qquad\qquad \times V(\mathbf{x},\mathbf{x}')\psi_{\beta,\lambda}(\mathbf{x}')\psi_{\alpha,\nu}(\mathbf{x})\,,
\end{align}
with Coulomb potential $V(\mathbf{x},\mathbf{x}')$. Applying a mean-field decoupling of the interaction part of Eq.~(\ref{eq:Ham_full}) leads to
\begin{align}
\label{eq:Ham_MF}
H&=\sum_{\nu,\alpha}\int\mathrm{d}^3\mathbf{x}\,\psi_{\alpha,\nu}^\dagger(\mathbf{x})\left[\xi_\nu(\mathbf{p}-e\mathbf{A}(\mathbf{x},t))+e\phi(\mathbf{x},t)\right.\nonumber\\
&\qquad\qquad\qquad\qquad\left.+\mu^\nu_\mathrm{H}(\mathbf{x})\right]\psi_{\alpha,\nu}(\mathbf{x}) \nonumber \\
&-\sum_\nu\int\mathrm{d}^3\mathbf{x}\left[\Delta_\nu(\mathbf{x})\psi^\dagger_{\uparrow,\nu}(\mathbf{x})\psi^\dagger_{\downarrow,\nu}(\mathbf{x})+\mathrm{h.c.}\right]+H_\mathrm{F}\,,
\end{align}
with SC order parameter $\Delta_\nu(\mathbf{x})$ and Hartree potential $\mu^\nu_\mathrm{H}(\mathbf{x})$ defined in Eqs.~(\ref{eq:gap_eq}) and (\ref{eq:mu_H}). Here, we have neglected interband contributions of the form $\langle\psi_{\alpha,\nu}^\dagger(\mathbf{x})\psi_{\beta,\lambda}(\mathbf{x})\rangle$ with $\nu\neq\lambda$ and replaced the Coulomb interaction $V(\mathbf{x},\mathbf{x}')$, except in the Hartree potential, by an effective electron--electron interaction $g_{\nu,\lambda}$. The Fock term in Eq.~(\ref{eq:Ham_MF}) is explicitly given by 
\begin{align}
H_\mathrm{F}=-\sum_{\alpha,\beta,\nu}\int\mathrm{d}^3\mathbf{x}\langle\psi^\dagger_{\alpha,\nu}(\mathbf{x})\psi_{\beta,\nu}(\mathbf{x})\rangle g_{\nu,\nu}\psi^\dagger_{\beta,\nu}(\mathbf{x})\psi_{\alpha,\nu}(\mathbf{x})\,,
    \label{eq:Hf}
\end{align}
which includes contributions with the same spin as well as terms with opposite spin. However, based on Refs.~\cite{Nambu,Anderson}
it is sufficient to include only the contributions with the same spin in the Fock field~(\ref{eq:Hf}) to guarantee charge conservation such that Hamiltonian~(\ref{eq:Ham_MF}) reduces to Eq.~(\ref{eq:Ham}).

Hamiltonian~(\ref{eq:Ham}) is gauge invariant under the general gauge transformation~\cite{Nambu} 
	\begin{equation}
		\label{eq:gauge_trafo}
		\Psi_\nu(\mathbf{x}) \,\rightarrow\, \mathrm{e}^{\mathrm{i}\sigma_3\Lambda(\mathbf{x})/2}\Psi_\nu(\mathbf{x})
	\end{equation}
	when the vector potential, scalar potential, and SC order parameter phases all transform as
	\begin{align}
		&\mathbf{A}(\mathbf{x})\,\rightarrow\,\mathbf{A}(\mathbf{x})+\frac{1}{2e}\nabla\Lambda(\mathbf{x})\,,\quad
		\phi(\mathbf{x})\,\rightarrow\,\phi(\mathbf{x})-\frac{1}{2e}\frac{\partial}{\partial t}\Lambda(\mathbf{x})\,,\nonumber \\
		&\theta_\nu(\mathbf{x})\,\rightarrow\,\theta_\nu(\mathbf{x})+\Lambda(\mathbf{x})\,.
	\end{align}
	In the above,  $\Psi_\nu(\mathbf{x})=(\psi_{\uparrow,\nu}(\mathbf{x}),\psi^\dagger_{\downarrow,\nu}(\mathbf{x}))^T$ is the field operator for band $\nu$ in Nambu space.  Unlike for  the Hamiltonian, the system's density matrix is not  invariant under gauge transformation. More specifically, the diagonal block of the  density matrix describing band $\nu$, $\rho^{(\nu)}(\mathbf{x},\mathbf{x}')=\langle \hat{\rho}^{(\nu)}(\mathbf{x},\mathbf{x}')\rangle=\langle \Psi^\dagger_\nu(\mathbf{x})\Psi_\nu(\mathbf{x}')\rangle$, depends on the specific choice of the gauge. 
As an alternative,  	
	 we introduce center-of-mass and relative coordinates $\mathbf{R}=(\mathbf{x}+\mathbf{x}')/2$ and $\mathbf{r}=\mathbf{x}-\mathbf{x}'$ and define the transformed density matrix~\cite{Stephen1965,Wu2017}
\begin{align}
\label{eq:rho_trans}
	&\tilde{\rho}^{(\nu)}(\mathbf{r},\mathbf{R})=\mathrm{exp}\left[-\mathrm{i}e\int_0^\frac{1}{2}\mathrm{d}\lambda\,\mathbf{A}(\mathbf{R}+\lambda\,\mathbf{r},t)\cdot\mathbf{r}\,\sigma_3\right] \nonumber \\
	&\times\rho^{(\nu)}(\mathbf{r},\mathbf{R})\, \mathrm{exp}\left[-\mathrm{i}e\int^0_{-\frac{1}{2}}\mathrm{d}\lambda\,\mathbf{A}(\mathbf{R}+\lambda\,\mathbf{r},t)\cdot\mathbf{r}\,\sigma_3\right]\,,
\end{align}
where $\rho^{(\nu)}(\mathbf{r},\mathbf{R})=\langle\Psi^\dagger_\nu(\mathbf{R}+\frac{\mathbf{r}}{2})\Psi_\nu(\mathbf{R}-\frac{\mathbf{r}}{2}))\rangle$ is the Wigner function. By applying the gauge transformation (\ref{eq:gauge_trafo}), we see that, unlike for the original density matrix,  $\tilde{\rho}^{(\nu)}(\mathbf{r},\mathbf{R})$ transforms as~\cite{Wu2017}
\begin{equation}
	\tilde{\rho}^{(\nu)}(\mathbf{r},\mathbf{R})\,\rightarrow\,\mathrm{exp}\left[\mathrm{i}\sigma_3\Lambda(\mathbf{R})/2\right]\tilde{\rho}^{(\nu)}(\mathbf{r},\mathbf{R})\mathrm{exp}\left[-\mathrm{i}\sigma_3\Lambda(\mathbf{R})/2\right]\,.
\end{equation}
Here, the phase $\Lambda(\mathbf{R})$ only depends on the center-of-mass coordinate and not on both coordinates $\mathbf{R}$ and $\mathbf{r}$. The latter  is the case for the phase of $\rho^{(\nu)}(\mathbf{r},\mathbf{R})$, which complicates a gauge-invariant description of the non-equilibrium dynamics of superconductors, especially in the nonlinear regime.

\section{Gauge-invariant spatially-dependent superconductor Bloch equations}
\label{eq:be_full}

We calculate the dynamics of the density matrix (\ref{eq:rho_trans}) by using the Heisenberg equation of motion
\begin{equation}
\mathrm{i}\frac{\partial}{\partial t}\tilde{\rho}^{(\nu)}=\langle \left[\tilde{\rho}^{(\nu)},H\right] \rangle\,.
\end{equation}
Here we  work with the Wigner function
\begin{equation}
	\tilde{\rho}^{(\nu)}(\mathbf{k},\mathbf{R})=\int\mathrm{d}^3\mathbf{r}\,\tilde{\rho}^{(\nu)}(\mathbf{r},\mathbf{R})\,\mathrm{e}^{-\mathrm{i}\mathbf{k}\cdot\mathbf{r}}\,.
\end{equation}
rather than $\tilde{\rho}^{(\nu)}(\mathbf{r},\mathbf{R})$,
obtained by Fourier transformation with respect to the relative coordinate $\mathbf{r}$.
In the case of weak order parameter spatial dependence, we can then 
expand contributions to the equations of motion of the form $\Delta_\nu(\mathbf{R}+\frac{\mathrm{i}}{2}\nabla_\mathbf{k})\tilde{\rho}^{(\nu)}(\mathbf{k},\mathbf{R})$  by applying the gradient expansion 
\begin{equation}
\label{eq:gradient}
	\Delta_\nu(\mathbf{R}+\frac{\mathrm{i}}{2}\nabla_\mathbf{k})=\sum_{n=0}^\infty\left(\frac{\mathrm{i}}{2}\right)^n\frac{(\nabla_\mathbf{R}\cdot\nabla_\mathbf{k})^n}{n!}\Delta_\nu(\mathbf{R})\,.
\end{equation}
Such expansions in powers of $\nabla_\mathbf{R}\cdot\nabla_\mathbf{k}$ are most useful  when the coherence length of the SC state is shorter  than the spatial variation of the  system.
 To simplify the equations of motion further, we apply the unitary transformation
\begin{align}
		\tilde{\rho}^{(\nu)}(\mathbf{k},\mathbf{R})=\mathrm{e}^{-\mathrm{i}\sigma_3\theta_{\nu_0}(\mathbf{R})/2}\tilde{\rho}^{(\nu)}(\mathbf{k},\mathbf{R})\mathrm{e}^{\mathrm{i}\sigma_3\theta_{\nu_0}(\mathbf{R})/2}\,,
		\label{eq:phase-trafo}
\end{align}
such that the phase of the SC order parameter $\Delta_{\nu_0}(\mathbf{R})$ only appears in an effective chemical potential as demonstrated below. We thus  obtain the most general gauge-invariant SC Bloch equations for multi-band spatially-dependent superconductors driven by both electric and magnetic fields. These equations include the effects of a spatially--dependent scalar potential or SC order parameter phase. The exact equations of motion for the 
general $\tilde{\rho}^{(\nu)}(\mathbf{k},\mathbf{R})$ are  as follows: 
\begin{widetext}
\begin{align}
\label{eq:eom-full1-f}
	\mathrm{i}\frac{\partial}{\partial t}\tilde{\rho}^{(\nu)}_{1,1}(\mathbf{k},\mathbf{R})&=\left[\xi_\nu\left(\mathbf{k}-\frac{\mathrm{i}}{2}\nabla_\mathbf{R}+\mathrm{i}\frac{e}{2}\sum_{n=0}^\infty\left(-\frac{1}{4}\right)^n\frac{(\nabla_\mathbf{k}\cdot\nabla_\mathbf{R})^{2n}}{(2n+1)!}\nabla_\mathbf{k}\times\mathbf{B}(\mathbf{R})\right.\right. \nonumber \\
	&\left.\left.\qquad\quad\;-e\sum_{n=1}^\infty 2n\left(-\frac{1}{4}\right)^n\frac{(\nabla_\mathbf{k}\cdot\nabla_\mathbf{R})^{2n-1}}{(2n+1)!}\nabla_\mathbf{k}\times\mathbf{B}(\mathbf{R})\right)\right. \nonumber \\ 
	&\quad\;\;\left.-\xi_\nu\left(\mathbf{k}+\frac{\mathrm{i}}{2}\nabla_\mathbf{R}-\mathrm{i}\frac{e}{2}\sum_{n=0}^\infty\left(-\frac{1}{4}\right)^n\frac{(\nabla_\mathbf{k}\cdot\nabla_\mathbf{R})^{2n}}{(2n+1)!}\nabla_\mathbf{k}\times\mathbf{B}(\mathbf{R})\right.\right. \nonumber \\
	&\left.\left.\qquad\quad\;-e\sum_{n=1}^\infty 2n\left(-\frac{1}{4}\right)^n\frac{(\nabla_\mathbf{k}\cdot\nabla_\mathbf{R})^{2n-1}}{(2n+1)!}\nabla_\mathbf{k}\times\mathbf{B}(\mathbf{R})\right)\right. \nonumber \\ 
	&\quad\;\;\left. -2\sum_{n=0}^\infty\frac{\left(\frac{\mathrm{i}}{2}\right)^{2n+1}(\nabla_\mathbf{k}\cdot\nabla_\mathbf{R})^{2n+1}}{(2n+1)!}\left(\mu^\nu_\mathrm{H}(\mathbf{R})+\mu_\mathrm{F}^{\uparrow,\nu}(\mathbf{R})\right)
	\right. \nonumber \\ 
	&\quad\;\;\left.-\mathrm{i}e\sum_{n=0}^\infty\frac{(\nabla_\mathbf{k}\cdot\nabla_\mathbf{R})^{2n}}{(2n+1)!}\left(-\frac{1}{4}\right)^n\,\mathbf{E}(\mathbf{R})\cdot\nabla_\mathbf{k}\right]\tilde{\rho}^{(\nu)}_{1,1}(\mathbf{k},\mathbf{R}) \nonumber \\
	&+\mathrm{exp}\left[\frac{\mathrm{i}}{2}\nabla_\mathbf{R}\cdot\nabla_\mathbf{k}\right]|\Delta_\nu(\mathbf{R})|\;\mathrm{exp}\left[-\mathrm{i}\,\delta\theta_\nu(\mathbf{R)}-\frac{1}{2}\sum_{n=0}^\infty\frac{(\nabla_\mathbf{k}\cdot\nabla_\mathbf{R})^n}{(n+1)!}\left(\frac{\mathrm{i}}{2}\right)^n\,\mathbf{p}^\nu_\mathrm{S}(\mathbf{R})\cdot\nabla_\mathbf{k}\right]\tilde{\rho}^{(\nu)}_{2,1}(\mathbf{k},\mathbf{R})\nonumber \\
	&-\mathrm{exp}\left[-\frac{\mathrm{i}}{2}\nabla_\mathbf{R}\cdot\nabla_\mathbf{k}\right]|\Delta_\nu(\mathbf{R})|\;\mathrm{exp}\left[\mathrm{i}\,\delta\theta_\nu(\mathbf{R})-\frac{1}{2}\sum_{n=0}^\infty\frac{(\nabla_\mathbf{k}\cdot\nabla_\mathbf{R})^n}{(n+1)!}\left(-\frac{\mathrm{i}}{2}\right)^n\,\mathbf{p}^\nu_\mathrm{S}(\mathbf{R})\cdot\nabla_\mathbf{k}\right]\tilde{\rho}^{(\nu)}_{1,2}(\mathbf{k},\mathbf{R})\,,
\end{align}
\begin{align}	
\label{eq:eom-full2-f}
	\mathrm{i}\frac{\partial}{\partial t}\tilde{\rho}^{(\nu)}_{2,2}(\mathbf{k},\mathbf{R})&=\left[\xi_\nu\left(-\mathbf{k}-\frac{\mathrm{i}}{2}\nabla_\mathbf{R}-\mathrm{i}\frac{e}{2}\sum_{n=0}^\infty\left(-\frac{1}{4}\right)^n\frac{(\nabla_\mathbf{k}\cdot\nabla_\mathbf{R})^{2n}}{(2n+1)!}\nabla_\mathbf{k}\times\mathbf{B}(\mathbf{R})\right.\right. \nonumber \\
	&\left.\left.\qquad\quad\;+e\sum_{n=1}^\infty 2n\left(-\frac{1}{4}\right)^n\frac{(\nabla_\mathbf{k}\cdot\nabla_\mathbf{R})^{2n-1}}{(2n+1)!}\nabla_\mathbf{k}\times\mathbf{B}(\mathbf{R})\right)\right. \nonumber \\ 
	&\quad\;\;\left.-\xi_\nu\left(-\mathbf{k}+\frac{\mathrm{i}}{2}\nabla_\mathbf{R}+\mathrm{i}\frac{e}{2}\sum_{n=0}^\infty\left(-\frac{1}{4}\right)^n\frac{(\nabla_\mathbf{k}\cdot\nabla_\mathbf{R})^{2n}}{(2n+1)!}\nabla_\mathbf{k}\times\mathbf{B}(\mathbf{R})\right.\right. \nonumber \\
	&\left.\left.\qquad\quad\;+e\sum_{n=1}^\infty 2n\left(-\frac{1}{4}\right)^n\frac{(\nabla_\mathbf{k}\cdot\nabla_\mathbf{R})^{2n-1}}{(2n+1)!}\nabla_\mathbf{k}\times\mathbf{B}(\mathbf{R})\right)\right. \nonumber \\ 
	&\quad\;\;\left. +2\sum_{n=0}^\infty\frac{\left(\frac{\mathrm{i}}{2}\right)^{2n+1}(\nabla_\mathbf{k}\cdot\nabla_\mathbf{R})^{2n+1}}{(2n+1)!}\left(\mu^\nu_\mathrm{H}(\mathbf{R})+\mu_\mathrm{F}^{\downarrow,\nu}(\mathbf{R})\right)\right. \nonumber \\	 
	&\quad\;\;\left.+\mathrm{i}e\sum_{n=0}^\infty\frac{(\nabla_\mathbf{k}\cdot\nabla_\mathbf{R})^{2n}}{(2n+1)!}\left(-\frac{1}{4}\right)^n\,\mathbf{E}(\mathbf{R})\cdot\nabla_\mathbf{k}\right]\tilde{\rho}^{(\nu)}_{2,2}(\mathbf{k},\mathbf{R}) \nonumber \\
	&-\mathrm{exp}\left[-\frac{\mathrm{i}}{2}\nabla_\mathbf{R}\cdot\nabla_\mathbf{k}\right]|\Delta_\nu(\mathbf{R})|\;\mathrm{exp}\left[-\mathrm{i}\,\delta\theta_\nu(\mathbf{R})+\frac{1}{2}\sum_{n=0}^\infty\frac{(\nabla_\mathbf{k}\cdot\nabla_\mathbf{R})^n}{(n+1)!}\left(-\frac{\mathrm{i}}{2}\right)^n\,\mathbf{p}^\nu_\mathrm{S}(\mathbf{R})\cdot\nabla_\mathbf{k}\right]\tilde{\rho}^{(\nu)}_{2,1}(\mathbf{k},\mathbf{R})\nonumber \\
	&+\mathrm{exp}\left[\frac{\mathrm{i}}{2}\nabla_\mathbf{R}\cdot\nabla_\mathbf{k}\right]|\Delta_\nu(\mathbf{R})|\;\mathrm{exp}\left[\mathrm{i}\,\delta\theta_\nu(\mathbf{R})+\frac{1}{2}\sum_{n=0}^\infty\frac{(\nabla_\mathbf{k}\cdot\nabla_\mathbf{R})^n}{(n+1)!}\left(\frac{\mathrm{i}}{2}\right)^n\,\mathbf{p}^\nu_\mathrm{S}(\mathbf{R})\cdot\nabla_\mathbf{k}\right]\tilde{\rho}^{(\nu)}_{1,2}(\mathbf{k},\mathbf{R})\,,
	\end{align}
\begin{align}
\label{eq:eom-full3-f}
	\mathrm{i}\frac{\partial}{\partial t}\tilde{\rho}^{(\nu)}_{1,2}(\mathbf{k},\mathbf{R})&=\left[-\xi_\nu\left(\mathbf{k}+\frac{\mathrm{i}}{2}\nabla_\mathbf{R}-e\sum_{n=0}^\infty(2n+1)\left(\frac{\mathrm{i}}{2}\right)^{2n+1}\frac{(\nabla_\mathbf{k}\cdot\nabla_\mathbf{R})^{2n}}{(2n+2)!}\nabla_\mathbf{k}\times\mathbf{B}(\mathbf{R})\right.\right. \nonumber \\
		&\qquad\qquad\left.\left.-\mathrm{i}\frac{e}{2}\sum_{n=0}^\infty\left(\frac{\mathrm{i}}{2}\right)^{2n+1}\frac{(\nabla_\mathbf{k}\cdot\nabla_\mathbf{R})^{2n+1}}{(2n+2)!}\nabla_\mathbf{k}\times\mathbf{B}(\mathbf{R})-\mathbf{p}_\mathrm{S}^{\nu_0}(\mathbf{R})/2\right)\right. \nonumber \\ 
	&\quad\;\;\left.-\xi_\nu\left(-\mathbf{k}+\frac{\mathrm{i}}{2}\nabla_\mathbf{R}+e\sum_{n=0}^\infty(2n+1)\left(\frac{\mathrm{i}}{2}\right)^{2n+1}\frac{(\nabla_\mathbf{k}\cdot\nabla_\mathbf{R})^{2n}}{(2n+2)!}\nabla_\mathbf{k}\times\mathbf{B}(\mathbf{R})\right.\right. \nonumber \\
		&\qquad\qquad\left.\left.+\mathrm{i}\frac{e}{2}\sum_{n=0}^\infty\left(\frac{\mathrm{i}}{2}\right)^{2n+1}\frac{(\nabla_\mathbf{k}\cdot\nabla_\mathbf{R})^{2n+1}}{(2n+2)!}\nabla_\mathbf{k}\times\mathbf{B}(\mathbf{R})-\mathbf{p}_\mathrm{S}^{\nu_0}(\mathbf{R})/2\right)-2\mu_\mathrm{eff}(\mathbf{R})\right. \nonumber \\
	&\quad\;\;\left. -2\sum_{n=0}^\infty\frac{\left(\frac{\mathrm{i}}{2}\right)^{2n}(\nabla_\mathbf{k}\cdot\nabla_\mathbf{R})^{2n}}{(2n)!}\mu^\nu_\mathrm{H}(\mathbf{R})-\sum_{n=0}^\infty\frac{\left(-\frac{\mathrm{i}}{2}\right)^n(\nabla_\mathbf{k}\cdot\nabla_\mathbf{R})^n}{n!}\left(\mu_\mathrm{F}^{\downarrow,\nu}(\mathbf{R})+(-1)^n\mu_\mathrm{F}^{\uparrow,\nu}(\mathbf{R})\right) \right. \nonumber \\
	&\quad\;\;\left. -\mathrm{i}\,e\,\sum_{n=0}^\infty\frac{(\nabla_\mathbf{k}\cdot\nabla_\mathbf{R})^{2n+1}}{(2n+2)!}\left(\frac{\mathrm{i}}{2}\right)^{2n+1}\,\mathbf{E}(\mathbf{R})\cdot\nabla_\mathbf{k}\right]\tilde{\rho}^{(\nu)}_{1,2}(\mathbf{k},\mathbf{R})\nonumber \\
	&+\mathrm{exp}\left[\frac{\mathrm{i}}{2}\nabla_\mathbf{R}\cdot\nabla_\mathbf{k}\right]|\Delta_\nu(\mathbf{R})|\mathrm{exp}\left[-\mathrm{i}\,\delta\theta_\nu(\mathbf{R})-\frac{1}{2}\sum_{n=0}^\infty\frac{(\nabla_\mathbf{k}\cdot\nabla_\mathbf{R})^n}{(n+1)!}\left(\frac{\mathrm{i}}{2}\right)^n\,\mathbf{p}^\nu_\mathrm{S}(\mathbf{R})\cdot\nabla_\mathbf{k}\right]\tilde{\rho}^{(\nu)}_{2,2}(\mathbf{k},\mathbf{R})\nonumber \\
	&-\mathrm{exp}\left[-\frac{\mathrm{i}}{2}\nabla_\mathbf{R}\cdot\nabla_\mathbf{k}\right]|\Delta_\nu(\mathbf{R})|\mathrm{exp}\left[-\mathrm{i}\,\delta\theta_\nu(\mathbf{R})+\frac{1}{2}\sum_{n=0}^\infty\frac{(\nabla_\mathbf{k}\cdot\nabla_\mathbf{R})^n}{(n+1)!}\left(-\frac{\mathrm{i}}{2}\right)^n\,\mathbf{p}^\nu_\mathrm{S}(\mathbf{R})\cdot\nabla_\mathbf{k}\right]\tilde{\rho}^{(\nu)}_{1,1}(\mathbf{k},\mathbf{R})\,.	
\end{align}
\end{widetext}
These full spatially--dependent equations of motion  do not contain the electromagnetic potentials. They only contain the gauge-invariant superfluid momentum and effective chemical potential, which are given by 
\begin{align}
		&\mathbf{p}^\nu_\mathrm{S}(\mathbf{R})=\nabla_\mathbf{R}\theta_\nu(\mathbf{R})-2e\mathbf{A}(\mathbf{R})\,,\nonumber \\
			&\mu_\mathrm{eff}(\mathbf{R})=e\,\phi(\mathbf{R})+\frac{1}{2}\frac{\partial}{\partial t}\theta_{\nu_0}(\mathbf{R})\,,
\end{align}
as well as  the electric and magnetic fields, 
\begin{equation}
	\mathbf{E}(\mathbf{R})=-\nabla_\mathbf{R}\phi(\mathbf{R})-\frac{\partial}{\partial t}\mathbf{A}(\mathbf{R})\,,\quad	  \mathbf{B}(\mathbf{R})=\nabla_\mathbf{R}\times\mathbf{A}(\mathbf{R})\,.
\end{equation}
The Higgs mode is determined by amplitude fluctuations of the SC order parameters in different bands. The latter amplitudes  are expressed as
\begin{widetext}
\begin{align}
|\Delta_\nu(\mathbf{R})|=-\sum_{\lambda,\mathbf{k}}g_{\nu,\lambda}\mathrm{exp}\left[-\mathrm{i}\delta_\lambda(\mathbf{R})-\sum_{n=0}^\infty\left(\frac{\mathrm{i}}{2}\right)^{2n+1}\frac{(\nabla_\mathbf{k}\cdot\nabla_\mathbf{R})^{2n+1}}{(2n+2)!}\mathbf{p}^{\nu_0}_\mathrm{s}(\mathbf{R})\cdot\nabla_\mathbf{k}\right]\tilde{\rho}^{(\lambda)}_{2,1}(\mathbf{k},\mathbf{R})\,.
\end{align}
\end{widetext}
The Leggett mode corresponds to fluctuations of the phase difference between different order parameters, 
\begin{align}
\delta\theta_\nu(\mathbf{R})=\theta_{\nu_0}(\mathbf{R})-\theta_\nu(\mathbf{R})\,.
\end{align}
The center-of-mass acceleration of the Cooper-pair condensate, determined by the electric and magnetic fields in the kinetic contributions of the equations of motion (\ref{eq:eom-full1-f})--(\ref{eq:eom-full3-f}), is described by  
\begin{align}
\frac{\partial}{\partial t}\mathbf{p}_\mathrm{S}^\nu(\mathbf{R})&=2\nabla_\mathbf{R}\mu_\mathrm{eff}(\mathbf{R})+2e\,\mathbf{E}(\mathbf{R})\,.
\end{align}
This light--induced Cooper pair total momentum  breaks the equilibrium inversion symmetry. The Fock contributions to the gauge-invariant equations of motion are given by
\begin{widetext}
\begin{align}
\label{eq:mu_f}
\mu_\mathrm{F}^{\uparrow,\nu}(\mathbf{R})=-g_{\nu,\nu}\sum_\mathbf{k}&\mathrm{exp}\left[-\frac{1}{2}\sum_{n=0}^\infty\left(\frac{\mathrm{i}}{2}\right)^{2n}\frac{(\nabla_\mathbf{k}\cdot\nabla_\mathbf{R})^{2n}}{(2n+1)!}\mathbf{p}_\mathrm{s}^{\nu_0}(\mathbf{R})\cdot\nabla_\mathbf{k}\right]\tilde{\rho}^{(\nu)}_{1,1}(\mathbf{k},\mathbf{R})\,, \nonumber \\
\mu_\mathrm{F}^{\downarrow,\nu}(\mathbf{R})=-g_{\nu,\nu}\sum_\mathbf{k}&\left(1-\mathrm{exp}\left[\frac{1}{2}\sum_{n=0}^\infty\left(\frac{\mathrm{i}}{2}\right)^{2n}\frac{(\nabla_\mathbf{k}\cdot\nabla_\mathbf{R})^{2n}}{(2n+1)!}\mathbf{p}_\mathrm{s}^{\nu_0}(\mathbf{R})\cdot\nabla_\mathbf{k}\right]\tilde{\rho}^{(\nu)}_{2,2}(\mathbf{k},\mathbf{R})\right)\,.
\end{align}
\end{widetext}
These Fock contributions ensure charge conservation of the SC system: the gauge-invariant current density,
\begin{align}
\label{eq:current}
\mathbf{J}(\mathbf{R})=\frac{e}{V}\sum_{\lambda,\mathbf{k}}\nabla_\mathbf{k}\xi_\nu(\mathbf{k})\left[\tilde{\rho}^{(\nu)}_{1,1}(\mathbf{k},\mathbf{R})+\tilde{\rho}_{2,2}^{(\nu)}(\mathbf{k},\mathbf{R})\right]\,,
\end{align}
and electron density,
\begin{align}
n(\mathbf{R})=\frac{1}{V}\sum_{\lambda,\mathbf{k}}\left[1+\tilde{\rho}^{(\nu)}_{1,1}(\mathbf{k},\mathbf{R})-\tilde{\rho}^{(\nu)}_{2,2}(\mathbf{k},\mathbf{R})\right]\,,
\end{align}
satisfy the continuity equation
\begin{align}
	e\frac{\partial}{\partial t}n(\mathbf{R})+\nabla_\mathbf{R}\cdot \mathbf{J}(\mathbf{R})=0\,.
\end{align}
This continuity/charge conservation equation directly follows from the equations of motion (\ref{eq:eom-full1-f})--(\ref{eq:eom-full3-f}).

We can simplify the full equations of motion by neglecting 
contributions of the form $(\nabla_\mathbf{k}\cdot\nabla_\mathbf{R})^{n}(\mathbf{p}_\mathrm{s}^\nu(\mathbf{R})\cdot\nabla_\mathbf{k})$ 
in the above equations which is a valid approximation for strong homogeneous electric fields. 
 We then  obtain the following simplified equations of motion for the density matrix to treat the case of 
spatial dependence induced by, e.~g., strong disorder:
\begin{widetext}
\begin{align}
\label{eq:eom-full1b}
	\mathrm{i}\frac{\partial}{\partial t}\tilde{\rho}^{(\nu)}_{1,1}(\mathbf{k},\mathbf{R})&=\left[\xi_\nu\left(\mathbf{k}-\frac{\mathrm{i}}{2}\nabla_\mathbf{R}+\mathrm{i}\frac{e}{2}\nabla_\mathbf{k}\times\mathbf{B}(\mathbf{R})\right)-\xi_\nu\left(\mathbf{k}+\frac{\mathrm{i}}{2}\nabla_\mathbf{R}-\mathrm{i}\frac{e}{2}\nabla_\mathbf{k}\times\mathbf{B}(\mathbf{R})\right)-\mathrm{i}e\,\mathbf{E}(\mathbf{R})\cdot\nabla_\mathbf{k}\right. \nonumber \\
 &\quad\left. -\mu^\nu_\mathrm{H}\left(\mathbf{R}+\frac{\mathrm{i}}{2}\nabla_\mathbf{k}\right)+\mu^\nu_\mathrm{H}\left(\mathbf{R}-\frac{\mathrm{i}}{2}\nabla_\mathbf{k}\right)-\mu^{\uparrow,\nu}_\mathrm{F}\left(\mathbf{R}+\frac{\mathrm{i}}{2}\nabla_\mathbf{k}\right)+\mu^{\uparrow,\nu}_\mathrm{F}\left(\mathbf{R}-\frac{\mathrm{i}}{2}\nabla_\mathbf{k}\right)\right]\tilde{\rho}^{(\nu)}_{1,1}(\mathbf{k},\mathbf{R}) \nonumber \\
	&+|\Delta_\nu(\mathbf{R}+\frac{\mathrm{i}}{2}\nabla_\mathbf{k})|\;\mathrm{exp}\left[-\mathrm{i}\,\delta\theta_\nu(\mathbf{R)}\right]\tilde{\rho}^{(\nu)}_{2,1}(\mathbf{k}-\mathbf{p}^\nu_\mathrm{S}/2,\mathbf{R})\nonumber \\
	&-|\Delta_\nu(\mathbf{R}-\frac{\mathrm{i}}{2}\nabla_\mathbf{k}))|\;\mathrm{exp}\left[\mathrm{i}\,\delta\theta_\nu(\mathbf{R})\right]\tilde{\rho}^{(\nu)}_{1,2}(\mathbf{k}-\mathbf{p}^\nu_\mathrm{S}/2,\mathbf{R})\,,
\end{align}
\begin{align}	
\label{eq:eom-full2b}
	\mathrm{i}\frac{\partial}{\partial t}\tilde{\rho}^{(\nu)}_{2,2}(\mathbf{k},\mathbf{R})&=\left[\xi_\nu\left(-\mathbf{k}-\frac{\mathrm{i}}{2}\nabla_\mathbf{R}-\mathrm{i}\frac{e}{2}\nabla_\mathbf{k}\times\mathbf{B}(\mathbf{R})\right)-\xi_\nu\left(-\mathbf{k}+\frac{\mathrm{i}}{2}\nabla_\mathbf{R}+\mathrm{i}\frac{e}{2}\nabla_\mathbf{k}\times\mathbf{B}(\mathbf{R})\right)+\mathrm{i}e\,\mathbf{E}(\mathbf{R})\cdot\nabla_\mathbf{k}\right. \nonumber \\
 &\quad\left. +\mu^\nu_\mathrm{H}\left(\mathbf{R}+\frac{\mathrm{i}}{2}\nabla_\mathbf{k}\right)-\mu^\nu_\mathrm{H}\left(\mathbf{R}-\frac{\mathrm{i}}{2}\nabla_\mathbf{k}\right)+\mu^{\downarrow,\nu}_\mathrm{F}\left(\mathbf{R}+\frac{\mathrm{i}}{2}\nabla_\mathbf{k}\right)-\mu^{\downarrow,\nu}_\mathrm{F}\left(\mathbf{R}-\frac{\mathrm{i}}{2}\nabla_\mathbf{k}\right)\right]\tilde{\rho}^{(\nu)}_{2,2}(\mathbf{k},\mathbf{R}) \nonumber \\
	&-|\Delta_\nu(\mathbf{R}-\frac{\mathrm{i}}{2}\nabla_\mathbf{k}))|\;\mathrm{exp}\left[-\mathrm{i}\,\delta\theta_\nu(\mathbf{R})\right]\tilde{\rho}^{(\nu)}_{2,1}(\mathbf{k}+\mathbf{p}^\nu_\mathrm{S}/2,\mathbf{R})\nonumber \\
	&+|\Delta_\nu(\mathbf{R}+\frac{\mathrm{i}}{2}\nabla_\mathbf{k}))|\;\mathrm{exp}\left[\mathrm{i}\,\delta\theta_\nu(\mathbf{R})\right]\tilde{\rho}^{(\nu)}_{1,2}(\mathbf{k}+\mathbf{p}^\nu_\mathrm{S}/2,\mathbf{R})\,,
	\end{align}
\begin{align}
\label{eq:eom-full3b}
	\mathrm{i}\frac{\partial}{\partial t}\tilde{\rho}^{(\nu)}_{1,2}(\mathbf{k},\mathbf{R})&=\left[-\xi_\nu\left(\mathbf{k}+\frac{\mathrm{i}}{4}\nabla_\mathbf{R}-e\frac{\mathrm{i}}{2}\nabla_\mathbf{k}\times\mathbf{B}(\mathbf{R})-\mathbf{p}_\mathrm{S}^{\nu_0}/2\right)-\xi_\nu\left(-\mathbf{k}+\frac{\mathrm{i}}{2}\nabla_\mathbf{R}+e\frac{\mathrm{i}}{4}\nabla_\mathbf{k}\times\mathbf{B}(\mathbf{R})-\mathbf{p}_\mathrm{S}^{\nu_0}/2\right)\right. \nonumber \\
	&\quad\;\;\left.-2\mu_\mathrm{eff}(\mathbf{R}) -\mu^\nu_\mathrm{H}\left(\mathbf{R}+\frac{\mathrm{i}}{2}\nabla_\mathbf{k}\right)-\mu^\nu_\mathrm{H}\left(\mathbf{R}-\frac{\mathrm{i}}{2}\nabla_\mathbf{k}\right)\right. \nonumber \\
	&\quad\;\;\left.-\mu_\mathrm{F}^{\downarrow,\nu}(\mathbf{R}-\frac{\mathrm{i}}{2}\nabla_\mathbf{k})-\mu_\mathrm{F}^{\uparrow,\nu}(\mathbf{R}+\frac{\mathrm{i}}{2}\nabla_\mathbf{k}) \right]\tilde{\rho}^{(\nu)}_{1,2}(\mathbf{k},\mathbf{R})\nonumber \\
	&+|\Delta_\nu(\mathbf{R}+\frac{\mathrm{i}}{2}\nabla_\mathbf{k}))|\mathrm{exp}\left[-\mathrm{i}\,\delta\theta_\nu(\mathbf{R})\right]\tilde{\rho}^{(\nu)}_{2,2}(\mathbf{k}-\mathbf{p}^\nu_\mathrm{S}/2,\mathbf{R})\nonumber \\
	&-|\Delta_\nu(\mathbf{R}-\frac{\mathrm{i}}{2}\nabla_\mathbf{k}))|\mathrm{exp}\left[-\mathrm{i}\,\delta\theta_\nu(\mathbf{R})\right]\tilde{\rho}^{(\nu)}_{1,1}(\mathbf{k}+\mathbf{p}^\nu_\mathrm{S}/2,\mathbf{R})\,.	
\end{align}
\end{widetext}

When the SC system is only weakly spatially-dependent, we can neglect all orders $\mathcal{O}(\nabla_\mathbf{k}\cdot\nabla_\mathbf{R})$ and higher in the above equations.  In the  case of a homogeneous system, we then obtain 
\begin{widetext}
\begin{align}
\label{eq:eoms}
	&\mathrm{i}\frac{\partial}{\partial t}\tilde{\rho}^{(\nu)}_{1,1}(\mathbf{k})=-\mathrm{i}\,e\,\mathbf{E}(t)\cdot\nabla_\mathbf{k}\tilde{\rho}^{(\nu)}_{1,1}(\mathbf{k}) 
	-|\Delta_\nu|\left[\mathrm{e}^{\mathrm{i}\,\delta\theta_\nu}\tilde{\rho}^{(\nu)}_{1,2}(\mathbf{k}-\mathbf{p}_\mathrm{S}/2)-\mathrm{e}^{-\mathrm{i}\,\delta\theta_\nu}\tilde{\rho}^{(\nu)}_{2,1}(\mathbf{k}-\mathbf{p}_\mathrm{S}/2)\right]\,, \nonumber \\
	&\mathrm{i}\frac{\partial}{\partial t}\tilde{\rho}^{(\nu)}_{2,2}(\mathbf{k})=\mathrm{i}\,e\,\mathbf{E}(t)\cdot\nabla_\mathbf{k}\tilde{\rho}^{(\nu)}_{2,2}(\mathbf{k})
	+|\Delta_\nu|\left[\mathrm{e}^{\mathrm{i}\,\delta\theta_\nu}\tilde{\rho}^{(\nu)}_{1,2}(\mathbf{k}+\mathbf{p}_\mathrm{S}/2)-\mathrm{e}^{-\mathrm{i}\,\delta\theta_\nu}\tilde{\rho}^{(\nu)}_{2,1}(\mathbf{k}+\mathbf{p}_\mathrm{S}/2)\right]\,, \nonumber \\ 
	&\mathrm{i}\frac{\partial}{\partial t}\tilde{\rho}^{(\nu)}_{1,2}(\mathbf{k})  =-[\xi_\nu(\mathbf{k}-\mathbf{p}_\mathrm{S}/2)+\xi_\nu(\mathbf{k}+\mathbf{p}_\mathrm{S}/2)+2(\mu_\mathrm{eff}+\mu^\nu_\mathrm{F})]\tilde{\rho}^{(\nu)}_{1,2}(\mathbf{k}) 
	 \nonumber \\&\qquad\qquad\quad
	+|\Delta_\nu|\mathrm{e}^{-\mathrm{i}\,\delta\theta_\nu}\left[\tilde{\rho}^{(\nu)}_{2,2}(\mathbf{k}-\mathbf{p}_\mathrm{S}/2)-\tilde{\rho}^{(\nu)}_{1,1}(\mathbf{k}+\mathbf{p}_\mathrm{S}/2)\right]\,,
\end{align}
\end{widetext}
with 
\begin{align}
&\mathbf{p}_\mathrm{S}\equiv\mathbf{p}^\nu_\mathrm{S}=-2\,e\,\mathbf{A}\,,\quad \mu_\mathrm{eff}=e\,\phi+\frac{1}{2}\frac{\partial}{\partial t}\theta_{\nu_0}\,,\nonumber \\
&\delta\theta_\nu=\theta_{\nu_0}-\theta_\nu\,,\qquad |\Delta_\nu|=-\mathrm{e}^{-\mathrm{i}\,\delta\theta_\nu}\sum_{\lambda,\mathbf{k}} g_{\nu,\lambda}\tilde{\rho}_{2,1}^{(\lambda)}(\mathbf{k})\,\nonumber \\  &\mu^\nu_\mathrm{F}\equiv\frac{1}{2}\left(\mu^{\downarrow,\nu}_\mathrm{F}+\mu^{\uparrow,\nu}_\mathrm{F}\right)=-g_{\nu,\nu}\sum_\mathbf{k}\left[1+\tilde{\rho}^{(\nu)}_{1,1}(\mathbf{k})-\tilde{\rho}^{(\nu)}_{2,2}(\mathbf{k})\right]\,.
\end{align}
The equations of motion of superfluid momentum $\mathbf{p}_\mathrm{S}$  simplifies to
$\frac{\partial}{\partial t}\mathbf{p}_\mathrm{S}=2e\,\mathbf{E}\,,$
while the SC order parameter phase $\theta_{\nu_0}$ is given by
\begin{widetext}
\begin{align}
\frac{\partial}{\partial t}\theta_{\nu_0}&=-2\,e\,\phi+\frac{1}{2|\Delta_{\nu_0}|}\sum_{\nu,\mathbf{k}} g_{\nu_0,\nu}\left[\xi_\nu(\mathbf{k}-\mathbf{p}_\mathrm{S}/2)+\xi_\nu(\mathbf{k}+\mathbf{p}_\mathrm{S}/2)+2\mu_\mathrm{F}^\nu\right]\left(\tilde{\rho}_{1,2}^{(\nu)}(\mathbf{k})+\tilde{\rho}_{2,1}^{(\nu)}(\mathbf{k})\right)\nonumber \\
&+\frac{1}{|\Delta_{\nu_0}|}\sum_{\nu,\mathbf{k}}g_{\nu_0,\nu}|\Delta_\nu|\left[\tilde{\rho}_{1,1}^{(\nu)}(\mathbf{k})-\tilde{\rho}_{2,2}^{(\nu)}(\mathbf{k})\right]\cos(\delta\theta_\nu)\,.
\end{align}
\end{widetext}
The above homogeneous equations of motion are equivalent to the conventional pseudo-spin model. To demonstrate this, we introduce the pseudo-spin
\begin{align}
   & \sigma_0^{(\nu)}(\mathbf{k})\equiv \frac{1}{2}\left[\tilde{\rho}^{(\nu)}_{1,1}\left(\mathbf{k}+\frac{\mathbf{p}_\mathrm{S}}{2}\right)+\tilde{\rho}^{(\nu)}_{2,2}\left(\mathbf{k}-\frac{\mathbf{p}_\mathrm{S}}{2}\right)\right]\,, \nonumber \\
   &\sigma_1^{(\nu)}(\mathbf{k})\equiv \frac{1}{2}\left[\tilde{\rho}^{(\nu)}_{1,2}\left(\mathbf{k}\right)e^{i\theta_{\nu_0}}+\tilde{\rho}^{(\nu)}_{2,1}\left(\mathbf{k}\right)e^{-i\theta_{\nu_0}}\right]\,, \nonumber \\ 
   &\sigma_2^{(\nu)}(\mathbf{k})\equiv \frac{1}{2i}\left[\tilde{\rho}^{(\nu)}_{1,2}\left(\mathbf{k}\right)e^{i\theta_{\nu_0}}-\tilde{\rho}^{(\nu)}_{2,1}\left(\mathbf{k}\right)e^{-i\theta_{\nu_0}}\right]\,, \nonumber \\ 
   & \sigma_3^{(\nu)}(\mathbf{k})\equiv \frac{1}{2}\left[\tilde{\rho}^{(\nu)}_{1,1}\left(\mathbf{k}+\frac{\mathbf{p}_\mathrm{S}}{2}\right)-\tilde{\rho}^{(\nu)}_{2,2}\left(\mathbf{k}-\frac{\mathbf{p}_\mathrm{S}}{2}\right)\right]\,, 
   \label{eq:sigmak}
\end{align}
such that Eq.~(\ref{eq:eoms}) transforms to the Bloch equations of the conventional pseudo-spin model
\begin{align}
    &\frac{\partial}{\partial t}\sigma_0^{(\nu)}(\mathbf{k})=0\,,\nonumber \\
    &\frac{\partial}{\partial t}\mathbf{\sigma}^{(\nu)}(\mathbf{k})=2\mathbf{b}^{(\nu)}(\mathbf{k})\times\mathbf{\sigma}^{(\nu)}(\mathbf{k})\,. \nonumber
\end{align}
Here, $\mathbf{\sigma}^{(\nu)}(\mathbf{k})=(\sigma^{(\nu)}_1(\mathbf{k}),\sigma^{(\nu)}_2(\mathbf{k}),\sigma^{(\nu)}_3(\mathbf{k}))$ is the Anderson pseudo-spin, and $\mathbf{b}^{(\nu)}(\mathbf{k})=(-\mathrm{Re}\Delta_\nu,\mathrm{Im}\Delta_\nu,[\xi_\nu(\mathbf{k}_{-})+\xi_\nu(\mathbf{k}_{+})]/2+e\phi+\mu^\nu_\mathrm{F})$ is the pseudo-magnetic field. Using $\sigma_i^{(\nu)}(\mathbf{k})$, the SC order parameter and the current density become
\begin{align}
&\Delta_\nu=-\sum_{\lambda,\mathbf{k}}g_{\nu,\lambda}\left[\sigma_x^{(\lambda)}(\mathbf{k})-i\sigma_y^{(\lambda)}(\mathbf{k})\right]\,, \nonumber \\
    &J=\frac{e}{V}\sum_{\lambda,\mathbf{k}}\{\sigma_0^{(\lambda)}(\mathbf{k})\nabla_\mathbf{k}[\xi_\lambda(\mathbf{k}_{+})+\xi_\lambda(\mathbf{k}_{-})] \nonumber \\
    &\qquad\qquad\quad+\sigma_3^{(\lambda)}(\mathbf{k})\nabla_\mathbf{k}[\xi_\lambda(\mathbf{k}_{+})-\xi_\lambda(\mathbf{k}_{-})]\}\,, \nonumber
\end{align}
respectively.

\section{Phase Coherent Pump--Probe Spectra} 
\label{sec:pp}

So far, we have discussed contributions arising from excitations by one pulse that are sensed by the other. In this appendix, we briefly  discuss the  wave mixing signals in the THz-MDCS spectra that are  generated by interference of {\em both} pump and probe excitations. As shown in  Ref.~\cite{Mootz2022}, these signals show up in the THz-MDCS for weak excitation, where  pulse A (B) can be considered as the probe of the non-equilibrium state driven by pulse B (A).   Reference~\cite{Mootz2022} showed that interference between pump and probe excitations  dominates the THz-MDCS spectral features  then, leading to correlated wave mixing peaks.
Unlike for the narrowband strong pump and broadband weak probe excitation protocol used there, here we consider two identical pulses that are both strong and broadband, so our results here relate to the results there only for  weak pulse-pair fields.
We can then decompose the density matrix driven by both pulses (pulse-pair) as
\begin{align}
\label{eq:rho-dev2}
	\tilde{\rho}^\mathrm{AB}(\mathbf{k}) =  \tilde{\rho}^\mathrm{A(B)}(\mathbf{k}) +\delta\tilde{\rho}^\mathrm{B(A)}(\mathbf{k})\,. 
\end{align} 
where $\delta\tilde{\rho}^\mathrm{B(A)}$ describes the small probe-induced change  in the density matrix of the non-equilibrium state driven by pulse A (B). The   nonlinear signal is then determined by the equation of motion obtained by linearizing in terms of the probe but not the pump~\cite{Mootz2022}: 
\begin{align}
\label{Deltarho_0_2} 
	&\partial_t\delta\tilde{\rho}_{0}^{\mathrm{B}}(\mathbf{k})=-e\,\mathbf{E}_\mathrm{A}\cdot\nabla_\mathbf{k}\,\delta\tilde{\rho}^{(\nu),\mathrm{B}}_3(\mathbf{k})-e\,\mathbf{E}_\mathrm{B}\cdot\nabla_\mathbf{k}\,\tilde{\rho}^{(\nu),\mathrm{A}}_3(\mathbf{k})\nonumber \\
&+|\Delta_\nu^\mathrm{A}|\delta\mathbf{p}_\mathrm{S}^\mathrm{B}\cdot\nabla_\mathbf{k}\Delta\tilde{\rho}_2^\mathrm{(\nu),A}(\mathbf{k})+|\Delta_\nu^\mathrm{A}|\Delta\theta^\mathrm{A}_\nu\delta\mathbf{p}_\mathrm{S}^\mathrm{B}\cdot\nabla_\mathbf{k}\tilde{\rho}_1^\mathrm{(\nu),A}(\mathbf{k})\nonumber \\
&+|\Delta_\nu^\mathrm{A}|\mathbf{p}_\mathrm{S}^\mathrm{A}\cdot\nabla_\mathbf{k}\,\delta\tilde{\rho}_2^\mathrm{(\nu),B}(\mathbf{k})+|\Delta_\nu^\mathrm{A}|\Delta\theta^\mathrm{A}_\nu\mathbf{p}_\mathrm{S}^\mathrm{A}\cdot\nabla_\mathbf{k}\,\delta\tilde{\rho}_1^\mathrm{(\nu),B}(\mathbf{k}) \nonumber \\	&+\delta|\Delta_\nu^\mathrm{B}|\Delta\theta_\nu^\mathrm{A}\mathbf{p}_\mathrm{S}^\mathrm{A}\cdot\nabla_\mathbf{k}\,\tilde{\rho}_1^\mathrm{(\nu),A}(\mathbf{k})+\delta|\Delta_\nu^\mathrm{B}|\mathbf{p}_\mathrm{S}^\mathrm{A}\cdot\nabla_\mathbf{k}\,\Delta\tilde{\rho}_2^\mathrm{(\nu),A}(\mathbf{k}) \nonumber \\
	&+\mathcal{O}(\mathbf{p}_\mathrm{A}^2)+\mathcal{O}((\Delta\theta_\nu)^2)+\mathcal{O}(\mathbf{E}_\mathrm{B}^2)\,.
\end{align}
Here, we expanded the equation of motion to first order in $\mathbf{p}_\mathrm{S}$ and $\Delta\theta_\nu$ and neglected all orders of $\mathcal{O}(\mathbf{E}_\mathrm{B}^2)$. Of importance in this case is the coherent  pump--probe modulation of the SC order parameter amplitude by {\em both} pulses. Such pump--probe order parameter modulation is controlled by the relative phase of the two fields through the time delay $\tau$ and should 
to be contrasted to the order parameter quantum quench by a single pulse that  dominates the signals  here.  
The coherent pump--probe modulation of the order parameter 
is given by $\delta|\Delta_\nu^\mathrm{B}|=|\Delta_\nu^\mathrm{AB}|-|\Delta_\nu^\mathrm{A}|$, while the probe-induced change of the superfluid momentum is given by $\delta\mathbf{p}_\mathrm{S}^\mathrm{B}=\mathbf{p}_\mathrm{S}^\mathrm{AB}-\mathbf{p}_\mathrm{S}^\mathrm{A}$. The equation of motion for $\partial_t\delta\tilde{\rho}_{0}^{\mathrm{A}}(\mathbf{k})$ is obtained by exchanging the labels A and B in Eq.~(\ref{Deltarho_0_2}). As demonstrated in Ref.~\cite{NatPhys}, the terms in the first line of Eq.~(\ref{Deltarho_0_2}) mainly drive the inversion symmetry breaking. The second line of Eq.~(\ref{Deltarho_0_2}) yields the pump--probe signal calculated in Ref.~\cite{hybrid-higgs}. It only contributes to the nonlinear response when the probe pulse arrives after the pump pulse, such that is does not rely on the temporal overlap between both pulses. The corresponding signals show up as harmonic sidebands determined by Higgs and Leggett mode frequencies  in the THz-MDCS spectra as demonstrated in Sections~\ref{sec:U=0} and \ref{sec:U>0}. In contrast to that, the contributions of lines three and four on the rhs of  Eq.~(\ref{Deltarho_0_2}) become significant when {\em pump and probe excitations overlap in time}. The conventional third-order nonlinear processes and fifth-order Raman processes discussed in Sections~\ref{sec:U=0} and \ref{sec:U>0} are mainly generated by the third line on the rhs of Eq.~(\ref{Deltarho_0_2})~\cite{Mootz2022}. These signals are determined by the equations of motion for $\delta\tilde{\rho}_2^\mathrm{(\nu),B}(\mathbf{k})$, obtained by linearizing Eq.~(\ref{eq:eom_t2}) with respect to the  probe field:
\begin{align}
	&\partial_t^2 \,
	\delta \tilde{\rho}^{(\nu),\mathrm{B}}_2(\mathbf{k}) + \left[ (E^\mathrm{A}_\nu(\mathbf{k}))^2 +
	4 |\Delta^\mathrm{A}_\nu|^2 \, \cos^2 \Delta \theta^\mathrm{A}_\nu 
	\right] \delta \tilde{\rho}^{(\nu),\mathrm{B}}_2(\mathbf{k}) 
	\nonumber \\
&	+\left[ -\partial_t E^\mathrm{A}_\nu(\mathbf{k}) +
	2 |\Delta_\nu^\mathrm{A}|^2 \, \sin 2 \Delta \theta^\mathrm{A}_\nu \right] \, 
\delta	 \tilde{\rho}_1^{(\nu),\mathrm{B}}(\mathbf{k})
	\nonumber \\
	&=\delta S^{(2),\mathrm{B}}_\nu(\mathbf{k})-\delta[E^2_\nu(\mathbf{k})]\,\Delta\tilde{\rho}_2^{(\nu),\mathrm{A}}(\mathbf{k})+\partial_t\delta [E_\nu(\mathbf{k})]\,\Delta\tilde{\rho}_1^{(\nu),\mathrm{A}}(\mathbf{k}) \nonumber \\
	&-\delta[\partial_t\delta\Delta^\prime_\nu+\delta\Delta^\dprime_\nu\,E_\nu(\mathbf{k})]\,N_\nu^\mathrm{A}(\mathbf{k})\nonumber \\
	&-[\partial_t\delta\Delta^{\prime,\mathrm{A}}_\nu+\delta\Delta^{\dprime,\mathrm{A}}_\nu\,E^\mathrm{A}_\nu(\mathbf{k})]\,\delta [N_\nu(\mathbf{k})]\nonumber \\
	&-8\,\delta\Delta^{\prime,\mathrm{B}}_\nu\Delta^{\prime,\mathrm{A}}_\nu\,\Delta\tilde{\rho}^{(\nu),\mathrm{A}}_2(\mathbf{k}) \nonumber \\
	&-4\Delta^{\prime,\mathrm{A}}_\nu\,\delta\Delta^{\dprime,\mathrm{B}}_\nu\,\Delta\tilde{\rho}^{(\nu),\mathrm{A}}_1(\mathbf{k})-4\Delta^{\dprime,\mathrm{A}}_\nu\,\delta\Delta^{\prime,\mathrm{B}}_\nu\,\Delta\tilde{\rho}^{(\nu),\mathrm{A}}_1(\mathbf{k})\,,
	\label{eq:drho2}
\end{align}
where the probe-induced changes are denoted by $\delta[...]$. In particular, the pump--probe modulation of the SC order parameter amplitude and phase are defined by
\begin{align}
		\delta\Delta^{\prime,\mathrm{B}}_\nu=\Delta^{\prime,\mathrm{AB}}_\nu-\Delta^{\prime,\mathrm{A}}_\nu,\qquad \delta\Delta^{\dprime,\mathrm{B}}_\nu=\Delta^{\dprime,\mathrm{AB}}_\nu-\Delta^{\dprime,\mathrm{A}}_\nu\,.
\end{align}
The first term on the rhs of Eq.~(\ref{eq:drho2}), $\delta S^{(2),\mathrm{B}}_\nu(\mathbf{k})=S^{(2),\mathrm{AB}}_\nu(\mathbf{k})-S^{(2),\mathrm{A}}_\nu(\mathbf{k})$, leads to pump--probe, four-wave mixing, and third harmonic generation signals which are generated by third-order nonlinear processes~\cite{Mootz2022}. 
The dominant contribution of this term is
\begin{align} 
\label{source-pp} 
\delta \delta S^{(2),\mathrm{B}}_\nu(\mathbf{k}) &\approx 
\tilde{\rho}_1^{(\nu),0}(\mathbf{k}) \,
\left[  (e\,{\bf E}_{\rm{A}} \cdot \nabla_{{\bf k}}) 
({\bf p^{\rm{B}}_{\rm{S}}} \cdot \nabla_{{\bf k}}) \right. \nonumber \\
&\left. + 
(e\,{\bf E}_{\rm{B}} \cdot \nabla_{{\bf k}}) 
({\bf p^{\rm{A}}_{\rm{S}}} \cdot \nabla_{{\bf k}}) 
\right] \xi_\nu({\bf k})\,,
\end{align}  
which drive the pseudo-spin oscillators  
via the  sum- and difference-frequency Raman  process
$\omega_\mathrm{A}\pm\omega_\mathrm{B}=(\omega_0\pm\omega_0,\mp\omega_0)$. Based on the fifth term on the rhs of Eq.~(\ref{Deltarho_0_2}), these processes lead to peaks at $\omega_\mathrm{A}\pm\omega_\mathrm{B}+\omega_\mathrm{A}$ and $\omega_\mathrm{A}\pm\omega_\mathrm{B}-\omega_\mathrm{A}$ in the THz-MDCS spectra. In particular, a pump--probe and four-wave mixing signals are observable at $(\omega_0,-\omega_0)$ and $(\omega_0,\omega_0)$, respectively,  while a third harmonic generation peak emerges at $(3\omega_0,-\omega_0)$.
Exchanging pulses A and B results in pump--probe, four-wave mixing, and third-harmonic generation signals at $(\omega_0,0)$, $(\omega_0,-2\omega_0)$, and $(3\omega_0,-2\omega_0)$, respectively. These signals are observable in the THz-MDCS spectra in the perturbative excitation regime, where the nonlinear response can be described by susceptibility expansion. 

The second and third term on the rhs of Eq.~(\ref{eq:drho2})  describe fifth-order difference-frequency-Raman processes which only slightly contribute to the THz-MDCS spectra~\cite{Mootz2022}. Probe-induced changes of collective modes and charge fluctuations are described by the second and third lines on the rhs of Eq.~(\ref{eq:drho2}). The fourth and fifth lines of Eq.~(\ref{eq:drho2}) generate strong seventh-order correlated wave-
mixing peaks when the system is excited with a strong narrowband pump pulse and sensed by a weak broadband probe pulse as discussed in Ref.~\cite{Mootz2022}. 
In particular,  pseudo-spin oscillators at different momenta ${\bf k}$ are parametrically driven by the time-dependent modulation of the order parameter $\delta |\Delta_\mathrm{A(B),h}| = |\Delta_\mathrm{AB,h}|-|\Delta_\mathrm{A(B),h}|$. This parametric driving leads to distinct high-order nonlinear peaks displaced from the conventional pump--probe peaks along the vertical $\omega_\tau$-axis. In the case of narrowband strong pump and weak broadband probe used in Ref.~\cite{Mootz2022}, the THz-MDCS spectra show four strong peaks which dominate over conventional third-order nonlinear signals and are generated by at least seventh-order nonlinear processes. Correlated pump--probe, four-wave mixing, and third-harmonic generation peaks separate from the corresponding third-order nonlinear signals while a fourth strong correlated wave-mixing peak emerges in a spectral region far separated from the conventional signals. These contributions are negligible in the pulse-pair excitation scheme with two strong broad pulses studied in this paper.

\section*{Acknowledgments}
The  work at Ames was supported by the Ames Laboratory and the US Department of Energy, Office of Science, Basic Energy
Sciences, Materials Science and Engineering Division under contract $\#$DE-AC02-07CH11358.



\begin{thebibliography}{70}%
\makeatletter
\providecommand \@ifxundefined [1]{%
 \@ifx{#1\undefined}
}%
\providecommand \@ifnum [1]{%
 \ifnum #1\expandafter \@firstoftwo
 \else \expandafter \@secondoftwo
 \fi
}%
\providecommand \@ifx [1]{%
 \ifx #1\expandafter \@firstoftwo
 \else \expandafter \@secondoftwo
 \fi
}%
\providecommand \natexlab [1]{#1}%
\providecommand \enquote  [1]{``#1''}%
\providecommand \bibnamefont  [1]{#1}%
\providecommand \bibfnamefont [1]{#1}%
\providecommand \citenamefont [1]{#1}%
\providecommand \href@noop [0]{\@secondoftwo}%
\providecommand \href [0]{\begingroup \@sanitize@url \@href}%
\providecommand \@href[1]{\@@startlink{#1}\@@href}%
\providecommand \@@href[1]{\endgroup#1\@@endlink}%
\providecommand \@sanitize@url [0]{\catcode `\\12\catcode `\$12\catcode `\&12\catcode `\#12\catcode `\^12\catcode `\_12\catcode `\%12\relax}%
\providecommand \@@startlink[1]{}%
\providecommand \@@endlink[0]{}%
\providecommand \url  [0]{\begingroup\@sanitize@url \@url }%
\providecommand \@url [1]{\endgroup\@href {#1}{\urlprefix }}%
\providecommand \urlprefix  [0]{URL }%
\providecommand \Eprint [0]{\href }%
\providecommand \doibase [0]{https://doi.org/}%
\providecommand \selectlanguage [0]{\@gobble}%
\providecommand \bibinfo  [0]{\@secondoftwo}%
\providecommand \bibfield  [0]{\@secondoftwo}%
\providecommand \translation [1]{[#1]}%
\providecommand \BibitemOpen [0]{}%
\providecommand \bibitemStop [0]{}%
\providecommand \bibitemNoStop [0]{.\EOS\space}%
\providecommand \EOS [0]{\spacefactor3000\relax}%
\providecommand \BibitemShut  [1]{\csname bibitem#1\endcsname}%
\let\auto@bib@innerbib\@empty
\bibitem [{\citenamefont {Kuehn}\ \emph {et~al.}(2009)\citenamefont {Kuehn}, \citenamefont {Reimann}, \citenamefont {Woerner},\ and\ \citenamefont {Elsaesser}}]{Kuehn2009}%
  \BibitemOpen
  \bibfield  {author} {\bibinfo {author} {\bibfnamefont {W.}~\bibnamefont {Kuehn}}, \bibinfo {author} {\bibfnamefont {K.}~\bibnamefont {Reimann}}, \bibinfo {author} {\bibfnamefont {M.}~\bibnamefont {Woerner}},\ and\ \bibinfo {author} {\bibfnamefont {T.}~\bibnamefont {Elsaesser}},\ }\bibfield  {title} {\bibinfo {title} {Phase-resolved two-dimensional spectroscopy based on collinear n-wave mixing in the ultrafast time domain},\ }\href@noop {} {\bibfield  {journal} {\bibinfo  {journal} {J. Chem. Phys.}\ }\textbf {\bibinfo {volume} {130}},\ \bibinfo {pages} {164503} (\bibinfo {year} {2009})}\BibitemShut {NoStop}%
\bibitem [{\citenamefont {Maag}\ \emph {et~al.}(2016)\citenamefont {Maag}, \citenamefont {Bayer}, \citenamefont {Baierl}, \citenamefont {Hohenleutner}, \citenamefont {Korn}, \citenamefont {Sch{\"u}ller}, \citenamefont {Schuh}, \citenamefont {Bougeard}, \citenamefont {Lange}, \citenamefont {Huber}, \citenamefont {Mootz}, \citenamefont {Sipe}, \citenamefont {Koch},\ and\ \citenamefont {Kira}}]{maag2016}%
  \BibitemOpen
  \bibfield  {author} {\bibinfo {author} {\bibfnamefont {T.}~\bibnamefont {Maag}}, \bibinfo {author} {\bibfnamefont {A.}~\bibnamefont {Bayer}}, \bibinfo {author} {\bibfnamefont {S.}~\bibnamefont {Baierl}}, \bibinfo {author} {\bibfnamefont {M.}~\bibnamefont {Hohenleutner}}, \bibinfo {author} {\bibfnamefont {T.}~\bibnamefont {Korn}}, \bibinfo {author} {\bibfnamefont {C.}~\bibnamefont {Sch{\"u}ller}}, \bibinfo {author} {\bibfnamefont {D.}~\bibnamefont {Schuh}}, \bibinfo {author} {\bibfnamefont {D.}~\bibnamefont {Bougeard}}, \bibinfo {author} {\bibfnamefont {C.}~\bibnamefont {Lange}}, \bibinfo {author} {\bibfnamefont {R.}~\bibnamefont {Huber}}, \bibinfo {author} {\bibfnamefont {M.}~\bibnamefont {Mootz}}, \bibinfo {author} {\bibfnamefont {J.~E.}\ \bibnamefont {Sipe}}, \bibinfo {author} {\bibfnamefont {S.~W.}\ \bibnamefont {Koch}},\ and\ \bibinfo {author} {\bibfnamefont {M.}~\bibnamefont {Kira}},\ }\bibfield  {title} {\bibinfo {title} {Coherent cyclotron motion beyond {Kohn}'s theorem},\ }\href
  {https://doi.org/10.1038/nphys3559} {\bibfield  {journal} {\bibinfo  {journal} {Nat. Phys.}\ }\textbf {\bibinfo {volume} {12}},\ \bibinfo {pages} {119} (\bibinfo {year} {2016})}\BibitemShut {NoStop}%
\bibitem [{\citenamefont {Junginger}\ \emph {et~al.}(2012)\citenamefont {Junginger}, \citenamefont {Mayer}, \citenamefont {Schmidt}, \citenamefont {Schubert}, \citenamefont {M\"ahrlein}, \citenamefont {Leitenstorfer}, \citenamefont {Huber},\ and\ \citenamefont {Pashkin}}]{Junginger2012}%
  \BibitemOpen
  \bibfield  {author} {\bibinfo {author} {\bibfnamefont {F.}~\bibnamefont {Junginger}}, \bibinfo {author} {\bibfnamefont {B.}~\bibnamefont {Mayer}}, \bibinfo {author} {\bibfnamefont {C.}~\bibnamefont {Schmidt}}, \bibinfo {author} {\bibfnamefont {O.}~\bibnamefont {Schubert}}, \bibinfo {author} {\bibfnamefont {S.}~\bibnamefont {M\"ahrlein}}, \bibinfo {author} {\bibfnamefont {A.}~\bibnamefont {Leitenstorfer}}, \bibinfo {author} {\bibfnamefont {R.}~\bibnamefont {Huber}},\ and\ \bibinfo {author} {\bibfnamefont {A.}~\bibnamefont {Pashkin}},\ }\bibfield  {title} {\bibinfo {title} {Nonperturbative interband response of a bulk {InSb} semiconductor driven off resonantly by terahertz electromagnetic few-cycle pulses},\ }\href@noop {} {\bibfield  {journal} {\bibinfo  {journal} {Phys. Rev. Lett.}\ }\textbf {\bibinfo {volume} {109}},\ \bibinfo {pages} {147403} (\bibinfo {year} {2012})}\BibitemShut {NoStop}%
\bibitem [{\citenamefont {Tarekegne}\ \emph {et~al.}(2020)\citenamefont {Tarekegne}, \citenamefont {Kaltenecker}, \citenamefont {Klarskov}, \citenamefont {Iwaszczuk}, \citenamefont {Lu}, \citenamefont {Ou}, \citenamefont {Norrman},\ and\ \citenamefont {Jepsen}}]{Tarekegne2020}%
  \BibitemOpen
  \bibfield  {author} {\bibinfo {author} {\bibfnamefont {A.~T.}\ \bibnamefont {Tarekegne}}, \bibinfo {author} {\bibfnamefont {K.~J.}\ \bibnamefont {Kaltenecker}}, \bibinfo {author} {\bibfnamefont {P.}~\bibnamefont {Klarskov}}, \bibinfo {author} {\bibfnamefont {K.}~\bibnamefont {Iwaszczuk}}, \bibinfo {author} {\bibfnamefont {W.}~\bibnamefont {Lu}}, \bibinfo {author} {\bibfnamefont {H.}~\bibnamefont {Ou}}, \bibinfo {author} {\bibfnamefont {K.}~\bibnamefont {Norrman}},\ and\ \bibinfo {author} {\bibfnamefont {P.~U.}\ \bibnamefont {Jepsen}},\ }\bibfield  {title} {\bibinfo {title} {Subcycle nonlinear response of doped {4H} silicon carbide revealed by two-dimensional terahertz spectroscopy},\ }\href@noop {} {\bibfield  {journal} {\bibinfo  {journal} {ACS Photonics}\ }\textbf {\bibinfo {volume} {7}},\ \bibinfo {pages} {221} (\bibinfo {year} {2020})}\BibitemShut {NoStop}%
\bibitem [{\citenamefont {Pal}\ \emph {et~al.}(2021)\citenamefont {Pal}, \citenamefont {Strkalj}, \citenamefont {Yang}, \citenamefont {Weber}, \citenamefont {Trassin}, \citenamefont {Woerner},\ and\ \citenamefont {Fiebig}}]{Pal2021}%
  \BibitemOpen
  \bibfield  {author} {\bibinfo {author} {\bibfnamefont {S.}~\bibnamefont {Pal}}, \bibinfo {author} {\bibfnamefont {N.}~\bibnamefont {Strkalj}}, \bibinfo {author} {\bibfnamefont {C.-J.}\ \bibnamefont {Yang}}, \bibinfo {author} {\bibfnamefont {M.~C.}\ \bibnamefont {Weber}}, \bibinfo {author} {\bibfnamefont {M.}~\bibnamefont {Trassin}}, \bibinfo {author} {\bibfnamefont {M.}~\bibnamefont {Woerner}},\ and\ \bibinfo {author} {\bibfnamefont {M.}~\bibnamefont {Fiebig}},\ }\bibfield  {title} {\bibinfo {title} {Origin of terahertz soft-mode nonlinearities in ferroelectric perovskites},\ }\href {https://doi.org/10.1103/PhysRevX.11.021023} {\bibfield  {journal} {\bibinfo  {journal} {Phys. Rev. X}\ }\textbf {\bibinfo {volume} {11}},\ \bibinfo {pages} {021023} (\bibinfo {year} {2021})}\BibitemShut {NoStop}%
\bibitem [{\citenamefont {Fuller}\ and\ \citenamefont {Ogilvie}(2015)}]{Fuller2015}%
  \BibitemOpen
  \bibfield  {author} {\bibinfo {author} {\bibfnamefont {F.~D.}\ \bibnamefont {Fuller}}\ and\ \bibinfo {author} {\bibfnamefont {J.~P.}\ \bibnamefont {Ogilvie}},\ }\bibfield  {title} {\bibinfo {title} {Experimental implementations of two-dimensional fourier transform electronic spectroscopy},\ }\href@noop {} {\bibfield  {journal} {\bibinfo  {journal} {Annu. Rev. Phys. Chem.}\ }\textbf {\bibinfo {volume} {66}},\ \bibinfo {pages} {667} (\bibinfo {year} {2015})}\BibitemShut {NoStop}%
\bibitem [{\citenamefont {Lu}\ \emph {et~al.}(2017)\citenamefont {Lu}, \citenamefont {Li}, \citenamefont {Hwang}, \citenamefont {Ofori-Okai}, \citenamefont {Kurihara}, \citenamefont {Suemoto},\ and\ \citenamefont {Nelson}}]{Nelson}%
  \BibitemOpen
  \bibfield  {author} {\bibinfo {author} {\bibfnamefont {J.}~\bibnamefont {Lu}}, \bibinfo {author} {\bibfnamefont {X.}~\bibnamefont {Li}}, \bibinfo {author} {\bibfnamefont {H.~Y.}\ \bibnamefont {Hwang}}, \bibinfo {author} {\bibfnamefont {B.~K.}\ \bibnamefont {Ofori-Okai}}, \bibinfo {author} {\bibfnamefont {T.}~\bibnamefont {Kurihara}}, \bibinfo {author} {\bibfnamefont {T.}~\bibnamefont {Suemoto}},\ and\ \bibinfo {author} {\bibfnamefont {K.~A.}\ \bibnamefont {Nelson}},\ }\bibfield  {title} {\bibinfo {title} {Coherent two-dimensional terahertz magnetic resonance spectroscopy of collective spin waves},\ }\href@noop {} {\bibfield  {journal} {\bibinfo  {journal} {Phys. Rev. Lett.}\ }\textbf {\bibinfo {volume} {118}},\ \bibinfo {pages} {207204} (\bibinfo {year} {2017})}\BibitemShut {NoStop}%
\bibitem [{\citenamefont {Johnson}\ \emph {et~al.}(2019)\citenamefont {Johnson}, \citenamefont {Knighton},\ and\ \citenamefont {Johnson}}]{Johnson2019}%
  \BibitemOpen
  \bibfield  {author} {\bibinfo {author} {\bibfnamefont {C.~L.}\ \bibnamefont {Johnson}}, \bibinfo {author} {\bibfnamefont {B.~E.}\ \bibnamefont {Knighton}},\ and\ \bibinfo {author} {\bibfnamefont {J.~A.}\ \bibnamefont {Johnson}},\ }\bibfield  {title} {\bibinfo {title} {Distinguishing nonlinear terahertz excitation pathways with two-dimensional spectroscopy},\ }\href@noop {} {\bibfield  {journal} {\bibinfo  {journal} {Phys. Rev. Lett.}\ }\textbf {\bibinfo {volume} {122}},\ \bibinfo {pages} {073901} (\bibinfo {year} {2019})}\BibitemShut {NoStop}%
\bibitem [{\citenamefont {Blank}\ \emph {et~al.}(2023)\citenamefont {Blank}, \citenamefont {Grishunin}, \citenamefont {Zvezdin}, \citenamefont {Hai}, \citenamefont {Wu}, \citenamefont {Su}, \citenamefont {Huang}, \citenamefont {Zvezdin},\ and\ \citenamefont {Kimel}}]{Blank2023}%
  \BibitemOpen
  \bibfield  {author} {\bibinfo {author} {\bibfnamefont {T.~G.~H.}\ \bibnamefont {Blank}}, \bibinfo {author} {\bibfnamefont {K.~A.}\ \bibnamefont {Grishunin}}, \bibinfo {author} {\bibfnamefont {K.~A.}\ \bibnamefont {Zvezdin}}, \bibinfo {author} {\bibfnamefont {N.~T.}\ \bibnamefont {Hai}}, \bibinfo {author} {\bibfnamefont {J.~C.}\ \bibnamefont {Wu}}, \bibinfo {author} {\bibfnamefont {S.-H.}\ \bibnamefont {Su}}, \bibinfo {author} {\bibfnamefont {J.-C.~A.}\ \bibnamefont {Huang}}, \bibinfo {author} {\bibfnamefont {A.~K.}\ \bibnamefont {Zvezdin}},\ and\ \bibinfo {author} {\bibfnamefont {A.~V.}\ \bibnamefont {Kimel}},\ }\bibfield  {title} {\bibinfo {title} {Two-dimensional terahertz spectroscopy of nonlinear phononics in the topological insulator {${\mathrm{MnBi}}_{2}{\mathrm{Te}}_{4}$}},\ }\href {https://doi.org/10.1103/PhysRevLett.131.026902} {\bibfield  {journal} {\bibinfo  {journal} {Phys. Rev. Lett.}\ }\textbf {\bibinfo {volume} {131}},\ \bibinfo {pages} {026902} (\bibinfo {year} {2023})}\BibitemShut {NoStop}%
\bibitem [{\citenamefont {Mootz}\ \emph {et~al.}(2022)\citenamefont {Mootz}, \citenamefont {Luo}, \citenamefont {Wang},\ and\ \citenamefont {Perakis}}]{Mootz2022}%
  \BibitemOpen
  \bibfield  {author} {\bibinfo {author} {\bibfnamefont {M.}~\bibnamefont {Mootz}}, \bibinfo {author} {\bibfnamefont {L.}~\bibnamefont {Luo}}, \bibinfo {author} {\bibfnamefont {J.}~\bibnamefont {Wang}},\ and\ \bibinfo {author} {\bibfnamefont {l.~E.}\ \bibnamefont {Perakis}},\ }\bibfield  {title} {\bibinfo {title} {Visualization and quantum control of light-accelerated condensates by terahertz multi-dimensional coherent spectroscopy},\ }\href@noop {} {\bibfield  {journal} {\bibinfo  {journal} {Commun. Phys.}\ }\textbf {\bibinfo {volume} {5}},\ \bibinfo {pages} {47} (\bibinfo {year} {2022})}\BibitemShut {NoStop}%
\bibitem [{\citenamefont {Luo}\ \emph {et~al.}(2023)\citenamefont {Luo}, \citenamefont {Mootz}, \citenamefont {Kang}, \citenamefont {Huang}, \citenamefont {Eom}, \citenamefont {Lee}, \citenamefont {Vaswani}, \citenamefont {Collantes}, \citenamefont {Hellstrom}, \citenamefont {Perakis}, \citenamefont {Eom},\ and\ \citenamefont {Wang}}]{NatPhys}%
  \BibitemOpen
  \bibfield  {author} {\bibinfo {author} {\bibfnamefont {L.}~\bibnamefont {Luo}}, \bibinfo {author} {\bibfnamefont {M.}~\bibnamefont {Mootz}}, \bibinfo {author} {\bibfnamefont {J.~H.}\ \bibnamefont {Kang}}, \bibinfo {author} {\bibfnamefont {C.}~\bibnamefont {Huang}}, \bibinfo {author} {\bibfnamefont {K.}~\bibnamefont {Eom}}, \bibinfo {author} {\bibfnamefont {J.~W.}\ \bibnamefont {Lee}}, \bibinfo {author} {\bibfnamefont {C.}~\bibnamefont {Vaswani}}, \bibinfo {author} {\bibfnamefont {Y.~G.}\ \bibnamefont {Collantes}}, \bibinfo {author} {\bibfnamefont {E.~E.}\ \bibnamefont {Hellstrom}}, \bibinfo {author} {\bibfnamefont {I.~E.}\ \bibnamefont {Perakis}}, \bibinfo {author} {\bibfnamefont {C.~B.}\ \bibnamefont {Eom}},\ and\ \bibinfo {author} {\bibfnamefont {J.}~\bibnamefont {Wang}},\ }\bibfield  {title} {\bibinfo {title} {Quantum coherence tomography of light-controlled superconductivity},\ }\href@noop {} {\bibfield  {journal} {\bibinfo  {journal} {Nat. Phys.}\ }\textbf {\bibinfo {volume} {19}},\ \bibinfo {pages}
  {201} (\bibinfo {year} {2023})}\BibitemShut {NoStop}%
\bibitem [{\citenamefont {Puviani}\ \emph {et~al.}(2023)\citenamefont {Puviani}, \citenamefont {Haenel},\ and\ \citenamefont {Manske}}]{Manske2023}%
  \BibitemOpen
  \bibfield  {author} {\bibinfo {author} {\bibfnamefont {M.}~\bibnamefont {Puviani}}, \bibinfo {author} {\bibfnamefont {R.}~\bibnamefont {Haenel}},\ and\ \bibinfo {author} {\bibfnamefont {D.}~\bibnamefont {Manske}},\ }\bibfield  {title} {\bibinfo {title} {Quench-drive spectroscopy and high-harmonic generation in {BCS} superconductors},\ }\href {https://doi.org/10.1103/PhysRevB.107.094501} {\bibfield  {journal} {\bibinfo  {journal} {Phys. Rev. B}\ }\textbf {\bibinfo {volume} {107}},\ \bibinfo {pages} {094501} (\bibinfo {year} {2023})}\BibitemShut {NoStop}%
\bibitem [{\citenamefont {Cheng}\ \emph {et~al.}(2023{\natexlab{a}})\citenamefont {Cheng}, \citenamefont {Cheng}, \citenamefont {Lee}, \citenamefont {Mootz}, \citenamefont {Huang}, \citenamefont {Luo}, \citenamefont {Chen}, \citenamefont {Lee}, \citenamefont {Wang}, \citenamefont {Perakis}, \citenamefont {Shen}, \citenamefont {Hwang},\ and\ \citenamefont {Wang}}]{cheng2023evidence}%
  \BibitemOpen
  \bibfield  {author} {\bibinfo {author} {\bibfnamefont {B.}~\bibnamefont {Cheng}}, \bibinfo {author} {\bibfnamefont {D.}~\bibnamefont {Cheng}}, \bibinfo {author} {\bibfnamefont {K.}~\bibnamefont {Lee}}, \bibinfo {author} {\bibfnamefont {M.}~\bibnamefont {Mootz}}, \bibinfo {author} {\bibfnamefont {C.}~\bibnamefont {Huang}}, \bibinfo {author} {\bibfnamefont {L.}~\bibnamefont {Luo}}, \bibinfo {author} {\bibfnamefont {.~Z.}\ \bibnamefont {Chen}}, \bibinfo {author} {\bibfnamefont {Y.}~\bibnamefont {Lee}}, \bibinfo {author} {\bibfnamefont {B.~Y.}\ \bibnamefont {Wang}}, \bibinfo {author} {\bibfnamefont {I.~E.}\ \bibnamefont {Perakis}}, \bibinfo {author} {\bibfnamefont {Z.-X.}\ \bibnamefont {Shen}}, \bibinfo {author} {\bibfnamefont {H.~Y.}\ \bibnamefont {Hwang}},\ and\ \bibinfo {author} {\bibfnamefont {J.}~\bibnamefont {Wang}},\ }\href@noop {} {\bibinfo {title} {Evidence for highly damped {Higgs} mode in infinite-layer nickelates}} (\bibinfo {year} {2023}{\natexlab{a}}),\ \Eprint {https://arxiv.org/abs/2310.02589}
  {arXiv:2310.02589} \BibitemShut {NoStop}%
\bibitem [{\citenamefont {Matsunaga}\ \emph {et~al.}(2014)\citenamefont {Matsunaga}, \citenamefont {Tsuji}, \citenamefont {Fujita}, \citenamefont {Sugioka}, \citenamefont {Makise}, \citenamefont {Uzawa}, \citenamefont {Terai}, \citenamefont {Wang}, \citenamefont {Aoki},\ and\ \citenamefont {Shimano}}]{matsunaga2014}%
  \BibitemOpen
  \bibfield  {author} {\bibinfo {author} {\bibfnamefont {R.}~\bibnamefont {Matsunaga}}, \bibinfo {author} {\bibfnamefont {N.}~\bibnamefont {Tsuji}}, \bibinfo {author} {\bibfnamefont {H.}~\bibnamefont {Fujita}}, \bibinfo {author} {\bibfnamefont {A.}~\bibnamefont {Sugioka}}, \bibinfo {author} {\bibfnamefont {K.}~\bibnamefont {Makise}}, \bibinfo {author} {\bibfnamefont {Y.}~\bibnamefont {Uzawa}}, \bibinfo {author} {\bibfnamefont {H.}~\bibnamefont {Terai}}, \bibinfo {author} {\bibfnamefont {Z.}~\bibnamefont {Wang}}, \bibinfo {author} {\bibfnamefont {H.}~\bibnamefont {Aoki}},\ and\ \bibinfo {author} {\bibfnamefont {R.}~\bibnamefont {Shimano}},\ }\bibfield  {title} {\bibinfo {title} {Light-induced collective pseudospin precession resonating with {Higgs} mode in a superconductor},\ }\href@noop {} {\bibfield  {journal} {\bibinfo  {journal} {Science}\ }\textbf {\bibinfo {volume} {345}},\ \bibinfo {pages} {1145} (\bibinfo {year} {2014})}\BibitemShut {NoStop}%
\bibitem [{\citenamefont {Matsunaga}\ \emph {et~al.}(2017)\citenamefont {Matsunaga}, \citenamefont {Tsuji}, \citenamefont {Makise}, \citenamefont {Terai}, \citenamefont {Aoki},\ and\ \citenamefont {Shimano}}]{Matsunaga2017}%
  \BibitemOpen
  \bibfield  {author} {\bibinfo {author} {\bibfnamefont {R.}~\bibnamefont {Matsunaga}}, \bibinfo {author} {\bibfnamefont {N.}~\bibnamefont {Tsuji}}, \bibinfo {author} {\bibfnamefont {K.}~\bibnamefont {Makise}}, \bibinfo {author} {\bibfnamefont {H.}~\bibnamefont {Terai}}, \bibinfo {author} {\bibfnamefont {H.}~\bibnamefont {Aoki}},\ and\ \bibinfo {author} {\bibfnamefont {R.}~\bibnamefont {Shimano}},\ }\bibfield  {title} {\bibinfo {title} {Polarization-resolved terahertz third-harmonic generation in a single-crystal superconductor {NbN}: Dominance of the {Higgs} mode beyond the {BCS} approximation},\ }\href@noop {} {\bibfield  {journal} {\bibinfo  {journal} {Phys. Rev. B}\ }\textbf {\bibinfo {volume} {96}},\ \bibinfo {pages} {020505(R)} (\bibinfo {year} {2017})}\BibitemShut {NoStop}%
\bibitem [{\citenamefont {Giorgianni}\ \emph {et~al.}(2019)\citenamefont {Giorgianni}, \citenamefont {Cea}, \citenamefont {Vicario}, \citenamefont {Hauri}, \citenamefont {Withanage}, \citenamefont {Xi},\ and\ \citenamefont {Benfatto}}]{Giorgianni2019}%
  \BibitemOpen
  \bibfield  {author} {\bibinfo {author} {\bibfnamefont {F.}~\bibnamefont {Giorgianni}}, \bibinfo {author} {\bibfnamefont {T.}~\bibnamefont {Cea}}, \bibinfo {author} {\bibfnamefont {C.}~\bibnamefont {Vicario}}, \bibinfo {author} {\bibfnamefont {C.~P.}\ \bibnamefont {Hauri}}, \bibinfo {author} {\bibfnamefont {W.~K.}\ \bibnamefont {Withanage}}, \bibinfo {author} {\bibfnamefont {X.}~\bibnamefont {Xi}},\ and\ \bibinfo {author} {\bibfnamefont {L.}~\bibnamefont {Benfatto}},\ }\bibfield  {title} {\bibinfo {title} {Leggett mode controlled by light pulses},\ }\href@noop {} {\bibfield  {journal} {\bibinfo  {journal} {Nat. Phys.}\ }\textbf {\bibinfo {volume} {15}},\ \bibinfo {pages} {341} (\bibinfo {year} {2019})}\BibitemShut {NoStop}%
\bibitem [{\citenamefont {Chu}\ \emph {et~al.}(2020)\citenamefont {Chu}, \citenamefont {Kim}, \citenamefont {Katsumi}, \citenamefont {Kovalev}, \citenamefont {Dawson}, \citenamefont {Schwarz}, \citenamefont {Yoshikawa}, \citenamefont {Kim}, \citenamefont {Putzky}, \citenamefont {Li}, \citenamefont {Raffy}, \citenamefont {Germanskiy}, \citenamefont {Deinert}, \citenamefont {Awari}, \citenamefont {Ilyakov}, \citenamefont {Green}, \citenamefont {Chen}, \citenamefont {Bawatna}, \citenamefont {Cristiani}, \citenamefont {Logvenov}, \citenamefont {Gallais}, \citenamefont {Boris}, \citenamefont {Keimer}, \citenamefont {Schnyder}, \citenamefont {Manske}, \citenamefont {Gensch}, \citenamefont {Wang}, \citenamefont {Shimano},\ and\ \citenamefont {Kaiser}}]{Chu2020}%
  \BibitemOpen
  \bibfield  {author} {\bibinfo {author} {\bibfnamefont {H.}~\bibnamefont {Chu}}, \bibinfo {author} {\bibfnamefont {M.-J.}\ \bibnamefont {Kim}}, \bibinfo {author} {\bibfnamefont {K.}~\bibnamefont {Katsumi}}, \bibinfo {author} {\bibfnamefont {S.}~\bibnamefont {Kovalev}}, \bibinfo {author} {\bibfnamefont {R.~D.}\ \bibnamefont {Dawson}}, \bibinfo {author} {\bibfnamefont {L.}~\bibnamefont {Schwarz}}, \bibinfo {author} {\bibfnamefont {N.}~\bibnamefont {Yoshikawa}}, \bibinfo {author} {\bibfnamefont {G.}~\bibnamefont {Kim}}, \bibinfo {author} {\bibfnamefont {D.}~\bibnamefont {Putzky}}, \bibinfo {author} {\bibfnamefont {Z.~Z.}\ \bibnamefont {Li}}, \bibinfo {author} {\bibfnamefont {H.}~\bibnamefont {Raffy}}, \bibinfo {author} {\bibfnamefont {S.}~\bibnamefont {Germanskiy}}, \bibinfo {author} {\bibfnamefont {J.-C.}\ \bibnamefont {Deinert}}, \bibinfo {author} {\bibfnamefont {N.}~\bibnamefont {Awari}}, \bibinfo {author} {\bibfnamefont {I.}~\bibnamefont {Ilyakov}}, \bibinfo {author} {\bibfnamefont {B.}~\bibnamefont
  {Green}}, \bibinfo {author} {\bibfnamefont {M.}~\bibnamefont {Chen}}, \bibinfo {author} {\bibfnamefont {M.}~\bibnamefont {Bawatna}}, \bibinfo {author} {\bibfnamefont {G.}~\bibnamefont {Cristiani}}, \bibinfo {author} {\bibfnamefont {G.}~\bibnamefont {Logvenov}}, \bibinfo {author} {\bibfnamefont {Y.}~\bibnamefont {Gallais}}, \bibinfo {author} {\bibfnamefont {A.~V.}\ \bibnamefont {Boris}}, \bibinfo {author} {\bibfnamefont {B.}~\bibnamefont {Keimer}}, \bibinfo {author} {\bibfnamefont {A.~P.}\ \bibnamefont {Schnyder}}, \bibinfo {author} {\bibfnamefont {D.}~\bibnamefont {Manske}}, \bibinfo {author} {\bibfnamefont {M.}~\bibnamefont {Gensch}}, \bibinfo {author} {\bibfnamefont {Z.}~\bibnamefont {Wang}}, \bibinfo {author} {\bibfnamefont {R.}~\bibnamefont {Shimano}},\ and\ \bibinfo {author} {\bibfnamefont {S.}~\bibnamefont {Kaiser}},\ }\bibfield  {title} {\bibinfo {title} {Phase-resolved {Higgs} response in superconducting cuprates},\ }\href@noop {} {\bibfield  {journal} {\bibinfo  {journal} {Nat. Commun.}\ }\textbf
  {\bibinfo {volume} {11}},\ \bibinfo {pages} {1793} (\bibinfo {year} {2020})}\BibitemShut {NoStop}%
\bibitem [{\citenamefont {Vaswani}\ \emph {et~al.}(2021)\citenamefont {Vaswani}, \citenamefont {Kang}, \citenamefont {Mootz}, \citenamefont {Luo}, \citenamefont {Yang}, \citenamefont {Sundahl}, \citenamefont {Cheng}, \citenamefont {Huang}, \citenamefont {Kim}, \citenamefont {Liu}, \citenamefont {Collantes}, \citenamefont {Hellstrom}, \citenamefont {Perakis}, \citenamefont {Eom},\ and\ \citenamefont {Wang}}]{hybrid-higgs}%
  \BibitemOpen
  \bibfield  {author} {\bibinfo {author} {\bibfnamefont {C.}~\bibnamefont {Vaswani}}, \bibinfo {author} {\bibfnamefont {J.~H.}\ \bibnamefont {Kang}}, \bibinfo {author} {\bibfnamefont {M.}~\bibnamefont {Mootz}}, \bibinfo {author} {\bibfnamefont {L.}~\bibnamefont {Luo}}, \bibinfo {author} {\bibfnamefont {X.}~\bibnamefont {Yang}}, \bibinfo {author} {\bibfnamefont {C.}~\bibnamefont {Sundahl}}, \bibinfo {author} {\bibfnamefont {D.}~\bibnamefont {Cheng}}, \bibinfo {author} {\bibfnamefont {C.}~\bibnamefont {Huang}}, \bibinfo {author} {\bibfnamefont {R.~H.~J.}\ \bibnamefont {Kim}}, \bibinfo {author} {\bibfnamefont {Z.}~\bibnamefont {Liu}}, \bibinfo {author} {\bibfnamefont {Y.~G.}\ \bibnamefont {Collantes}}, \bibinfo {author} {\bibfnamefont {E.~E.}\ \bibnamefont {Hellstrom}}, \bibinfo {author} {\bibfnamefont {I.~E.}\ \bibnamefont {Perakis}}, \bibinfo {author} {\bibfnamefont {C.~B.}\ \bibnamefont {Eom}},\ and\ \bibinfo {author} {\bibfnamefont {J.}~\bibnamefont {Wang}},\ }\bibfield  {title} {\bibinfo {title} {Light
  quantum control of persisting {Higgs} modes in iron-based superconductors},\ }\href@noop {} {\bibfield  {journal} {\bibinfo  {journal} {Nat. Commun.}\ }\textbf {\bibinfo {volume} {12}},\ \bibinfo {pages} {258} (\bibinfo {year} {2021})}\BibitemShut {NoStop}%
\bibitem [{\citenamefont {Cheng}\ \emph {et~al.}(2023{\natexlab{b}})\citenamefont {Cheng}, \citenamefont {Cheng}, \citenamefont {Lee}, \citenamefont {Luo}, \citenamefont {Chen}, \citenamefont {Lee}, \citenamefont {Wang}, \citenamefont {Mootz}, \citenamefont {Perakis}, \citenamefont {Shen}, \citenamefont {Hwang},\ and\ \citenamefont {Wang}}]{cheng2023lowenergy}%
  \BibitemOpen
  \bibfield  {author} {\bibinfo {author} {\bibfnamefont {B.}~\bibnamefont {Cheng}}, \bibinfo {author} {\bibfnamefont {D.}~\bibnamefont {Cheng}}, \bibinfo {author} {\bibfnamefont {K.}~\bibnamefont {Lee}}, \bibinfo {author} {\bibfnamefont {L.}~\bibnamefont {Luo}}, \bibinfo {author} {\bibfnamefont {Z.}~\bibnamefont {Chen}}, \bibinfo {author} {\bibfnamefont {Y.}~\bibnamefont {Lee}}, \bibinfo {author} {\bibfnamefont {B.~Y.}\ \bibnamefont {Wang}}, \bibinfo {author} {\bibfnamefont {M.}~\bibnamefont {Mootz}}, \bibinfo {author} {\bibfnamefont {I.~E.}\ \bibnamefont {Perakis}}, \bibinfo {author} {\bibfnamefont {Z.-X.}\ \bibnamefont {Shen}}, \bibinfo {author} {\bibfnamefont {H.~Y.}\ \bibnamefont {Hwang}},\ and\ \bibinfo {author} {\bibfnamefont {J.}~\bibnamefont {Wang}},\ }\href@noop {} {\bibinfo {title} {Low-energy electrodynamics of infinite-layer nickelates: evidence for d-wave superconductivity in the dirty limit}} (\bibinfo {year} {2023}{\natexlab{b}}),\ \Eprint {https://arxiv.org/abs/2310.02586} {arXiv:2310.02586}
  \BibitemShut {NoStop}%
\bibitem [{\citenamefont {Tsuji}\ and\ \citenamefont {Aoki}(2015)}]{Aoki2015}%
  \BibitemOpen
  \bibfield  {author} {\bibinfo {author} {\bibfnamefont {N.}~\bibnamefont {Tsuji}}\ and\ \bibinfo {author} {\bibfnamefont {H.}~\bibnamefont {Aoki}},\ }\bibfield  {title} {\bibinfo {title} {Theory of {Anderson} pseudospin resonance with {Higgs} mode in superconductors},\ }\href@noop {} {\bibfield  {journal} {\bibinfo  {journal} {Phys. Rev. B}\ }\textbf {\bibinfo {volume} {92}},\ \bibinfo {pages} {064508} (\bibinfo {year} {2015})}\BibitemShut {NoStop}%
\bibitem [{\citenamefont {Murotani}\ \emph {et~al.}(2017)\citenamefont {Murotani}, \citenamefont {Tsuji},\ and\ \citenamefont {Aoki}}]{Aoki2017}%
  \BibitemOpen
  \bibfield  {author} {\bibinfo {author} {\bibfnamefont {Y.}~\bibnamefont {Murotani}}, \bibinfo {author} {\bibfnamefont {N.}~\bibnamefont {Tsuji}},\ and\ \bibinfo {author} {\bibfnamefont {H.}~\bibnamefont {Aoki}},\ }\bibfield  {title} {\bibinfo {title} {Theory of light-induced resonances with collective {Higgs} and {Leggett} modes in multiband superconductors},\ }\href@noop {} {\bibfield  {journal} {\bibinfo  {journal} {Phys. Rev. B}\ }\textbf {\bibinfo {volume} {95}},\ \bibinfo {pages} {104503} (\bibinfo {year} {2017})}\BibitemShut {NoStop}%
\bibitem [{\citenamefont {Udina}\ \emph {et~al.}(2019)\citenamefont {Udina}, \citenamefont {Cea},\ and\ \citenamefont {Benfatto}}]{Benfatto2019}%
  \BibitemOpen
  \bibfield  {author} {\bibinfo {author} {\bibfnamefont {M.}~\bibnamefont {Udina}}, \bibinfo {author} {\bibfnamefont {T.}~\bibnamefont {Cea}},\ and\ \bibinfo {author} {\bibfnamefont {L.}~\bibnamefont {Benfatto}},\ }\bibfield  {title} {\bibinfo {title} {Theory of coherent-oscillations generation in terahertz pump-probe spectroscopy: From phonons to electronic collective modes},\ }\href@noop {} {\bibfield  {journal} {\bibinfo  {journal} {Phys. Rev. B}\ }\textbf {\bibinfo {volume} {100}},\ \bibinfo {pages} {165131} (\bibinfo {year} {2019})}\BibitemShut {NoStop}%
\bibitem [{\citenamefont {Udina}\ \emph {et~al.}(2022)\citenamefont {Udina}, \citenamefont {Fiore}, \citenamefont {Cea}, \citenamefont {Castellani}, \citenamefont {Seibold},\ and\ \citenamefont {Benfatto}}]{udina2022thz}%
  \BibitemOpen
  \bibfield  {author} {\bibinfo {author} {\bibfnamefont {M.}~\bibnamefont {Udina}}, \bibinfo {author} {\bibfnamefont {J.}~\bibnamefont {Fiore}}, \bibinfo {author} {\bibfnamefont {T.}~\bibnamefont {Cea}}, \bibinfo {author} {\bibfnamefont {C.}~\bibnamefont {Castellani}}, \bibinfo {author} {\bibfnamefont {G.}~\bibnamefont {Seibold}},\ and\ \bibinfo {author} {\bibfnamefont {L.}~\bibnamefont {Benfatto}},\ }\bibfield  {title} {\bibinfo {title} {Thz non-linear optical response in cuprates: predominance of the bcs response over the higgs mode},\ }\href@noop {} {\bibfield  {journal} {\bibinfo  {journal} {Faraday Discussions}\ }\textbf {\bibinfo {volume} {237}},\ \bibinfo {pages} {168} (\bibinfo {year} {2022})}\BibitemShut {NoStop}%
\bibitem [{\citenamefont {Papenkort}\ \emph {et~al.}(2007)\citenamefont {Papenkort}, \citenamefont {Axt},\ and\ \citenamefont {Kuhn}}]{Axt2007}%
  \BibitemOpen
  \bibfield  {author} {\bibinfo {author} {\bibfnamefont {T.}~\bibnamefont {Papenkort}}, \bibinfo {author} {\bibfnamefont {V.~M.}\ \bibnamefont {Axt}},\ and\ \bibinfo {author} {\bibfnamefont {T.}~\bibnamefont {Kuhn}},\ }\bibfield  {title} {\bibinfo {title} {Coherent dynamics and pump-probe spectra of {BCS} superconductors},\ }\href@noop {} {\bibfield  {journal} {\bibinfo  {journal} {Phys. Rev. B}\ }\textbf {\bibinfo {volume} {76}},\ \bibinfo {pages} {224522} (\bibinfo {year} {2007})}\BibitemShut {NoStop}%
\bibitem [{\citenamefont {Krull}\ \emph {et~al.}(2016)\citenamefont {Krull}, \citenamefont {Bittner}, \citenamefont {Uhrig}, \citenamefont {Manske},\ and\ \citenamefont {Schnyder}}]{krull2016}%
  \BibitemOpen
  \bibfield  {author} {\bibinfo {author} {\bibfnamefont {H.}~\bibnamefont {Krull}}, \bibinfo {author} {\bibfnamefont {N.}~\bibnamefont {Bittner}}, \bibinfo {author} {\bibfnamefont {G.}~\bibnamefont {Uhrig}}, \bibinfo {author} {\bibfnamefont {D.}~\bibnamefont {Manske}},\ and\ \bibinfo {author} {\bibfnamefont {A.}~\bibnamefont {Schnyder}},\ }\bibfield  {title} {\bibinfo {title} {Coupling of {Higgs} and {Leggett} modes in non-equilibrium superconductors},\ }\href@noop {} {\bibfield  {journal} {\bibinfo  {journal} {Nat. Commun.}\ }\textbf {\bibinfo {volume} {7}},\ \bibinfo {pages} {11921} (\bibinfo {year} {2016})}\BibitemShut {NoStop}%
\bibitem [{\citenamefont {Chou}\ \emph {et~al.}(2017)\citenamefont {Chou}, \citenamefont {Liao},\ and\ \citenamefont {Foster}}]{Forster2017}%
  \BibitemOpen
  \bibfield  {author} {\bibinfo {author} {\bibfnamefont {Y.-Z.}\ \bibnamefont {Chou}}, \bibinfo {author} {\bibfnamefont {Y.}~\bibnamefont {Liao}},\ and\ \bibinfo {author} {\bibfnamefont {M.~S.}\ \bibnamefont {Foster}},\ }\bibfield  {title} {\bibinfo {title} {Twisting {Anderson} pseudospins with light: Quench dynamics in terahertz-pumped {BCS} superconductors},\ }\href@noop {} {\bibfield  {journal} {\bibinfo  {journal} {Phys. Rev. B}\ }\textbf {\bibinfo {volume} {95}},\ \bibinfo {pages} {104507} (\bibinfo {year} {2017})}\BibitemShut {NoStop}%
\bibitem [{\citenamefont {Schwarz}\ \emph {et~al.}(2020)\citenamefont {Schwarz}, \citenamefont {Fauseweh},\ and\ \citenamefont {Tsuji}}]{Schwarz2020}%
  \BibitemOpen
  \bibfield  {author} {\bibinfo {author} {\bibfnamefont {L.}~\bibnamefont {Schwarz}}, \bibinfo {author} {\bibfnamefont {B.}~\bibnamefont {Fauseweh}},\ and\ \bibinfo {author} {\bibfnamefont {N.~e.~a.}\ \bibnamefont {Tsuji}},\ }\bibfield  {title} {\bibinfo {title} {Classification and characterization of nonequilibrium {Higgs} modes in unconventional superconductors.},\ }\href@noop {} {\bibfield  {journal} {\bibinfo  {journal} {Nat. Commun.}\ }\textbf {\bibinfo {volume} {11}},\ \bibinfo {pages} {287} (\bibinfo {year} {2020})}\BibitemShut {NoStop}%
\bibitem [{\citenamefont {Mootz}\ \emph {et~al.}(2020)\citenamefont {Mootz}, \citenamefont {Wang},\ and\ \citenamefont {Perakis}}]{Mootz2020}%
  \BibitemOpen
  \bibfield  {author} {\bibinfo {author} {\bibfnamefont {M.}~\bibnamefont {Mootz}}, \bibinfo {author} {\bibfnamefont {J.}~\bibnamefont {Wang}},\ and\ \bibinfo {author} {\bibfnamefont {I.~E.}\ \bibnamefont {Perakis}},\ }\bibfield  {title} {\bibinfo {title} {Lightwave terahertz quantum manipulation of nonequilibrium superconductor phases and their collective modes},\ }\href@noop {} {\bibfield  {journal} {\bibinfo  {journal} {Phys. Rev. B}\ }\textbf {\bibinfo {volume} {102}},\ \bibinfo {pages} {054517} (\bibinfo {year} {2020})}\BibitemShut {NoStop}%
\bibitem [{\citenamefont {Stephen}(1965)}]{Stephen1965}%
  \BibitemOpen
  \bibfield  {author} {\bibinfo {author} {\bibfnamefont {M.~J.}\ \bibnamefont {Stephen}},\ }\bibfield  {title} {\bibinfo {title} {Transport equations for superconductors},\ }\href@noop {} {\bibfield  {journal} {\bibinfo  {journal} {Phys. Rev.}\ }\textbf {\bibinfo {volume} {139}},\ \bibinfo {pages} {A197} (\bibinfo {year} {1965})}\BibitemShut {NoStop}%
\bibitem [{\citenamefont {Yu}\ and\ \citenamefont {Wu}(2017)}]{Wu2017}%
  \BibitemOpen
  \bibfield  {author} {\bibinfo {author} {\bibfnamefont {T.}~\bibnamefont {Yu}}\ and\ \bibinfo {author} {\bibfnamefont {M.~W.}\ \bibnamefont {Wu}},\ }\bibfield  {title} {\bibinfo {title} {Gauge-invariant theory of quasiparticle and condensate dynamics in response to terahertz optical pulses in superconducting semiconductor quantum wells. i. $s$-wave superconductivity in the weak spin-orbit coupling limit},\ }\href@noop {} {\bibfield  {journal} {\bibinfo  {journal} {Phys. Rev. B}\ }\textbf {\bibinfo {volume} {96}},\ \bibinfo {pages} {155311} (\bibinfo {year} {2017})}\BibitemShut {NoStop}%
\bibitem [{\citenamefont {Yang}\ and\ \citenamefont {Wu}(2019)}]{Wu2019}%
  \BibitemOpen
  \bibfield  {author} {\bibinfo {author} {\bibfnamefont {F.}~\bibnamefont {Yang}}\ and\ \bibinfo {author} {\bibfnamefont {M.~W.}\ \bibnamefont {Wu}},\ }\bibfield  {title} {\bibinfo {title} {Gauge-invariant microscopic kinetic theory of superconductivity: Application to the optical response of {Nambu-Goldstone} and {Higgs} modes},\ }\href@noop {} {\bibfield  {journal} {\bibinfo  {journal} {Phys. Rev. B}\ }\textbf {\bibinfo {volume} {100}},\ \bibinfo {pages} {104513} (\bibinfo {year} {2019})}\BibitemShut {NoStop}%
\bibitem [{\citenamefont {Murotani}\ and\ \citenamefont {Shimano}(2019)}]{Murotani}%
  \BibitemOpen
  \bibfield  {author} {\bibinfo {author} {\bibfnamefont {Y.}~\bibnamefont {Murotani}}\ and\ \bibinfo {author} {\bibfnamefont {R.}~\bibnamefont {Shimano}},\ }\bibfield  {title} {\bibinfo {title} {Nonlinear optical response of collective modes in multiband superconductors assisted by nonmagnetic impurities},\ }\href@noop {} {\bibfield  {journal} {\bibinfo  {journal} {Phys. Rev. B}\ }\textbf {\bibinfo {volume} {99}},\ \bibinfo {pages} {224510} (\bibinfo {year} {2019})}\BibitemShut {NoStop}%
\bibitem [{\citenamefont {Seibold}\ \emph {et~al.}(2021)\citenamefont {Seibold}, \citenamefont {Udina}, \citenamefont {Castellani},\ and\ \citenamefont {Benfatto}}]{seibold2021}%
  \BibitemOpen
  \bibfield  {author} {\bibinfo {author} {\bibfnamefont {G.}~\bibnamefont {Seibold}}, \bibinfo {author} {\bibfnamefont {M.}~\bibnamefont {Udina}}, \bibinfo {author} {\bibfnamefont {C.}~\bibnamefont {Castellani}},\ and\ \bibinfo {author} {\bibfnamefont {L.}~\bibnamefont {Benfatto}},\ }\bibfield  {title} {\bibinfo {title} {Third harmonic generation from collective modes in disordered superconductors},\ }\href@noop {} {\bibfield  {journal} {\bibinfo  {journal} {Phys. Rev. B}\ }\textbf {\bibinfo {volume} {103}},\ \bibinfo {pages} {014512} (\bibinfo {year} {2021})}\BibitemShut {NoStop}%
\bibitem [{\citenamefont {Benfatto}\ \emph {et~al.}(2023)\citenamefont {Benfatto}, \citenamefont {Castellani},\ and\ \citenamefont {Seibold}}]{Benfatto2023}%
  \BibitemOpen
  \bibfield  {author} {\bibinfo {author} {\bibfnamefont {L.}~\bibnamefont {Benfatto}}, \bibinfo {author} {\bibfnamefont {C.}~\bibnamefont {Castellani}},\ and\ \bibinfo {author} {\bibfnamefont {G.}~\bibnamefont {Seibold}},\ }\bibfield  {title} {\bibinfo {title} {Linear and nonlinear current response in disordered $d$-wave superconductors},\ }\href@noop {} {\bibfield  {journal} {\bibinfo  {journal} {Phys. Rev. B}\ }\textbf {\bibinfo {volume} {108}},\ \bibinfo {pages} {134508} (\bibinfo {year} {2023})}\BibitemShut {NoStop}%
\bibitem [{\citenamefont {Yang}\ \emph {et~al.}(2018{\natexlab{a}})\citenamefont {Yang}, \citenamefont {Vaswani}, \citenamefont {Sundahl}, \citenamefont {Mootz}, \citenamefont {Gagel}, \citenamefont {Luo}, \citenamefont {Kang}, \citenamefont {Orth}, \citenamefont {Perakis}, \citenamefont {Eom} \emph {et~al.}}]{yang2018}%
  \BibitemOpen
  \bibfield  {author} {\bibinfo {author} {\bibfnamefont {X.}~\bibnamefont {Yang}}, \bibinfo {author} {\bibfnamefont {C.}~\bibnamefont {Vaswani}}, \bibinfo {author} {\bibfnamefont {C.}~\bibnamefont {Sundahl}}, \bibinfo {author} {\bibfnamefont {M.}~\bibnamefont {Mootz}}, \bibinfo {author} {\bibfnamefont {P.}~\bibnamefont {Gagel}}, \bibinfo {author} {\bibfnamefont {L.}~\bibnamefont {Luo}}, \bibinfo {author} {\bibfnamefont {J.}~\bibnamefont {Kang}}, \bibinfo {author} {\bibfnamefont {P.}~\bibnamefont {Orth}}, \bibinfo {author} {\bibfnamefont {I.}~\bibnamefont {Perakis}}, \bibinfo {author} {\bibfnamefont {C.}~\bibnamefont {Eom}}, \emph {et~al.},\ }\bibfield  {title} {\bibinfo {title} {Terahertz-light quantum tuning of a metastable emergent phase hidden by superconductivity},\ }\href@noop {} {\bibfield  {journal} {\bibinfo  {journal} {Nat. Mater.}\ }\textbf {\bibinfo {volume} {17}},\ \bibinfo {pages} {586} (\bibinfo {year} {2018}{\natexlab{a}})}\BibitemShut {NoStop}%
\bibitem [{\citenamefont {Yang}\ \emph {et~al.}(2019)\citenamefont {Yang}, \citenamefont {Vaswani}, \citenamefont {Sundahl}, \citenamefont {Mootz}, \citenamefont {Luo}, \citenamefont {Kang}, \citenamefont {Perakis}, \citenamefont {Eom},\ and\ \citenamefont {Wang}}]{yang2019lightwave}%
  \BibitemOpen
  \bibfield  {author} {\bibinfo {author} {\bibfnamefont {X.}~\bibnamefont {Yang}}, \bibinfo {author} {\bibfnamefont {C.}~\bibnamefont {Vaswani}}, \bibinfo {author} {\bibfnamefont {C.}~\bibnamefont {Sundahl}}, \bibinfo {author} {\bibfnamefont {M.}~\bibnamefont {Mootz}}, \bibinfo {author} {\bibfnamefont {L.}~\bibnamefont {Luo}}, \bibinfo {author} {\bibfnamefont {J.}~\bibnamefont {Kang}}, \bibinfo {author} {\bibfnamefont {I.}~\bibnamefont {Perakis}}, \bibinfo {author} {\bibfnamefont {C.}~\bibnamefont {Eom}},\ and\ \bibinfo {author} {\bibfnamefont {J.}~\bibnamefont {Wang}},\ }\bibfield  {title} {\bibinfo {title} {Lightwave-driven gapless superconductivity and forbidden quantum beats by terahertz symmetry breaking},\ }\href@noop {} {\bibfield  {journal} {\bibinfo  {journal} {Nat. Photon.}\ }\textbf {\bibinfo {volume} {13}},\ \bibinfo {pages} {707} (\bibinfo {year} {2019})}\BibitemShut {NoStop}%
\bibitem [{\citenamefont {Anderson}(1958)}]{Anderson}%
  \BibitemOpen
  \bibfield  {author} {\bibinfo {author} {\bibfnamefont {P.~W.}\ \bibnamefont {Anderson}},\ }\bibfield  {title} {\bibinfo {title} {Random-phase approximation in the theory of superconductivity},\ }\href@noop {} {\bibfield  {journal} {\bibinfo  {journal} {Phys. Rev.}\ }\textbf {\bibinfo {volume} {112}},\ \bibinfo {pages} {1900} (\bibinfo {year} {1958})}\BibitemShut {NoStop}%
\bibitem [{\citenamefont {Sooryakumar}\ and\ \citenamefont {Klein}(1980)}]{Klein1980}%
  \BibitemOpen
  \bibfield  {author} {\bibinfo {author} {\bibfnamefont {R.}~\bibnamefont {Sooryakumar}}\ and\ \bibinfo {author} {\bibfnamefont {M.~V.}\ \bibnamefont {Klein}},\ }\bibfield  {title} {\bibinfo {title} {Raman scattering by superconducting-gap excitations and their coupling to charge-density waves},\ }\href@noop {} {\bibfield  {journal} {\bibinfo  {journal} {Phys. Rev. Lett.}\ }\textbf {\bibinfo {volume} {45}},\ \bibinfo {pages} {660} (\bibinfo {year} {1980})}\BibitemShut {NoStop}%
\bibitem [{\citenamefont {Littlewood}\ and\ \citenamefont {Varma}(1981)}]{Littlewood1981}%
  \BibitemOpen
  \bibfield  {author} {\bibinfo {author} {\bibfnamefont {P.~B.}\ \bibnamefont {Littlewood}}\ and\ \bibinfo {author} {\bibfnamefont {C.~M.}\ \bibnamefont {Varma}},\ }\bibfield  {title} {\bibinfo {title} {Gauge-invariant theory of the dynamical interaction of charge density waves and superconductivity},\ }\href@noop {} {\bibfield  {journal} {\bibinfo  {journal} {Phys. Rev. Lett.}\ }\textbf {\bibinfo {volume} {47}},\ \bibinfo {pages} {811} (\bibinfo {year} {1981})}\BibitemShut {NoStop}%
\bibitem [{\citenamefont {Podolsky}\ \emph {et~al.}(2011)\citenamefont {Podolsky}, \citenamefont {Auerbach},\ and\ \citenamefont {Arovas}}]{Podolsky2011}%
  \BibitemOpen
  \bibfield  {author} {\bibinfo {author} {\bibfnamefont {D.}~\bibnamefont {Podolsky}}, \bibinfo {author} {\bibfnamefont {A.}~\bibnamefont {Auerbach}},\ and\ \bibinfo {author} {\bibfnamefont {D.~P.}\ \bibnamefont {Arovas}},\ }\bibfield  {title} {\bibinfo {title} {Visibility of the amplitude ({Higgs}) mode in condensed matter},\ }\href@noop {} {\bibfield  {journal} {\bibinfo  {journal} {Phys. Rev. B}\ }\textbf {\bibinfo {volume} {84}},\ \bibinfo {pages} {174522} (\bibinfo {year} {2011})}\BibitemShut {NoStop}%
\bibitem [{\citenamefont {Moor}\ \emph {et~al.}(2017)\citenamefont {Moor}, \citenamefont {Volkov},\ and\ \citenamefont {Efetov}}]{Moor2017}%
  \BibitemOpen
  \bibfield  {author} {\bibinfo {author} {\bibfnamefont {A.}~\bibnamefont {Moor}}, \bibinfo {author} {\bibfnamefont {A.~F.}\ \bibnamefont {Volkov}},\ and\ \bibinfo {author} {\bibfnamefont {K.~B.}\ \bibnamefont {Efetov}},\ }\bibfield  {title} {\bibinfo {title} {Amplitude {H}iggs mode and admittance in superconductors with a moving condensate},\ }\href@noop {} {\bibfield  {journal} {\bibinfo  {journal} {Phys. Rev. Lett.}\ }\textbf {\bibinfo {volume} {118}},\ \bibinfo {pages} {047001} (\bibinfo {year} {2017})}\BibitemShut {NoStop}%
\bibitem [{\citenamefont {Nakamura}\ \emph {et~al.}(2019)\citenamefont {Nakamura}, \citenamefont {Iida}, \citenamefont {Murotani}, \citenamefont {Matsunaga}, \citenamefont {Terai},\ and\ \citenamefont {Shimano}}]{Shimano2019}%
  \BibitemOpen
  \bibfield  {author} {\bibinfo {author} {\bibfnamefont {S.}~\bibnamefont {Nakamura}}, \bibinfo {author} {\bibfnamefont {Y.}~\bibnamefont {Iida}}, \bibinfo {author} {\bibfnamefont {Y.}~\bibnamefont {Murotani}}, \bibinfo {author} {\bibfnamefont {R.}~\bibnamefont {Matsunaga}}, \bibinfo {author} {\bibfnamefont {H.}~\bibnamefont {Terai}},\ and\ \bibinfo {author} {\bibfnamefont {R.}~\bibnamefont {Shimano}},\ }\bibfield  {title} {\bibinfo {title} {Infrared activation of the {Higgs} mode by supercurrent injection in superconducting {NbN}},\ }\href@noop {} {\bibfield  {journal} {\bibinfo  {journal} {Phys. Rev. Lett.}\ }\textbf {\bibinfo {volume} {122}},\ \bibinfo {pages} {257001} (\bibinfo {year} {2019})}\BibitemShut {NoStop}%
\bibitem [{\citenamefont {Yang}\ \emph {et~al.}(2018{\natexlab{b}})\citenamefont {Yang}, \citenamefont {Luo}, \citenamefont {Mootz}, \citenamefont {Patz}, \citenamefont {Bud'ko}, \citenamefont {Canfield}, \citenamefont {Perakis},\ and\ \citenamefont {Wang}}]{yangPRL}%
  \BibitemOpen
  \bibfield  {author} {\bibinfo {author} {\bibfnamefont {X.}~\bibnamefont {Yang}}, \bibinfo {author} {\bibfnamefont {L.}~\bibnamefont {Luo}}, \bibinfo {author} {\bibfnamefont {M.}~\bibnamefont {Mootz}}, \bibinfo {author} {\bibfnamefont {A.}~\bibnamefont {Patz}}, \bibinfo {author} {\bibfnamefont {S.~L.}\ \bibnamefont {Bud'ko}}, \bibinfo {author} {\bibfnamefont {P.~C.}\ \bibnamefont {Canfield}}, \bibinfo {author} {\bibfnamefont {I.~E.}\ \bibnamefont {Perakis}},\ and\ \bibinfo {author} {\bibfnamefont {J.}~\bibnamefont {Wang}},\ }\bibfield  {title} {\bibinfo {title} {Nonequilibrium pair breaking in {$\mathrm{Ba}({\mathrm{Fe}}_{1\ensuremath{-}x}{\mathrm{Co}}_{x}{)}_{2}{\mathrm{As}}_{2}$} superconductors: Evidence for formation of a photoinduced excitonic state},\ }\href@noop {} {\bibfield  {journal} {\bibinfo  {journal} {Phys. Rev. Lett.}\ }\textbf {\bibinfo {volume} {121}},\ \bibinfo {pages} {267001} (\bibinfo {year} {2018}{\natexlab{b}})}\BibitemShut {NoStop}%
\bibitem [{\citenamefont {Puviani}\ \emph {et~al.}(2020)\citenamefont {Puviani}, \citenamefont {Schwarz}, \citenamefont {Zhang}, \citenamefont {Kaiser},\ and\ \citenamefont {Manske}}]{Puviani2020}%
  \BibitemOpen
  \bibfield  {author} {\bibinfo {author} {\bibfnamefont {M.}~\bibnamefont {Puviani}}, \bibinfo {author} {\bibfnamefont {L.}~\bibnamefont {Schwarz}}, \bibinfo {author} {\bibfnamefont {X.-X.}\ \bibnamefont {Zhang}}, \bibinfo {author} {\bibfnamefont {S.}~\bibnamefont {Kaiser}},\ and\ \bibinfo {author} {\bibfnamefont {D.}~\bibnamefont {Manske}},\ }\bibfield  {title} {\bibinfo {title} {Current-assisted {Raman} activation of the {Higgs} mode in superconductors},\ }\href@noop {} {\bibfield  {journal} {\bibinfo  {journal} {Phys. Rev. B}\ }\textbf {\bibinfo {volume} {101}},\ \bibinfo {pages} {220507(R)} (\bibinfo {year} {2020})}\BibitemShut {NoStop}%
\bibitem [{\citenamefont {Cea}\ \emph {et~al.}(2016)\citenamefont {Cea}, \citenamefont {Castellani},\ and\ \citenamefont {Benfatto}}]{Cea2016}%
  \BibitemOpen
  \bibfield  {author} {\bibinfo {author} {\bibfnamefont {T.}~\bibnamefont {Cea}}, \bibinfo {author} {\bibfnamefont {C.}~\bibnamefont {Castellani}},\ and\ \bibinfo {author} {\bibfnamefont {L.}~\bibnamefont {Benfatto}},\ }\bibfield  {title} {\bibinfo {title} {Nonlinear optical effects and third-harmonic generation in superconductors: {Cooper} pairs versus {Higgs} mode contribution},\ }\href@noop {} {\bibfield  {journal} {\bibinfo  {journal} {Phys. Rev. B}\ }\textbf {\bibinfo {volume} {93}},\ \bibinfo {pages} {180507(R)} (\bibinfo {year} {2016})}\BibitemShut {NoStop}%
\bibitem [{\citenamefont {Vaswani}\ \emph {et~al.}(2020{\natexlab{a}})\citenamefont {Vaswani}, \citenamefont {Mootz}, \citenamefont {Sundahl}, \citenamefont {Mudiyanselage}, \citenamefont {Kang}, \citenamefont {Yang}, \citenamefont {Cheng}, \citenamefont {Huang}, \citenamefont {Kim}, \citenamefont {Liu}, \citenamefont {Luo}, \citenamefont {Perakis}, \citenamefont {Eom},\ and\ \citenamefont {Wang}}]{vaswani2019discovery}%
  \BibitemOpen
  \bibfield  {author} {\bibinfo {author} {\bibfnamefont {C.}~\bibnamefont {Vaswani}}, \bibinfo {author} {\bibfnamefont {M.}~\bibnamefont {Mootz}}, \bibinfo {author} {\bibfnamefont {C.}~\bibnamefont {Sundahl}}, \bibinfo {author} {\bibfnamefont {D.~H.}\ \bibnamefont {Mudiyanselage}}, \bibinfo {author} {\bibfnamefont {J.~H.}\ \bibnamefont {Kang}}, \bibinfo {author} {\bibfnamefont {X.}~\bibnamefont {Yang}}, \bibinfo {author} {\bibfnamefont {D.}~\bibnamefont {Cheng}}, \bibinfo {author} {\bibfnamefont {C.}~\bibnamefont {Huang}}, \bibinfo {author} {\bibfnamefont {R.~H.~J.}\ \bibnamefont {Kim}}, \bibinfo {author} {\bibfnamefont {Z.}~\bibnamefont {Liu}}, \bibinfo {author} {\bibfnamefont {L.}~\bibnamefont {Luo}}, \bibinfo {author} {\bibfnamefont {I.~E.}\ \bibnamefont {Perakis}}, \bibinfo {author} {\bibfnamefont {C.~B.}\ \bibnamefont {Eom}},\ and\ \bibinfo {author} {\bibfnamefont {J.}~\bibnamefont {Wang}},\ }\bibfield  {title} {\bibinfo {title} {Terahertz second-harmonic generation from lightwave acceleration of
  symmetry-breaking nonlinear supercurrents},\ }\href@noop {} {\bibfield  {journal} {\bibinfo  {journal} {Phys. Rev. Lett.}\ }\textbf {\bibinfo {volume} {124}},\ \bibinfo {pages} {207003} (\bibinfo {year} {2020}{\natexlab{a}})}\BibitemShut {NoStop}%
\bibitem [{\citenamefont {Nambu}(1960)}]{Nambu}%
  \BibitemOpen
  \bibfield  {author} {\bibinfo {author} {\bibfnamefont {Y.}~\bibnamefont {Nambu}},\ }\bibfield  {title} {\bibinfo {title} {Quasi-particles and gauge invariance in the theory of superconductivity},\ }\href@noop {} {\bibfield  {journal} {\bibinfo  {journal} {Phys. Rev.}\ }\textbf {\bibinfo {volume} {117}},\ \bibinfo {pages} {648} (\bibinfo {year} {1960})}\BibitemShut {NoStop}%
\bibitem [{\citenamefont {Liu}\ \emph {et~al.}(2010)\citenamefont {Liu}, \citenamefont {Kondo}, \citenamefont {Fernandes}, \citenamefont {Palczewski}, \citenamefont {Mun}, \citenamefont {Ni}, \citenamefont {Thaler}, \citenamefont {Bostwick}, \citenamefont {Rotenberg}, \citenamefont {Schmalian}, \citenamefont {Bud'ko}, \citenamefont {Canfield},\ and\ \citenamefont {Kaminski}}]{Liu2010}%
  \BibitemOpen
  \bibfield  {author} {\bibinfo {author} {\bibfnamefont {C.}~\bibnamefont {Liu}}, \bibinfo {author} {\bibfnamefont {T.}~\bibnamefont {Kondo}}, \bibinfo {author} {\bibfnamefont {R.~M.}\ \bibnamefont {Fernandes}}, \bibinfo {author} {\bibfnamefont {A.~D.}\ \bibnamefont {Palczewski}}, \bibinfo {author} {\bibfnamefont {E.~D.}\ \bibnamefont {Mun}}, \bibinfo {author} {\bibfnamefont {N.}~\bibnamefont {Ni}}, \bibinfo {author} {\bibfnamefont {A.~N.}\ \bibnamefont {Thaler}}, \bibinfo {author} {\bibfnamefont {A.}~\bibnamefont {Bostwick}}, \bibinfo {author} {\bibfnamefont {E.}~\bibnamefont {Rotenberg}}, \bibinfo {author} {\bibfnamefont {J.}~\bibnamefont {Schmalian}}, \bibinfo {author} {\bibfnamefont {S.~L.}\ \bibnamefont {Bud'ko}}, \bibinfo {author} {\bibfnamefont {P.~C.}\ \bibnamefont {Canfield}},\ and\ \bibinfo {author} {\bibfnamefont {A.}~\bibnamefont {Kaminski}},\ }\bibfield  {title} {\bibinfo {title} {Evidence for a {Lifshitz} transition in electron-doped iron arsenic superconductors at the onset of
  superconductivity},\ }\href@noop {} {\bibfield  {journal} {\bibinfo  {journal} {Nat. Phys.}\ }\textbf {\bibinfo {volume} {6}},\ \bibinfo {pages} {419} (\bibinfo {year} {2010})}\BibitemShut {NoStop}%
\bibitem [{\citenamefont {Fernandes}\ and\ \citenamefont {Schmalian}(2010)}]{Fernandes2010}%
  \BibitemOpen
  \bibfield  {author} {\bibinfo {author} {\bibfnamefont {R.~M.}\ \bibnamefont {Fernandes}}\ and\ \bibinfo {author} {\bibfnamefont {J.}~\bibnamefont {Schmalian}},\ }\bibfield  {title} {\bibinfo {title} {Competing order and nature of the pairing state in the iron pnictides},\ }\href@noop {} {\bibfield  {journal} {\bibinfo  {journal} {Phys. Rev. B}\ }\textbf {\bibinfo {volume} {82}},\ \bibinfo {pages} {014521} (\bibinfo {year} {2010})}\BibitemShut {NoStop}%
\bibitem [{\citenamefont {Kuehn}\ \emph {et~al.}(2011)\citenamefont {Kuehn}, \citenamefont {Reimann}, \citenamefont {Woerner}, \citenamefont {Elsaesser},\ and\ \citenamefont {Hey}}]{Kuehn2011}%
  \BibitemOpen
  \bibfield  {author} {\bibinfo {author} {\bibfnamefont {W.}~\bibnamefont {Kuehn}}, \bibinfo {author} {\bibfnamefont {K.}~\bibnamefont {Reimann}}, \bibinfo {author} {\bibfnamefont {M.}~\bibnamefont {Woerner}}, \bibinfo {author} {\bibfnamefont {T.}~\bibnamefont {Elsaesser}},\ and\ \bibinfo {author} {\bibfnamefont {R.}~\bibnamefont {Hey}},\ }\bibfield  {title} {\bibinfo {title} {Two-dimensional terahertz correlation spectra of electronic excitations in semiconductor quantum wells},\ }\href@noop {} {\bibfield  {journal} {\bibinfo  {journal} {J. Phys. Chem. B}\ }\textbf {\bibinfo {volume} {115}},\ \bibinfo {pages} {5448} (\bibinfo {year} {2011})}\BibitemShut {NoStop}%
\bibitem [{\citenamefont {Blumberg}\ \emph {et~al.}(2007)\citenamefont {Blumberg}, \citenamefont {Mialitsin}, \citenamefont {Dennis}, \citenamefont {Klein}, \citenamefont {Zhigadlo},\ and\ \citenamefont {Karpinski}}]{Blumberg}%
  \BibitemOpen
  \bibfield  {author} {\bibinfo {author} {\bibfnamefont {G.}~\bibnamefont {Blumberg}}, \bibinfo {author} {\bibfnamefont {A.}~\bibnamefont {Mialitsin}}, \bibinfo {author} {\bibfnamefont {B.~S.}\ \bibnamefont {Dennis}}, \bibinfo {author} {\bibfnamefont {M.~V.}\ \bibnamefont {Klein}}, \bibinfo {author} {\bibfnamefont {N.~D.}\ \bibnamefont {Zhigadlo}},\ and\ \bibinfo {author} {\bibfnamefont {J.}~\bibnamefont {Karpinski}},\ }\bibfield  {title} {\bibinfo {title} {Observation of {Leggett's} collective mode in a multiband {${\mathrm{MgB}}_{2}$} superconductor},\ }\href@noop {} {\bibfield  {journal} {\bibinfo  {journal} {Phys. Rev. Lett.}\ }\textbf {\bibinfo {volume} {99}},\ \bibinfo {pages} {227002} (\bibinfo {year} {2007})}\BibitemShut {NoStop}%
\bibitem [{\citenamefont {Klein}(2010)}]{Klein2010}%
  \BibitemOpen
  \bibfield  {author} {\bibinfo {author} {\bibfnamefont {M.~V.}\ \bibnamefont {Klein}},\ }\bibfield  {title} {\bibinfo {title} {Theory of raman scattering from {Leggett's} collective mode in a multiband superconductor: Application to {${\text{MgB}}_{2}$}},\ }\href@noop {} {\bibfield  {journal} {\bibinfo  {journal} {Phys. Rev. B}\ }\textbf {\bibinfo {volume} {82}},\ \bibinfo {pages} {014507} (\bibinfo {year} {2010})}\BibitemShut {NoStop}%
\bibitem [{\citenamefont {Cea}\ and\ \citenamefont {Benfatto}(2016)}]{Cea2016b}%
  \BibitemOpen
  \bibfield  {author} {\bibinfo {author} {\bibfnamefont {T.}~\bibnamefont {Cea}}\ and\ \bibinfo {author} {\bibfnamefont {L.}~\bibnamefont {Benfatto}},\ }\bibfield  {title} {\bibinfo {title} {Signature of the {Leggett} mode in the {${A}_{1g}$} {Raman} response: From {${\text{MgB}}_{2}$} to iron-based superconductors},\ }\href@noop {} {\bibfield  {journal} {\bibinfo  {journal} {Phys. Rev. B}\ }\textbf {\bibinfo {volume} {94}},\ \bibinfo {pages} {064512} (\bibinfo {year} {2016})}\BibitemShut {NoStop}%
\bibitem [{\citenamefont {Ortolani}\ \emph {et~al.}(2008)\citenamefont {Ortolani}, \citenamefont {Dore}, \citenamefont {Di~Castro}, \citenamefont {Perucchi}, \citenamefont {Lupi}, \citenamefont {Ferrando}, \citenamefont {Putti}, \citenamefont {Pallecchi}, \citenamefont {Ferdeghini},\ and\ \citenamefont {Xi}}]{Ortolani}%
  \BibitemOpen
  \bibfield  {author} {\bibinfo {author} {\bibfnamefont {M.}~\bibnamefont {Ortolani}}, \bibinfo {author} {\bibfnamefont {P.}~\bibnamefont {Dore}}, \bibinfo {author} {\bibfnamefont {D.}~\bibnamefont {Di~Castro}}, \bibinfo {author} {\bibfnamefont {A.}~\bibnamefont {Perucchi}}, \bibinfo {author} {\bibfnamefont {S.}~\bibnamefont {Lupi}}, \bibinfo {author} {\bibfnamefont {V.}~\bibnamefont {Ferrando}}, \bibinfo {author} {\bibfnamefont {M.}~\bibnamefont {Putti}}, \bibinfo {author} {\bibfnamefont {I.}~\bibnamefont {Pallecchi}}, \bibinfo {author} {\bibfnamefont {C.}~\bibnamefont {Ferdeghini}},\ and\ \bibinfo {author} {\bibfnamefont {X.~X.}\ \bibnamefont {Xi}},\ }\bibfield  {title} {\bibinfo {title} {Two-band parallel conductivity at terahertz frequencies in the superconducting state of {$\mathrm{Mg}{\mathrm{B}}_{2}$}},\ }\href@noop {} {\bibfield  {journal} {\bibinfo  {journal} {Phys. Rev. B}\ }\textbf {\bibinfo {volume} {77}},\ \bibinfo {pages} {100507(R)} (\bibinfo {year} {2008})}\BibitemShut {NoStop}%
\bibitem [{\citenamefont {Patz}\ \emph {et~al.}(2014)\citenamefont {Patz}, \citenamefont {Li}, \citenamefont {Ran}, \citenamefont {Fernandes}, \citenamefont {Schmalian}, \citenamefont {Bud'ko}, \citenamefont {Canfield}, \citenamefont {Perakis},\ and\ \citenamefont {Wang}}]{Patz2014}%
  \BibitemOpen
  \bibfield  {author} {\bibinfo {author} {\bibfnamefont {A.}~\bibnamefont {Patz}}, \bibinfo {author} {\bibfnamefont {T.}~\bibnamefont {Li}}, \bibinfo {author} {\bibfnamefont {S.}~\bibnamefont {Ran}}, \bibinfo {author} {\bibfnamefont {R.~M.}\ \bibnamefont {Fernandes}}, \bibinfo {author} {\bibfnamefont {J.}~\bibnamefont {Schmalian}}, \bibinfo {author} {\bibfnamefont {S.~L.}\ \bibnamefont {Bud'ko}}, \bibinfo {author} {\bibfnamefont {P.~C.}\ \bibnamefont {Canfield}}, \bibinfo {author} {\bibfnamefont {I.~E.}\ \bibnamefont {Perakis}},\ and\ \bibinfo {author} {\bibfnamefont {J.}~\bibnamefont {Wang}},\ }\bibfield  {title} {\bibinfo {title} {Ultrafast observation of critical nematic fluctuations and giant magnetoelastic coupling in iron pnictides},\ }\href@noop {} {\bibfield  {journal} {\bibinfo  {journal} {Nat. Commun.}\ }\textbf {\bibinfo {volume} {5}},\ \bibinfo {pages} {3229} (\bibinfo {year} {2014})}\BibitemShut {NoStop}%
\bibitem [{\citenamefont {Patz}\ \emph {et~al.}(2017)\citenamefont {Patz}, \citenamefont {Li}, \citenamefont {Luo}, \citenamefont {Yang}, \citenamefont {Bud'ko}, \citenamefont {Canfield}, \citenamefont {Perakis},\ and\ \citenamefont {Wang}}]{Patz2017}%
  \BibitemOpen
  \bibfield  {author} {\bibinfo {author} {\bibfnamefont {A.}~\bibnamefont {Patz}}, \bibinfo {author} {\bibfnamefont {T.}~\bibnamefont {Li}}, \bibinfo {author} {\bibfnamefont {L.}~\bibnamefont {Luo}}, \bibinfo {author} {\bibfnamefont {X.}~\bibnamefont {Yang}}, \bibinfo {author} {\bibfnamefont {S.}~\bibnamefont {Bud'ko}}, \bibinfo {author} {\bibfnamefont {P.~C.}\ \bibnamefont {Canfield}}, \bibinfo {author} {\bibfnamefont {I.~E.}\ \bibnamefont {Perakis}},\ and\ \bibinfo {author} {\bibfnamefont {J.}~\bibnamefont {Wang}},\ }\bibfield  {title} {\bibinfo {title} {Critical speeding up of nonequilibrium electronic relaxation near nematic phase transition in unstrained {Ba(${\mathrm{Fe}}_{1\ensuremath{-}x}{\mathrm{Co}}_{x}$)${}_{2}{\mathrm{As}}_{2}$}},\ }\href@noop {} {\bibfield  {journal} {\bibinfo  {journal} {Phys. Rev. B}\ }\textbf {\bibinfo {volume} {95}},\ \bibinfo {pages} {165122} (\bibinfo {year} {2017})}\BibitemShut {NoStop}%
\bibitem [{\citenamefont {Shahbazyan}\ \emph {et~al.}(2000)\citenamefont {Shahbazyan}, \citenamefont {Primozich}, \citenamefont {Perakis},\ and\ \citenamefont {Chemla}}]{shah}%
  \BibitemOpen
  \bibfield  {author} {\bibinfo {author} {\bibfnamefont {T.~V.}\ \bibnamefont {Shahbazyan}}, \bibinfo {author} {\bibfnamefont {N.}~\bibnamefont {Primozich}}, \bibinfo {author} {\bibfnamefont {I.~E.}\ \bibnamefont {Perakis}},\ and\ \bibinfo {author} {\bibfnamefont {D.~S.}\ \bibnamefont {Chemla}},\ }\bibfield  {title} {\bibinfo {title} {Femtosecond coherent dynamics of the {Fermi}-edge singularity and exciton hybrid},\ }\href@noop {} {\bibfield  {journal} {\bibinfo  {journal} {Phys. Rev. Lett.}\ }\textbf {\bibinfo {volume} {84}},\ \bibinfo {pages} {2006} (\bibinfo {year} {2000})}\BibitemShut {NoStop}%
\bibitem [{\citenamefont {Lingos}\ \emph {et~al.}(2017)\citenamefont {Lingos}, \citenamefont {Patz}, \citenamefont {Li}, \citenamefont {Barmparis}, \citenamefont {Keliri}, \citenamefont {Kapetanakis}, \citenamefont {Li}, \citenamefont {Yan}, \citenamefont {Wang},\ and\ \citenamefont {Perakis}}]{Lingos2017}%
  \BibitemOpen
  \bibfield  {author} {\bibinfo {author} {\bibfnamefont {P.~C.}\ \bibnamefont {Lingos}}, \bibinfo {author} {\bibfnamefont {A.}~\bibnamefont {Patz}}, \bibinfo {author} {\bibfnamefont {T.}~\bibnamefont {Li}}, \bibinfo {author} {\bibfnamefont {G.~D.}\ \bibnamefont {Barmparis}}, \bibinfo {author} {\bibfnamefont {A.}~\bibnamefont {Keliri}}, \bibinfo {author} {\bibfnamefont {M.~D.}\ \bibnamefont {Kapetanakis}}, \bibinfo {author} {\bibfnamefont {L.}~\bibnamefont {Li}}, \bibinfo {author} {\bibfnamefont {J.}~\bibnamefont {Yan}}, \bibinfo {author} {\bibfnamefont {J.}~\bibnamefont {Wang}},\ and\ \bibinfo {author} {\bibfnamefont {I.~E.}\ \bibnamefont {Perakis}},\ }\bibfield  {title} {\bibinfo {title} {Correlating quasiparticle excitations with quantum femtosecond magnetism in photoexcited nonequilibrium states of insulating antiferromagnetic manganites},\ }\href {https://doi.org/10.1103/PhysRevB.95.224432} {\bibfield  {journal} {\bibinfo  {journal} {Phys. Rev. B}\ }\textbf {\bibinfo {volume} {95}},\ \bibinfo {pages}
  {224432} (\bibinfo {year} {2017})}\BibitemShut {NoStop}%
\bibitem [{\citenamefont {Lingos}\ \emph {et~al.}(2021)\citenamefont {Lingos}, \citenamefont {Kapetanakis}, \citenamefont {Wang},\ and\ \citenamefont {Perakis}}]{Lingos2021}%
  \BibitemOpen
  \bibfield  {author} {\bibinfo {author} {\bibfnamefont {P.~C.}\ \bibnamefont {Lingos}}, \bibinfo {author} {\bibfnamefont {M.~D.}\ \bibnamefont {Kapetanakis}}, \bibinfo {author} {\bibfnamefont {J.}~\bibnamefont {Wang}},\ and\ \bibinfo {author} {\bibfnamefont {I.~E.}\ \bibnamefont {Perakis}},\ }\bibfield  {title} {\bibinfo {title} {Light-wave control of correlated materials using quantum magnetism during time-periodic modulation of coherent transport},\ }\href {https://doi.org/10.1038/s42005-021-00561-z} {\bibfield  {journal} {\bibinfo  {journal} {Commun. Phys.}\ }\textbf {\bibinfo {volume} {4}},\ \bibinfo {pages} {60} (\bibinfo {year} {2021})}\BibitemShut {NoStop}%
\bibitem [{\citenamefont {Yuzbashyan}\ \emph {et~al.}(2006)\citenamefont {Yuzbashyan}, \citenamefont {Tsyplyatyev},\ and\ \citenamefont {Altshuler}}]{Yuzbashyan:2006}%
  \BibitemOpen
  \bibfield  {author} {\bibinfo {author} {\bibfnamefont {E.~A.}\ \bibnamefont {Yuzbashyan}}, \bibinfo {author} {\bibfnamefont {O.}~\bibnamefont {Tsyplyatyev}},\ and\ \bibinfo {author} {\bibfnamefont {B.~L.}\ \bibnamefont {Altshuler}},\ }\bibfield  {title} {\bibinfo {title} {Relaxation and persistent oscillations of the order parameter in fermionic condensates},\ }\href@noop {} {\bibfield  {journal} {\bibinfo  {journal} {Phys. Rev. Lett.}\ }\textbf {\bibinfo {volume} {96}},\ \bibinfo {pages} {097005} (\bibinfo {year} {2006})}\BibitemShut {NoStop}%
\bibitem [{\citenamefont {Yuzbashyan}(2008)}]{Yuzbashyan2008}%
  \BibitemOpen
  \bibfield  {author} {\bibinfo {author} {\bibfnamefont {E.~A.}\ \bibnamefont {Yuzbashyan}},\ }\bibfield  {title} {\bibinfo {title} {Normal and anomalous solitons in the theory of dynamical {Cooper} pairing},\ }\href {https://doi.org/10.1103/PhysRevB.78.184507} {\bibfield  {journal} {\bibinfo  {journal} {Phys. Rev. B}\ }\textbf {\bibinfo {volume} {78}},\ \bibinfo {pages} {184507} (\bibinfo {year} {2008})}\BibitemShut {NoStop}%
\bibitem [{\citenamefont {Collado}\ \emph {et~al.}(2018)\citenamefont {Collado}, \citenamefont {Lorenzana}, \citenamefont {Usaj},\ and\ \citenamefont {Balseiro}}]{Balseiro}%
  \BibitemOpen
  \bibfield  {author} {\bibinfo {author} {\bibfnamefont {H.~P.~O.}\ \bibnamefont {Collado}}, \bibinfo {author} {\bibfnamefont {J.}~\bibnamefont {Lorenzana}}, \bibinfo {author} {\bibfnamefont {G.}~\bibnamefont {Usaj}},\ and\ \bibinfo {author} {\bibfnamefont {C.~A.}\ \bibnamefont {Balseiro}},\ }\bibfield  {title} {\bibinfo {title} {Population inversion and dynamical phase transitions in a driven superconductor},\ }\href@noop {} {\bibfield  {journal} {\bibinfo  {journal} {Phys. Rev. B}\ }\textbf {\bibinfo {volume} {98}},\ \bibinfo {pages} {214519} (\bibinfo {year} {2018})}\BibitemShut {NoStop}%
\bibitem [{\citenamefont {Luo}\ \emph {et~al.}(2021)\citenamefont {Luo}, \citenamefont {Cheng}, \citenamefont {Song}, \citenamefont {Wang}, \citenamefont {Vaswani}, \citenamefont {Lozano}, \citenamefont {Gu}, \citenamefont {Huang}, \citenamefont {Kim}, \citenamefont {Liu}, \citenamefont {Park}, \citenamefont {Yao}, \citenamefont {Ho}, \citenamefont {Perakis}, \citenamefont {Li},\ and\ \citenamefont {Wang}}]{Luo}%
  \BibitemOpen
  \bibfield  {author} {\bibinfo {author} {\bibfnamefont {L.}~\bibnamefont {Luo}}, \bibinfo {author} {\bibfnamefont {D.}~\bibnamefont {Cheng}}, \bibinfo {author} {\bibfnamefont {B.}~\bibnamefont {Song}}, \bibinfo {author} {\bibfnamefont {L.-L.}\ \bibnamefont {Wang}}, \bibinfo {author} {\bibfnamefont {C.}~\bibnamefont {Vaswani}}, \bibinfo {author} {\bibfnamefont {P.~M.}\ \bibnamefont {Lozano}}, \bibinfo {author} {\bibfnamefont {G.}~\bibnamefont {Gu}}, \bibinfo {author} {\bibfnamefont {C.}~\bibnamefont {Huang}}, \bibinfo {author} {\bibfnamefont {R.~H.~J.}\ \bibnamefont {Kim}}, \bibinfo {author} {\bibfnamefont {Z.}~\bibnamefont {Liu}}, \bibinfo {author} {\bibfnamefont {J.-M.}\ \bibnamefont {Park}}, \bibinfo {author} {\bibfnamefont {Y.}~\bibnamefont {Yao}}, \bibinfo {author} {\bibfnamefont {K.}~\bibnamefont {Ho}}, \bibinfo {author} {\bibfnamefont {I.~E.}\ \bibnamefont {Perakis}}, \bibinfo {author} {\bibfnamefont {Q.}~\bibnamefont {Li}},\ and\ \bibinfo {author} {\bibfnamefont {J.}~\bibnamefont {Wang}},\ }\bibfield
  {title} {\bibinfo {title} {A light-induced phononic symmetry switch and giant dissipationless topological photocurrent in {ZrTe$_5$}},\ }\href@noop {} {\bibfield  {journal} {\bibinfo  {journal} {Nat. Mater.}\ }\textbf {\bibinfo {volume} {20}},\ \bibinfo {pages} {329} (\bibinfo {year} {2021})}\BibitemShut {NoStop}%
\bibitem [{\citenamefont {Luo}\ \emph {et~al.}(2019)\citenamefont {Luo}, \citenamefont {Yang}, \citenamefont {Liu}, \citenamefont {Liu}, \citenamefont {Vaswani}, \citenamefont {Cheng}, \citenamefont {Mootz}, \citenamefont {Zhao}, \citenamefont {Yao}, \citenamefont {Wang}, \citenamefont {Ho}, \citenamefont {Perakis}, \citenamefont {Dobrowolska}, \citenamefont {Furdyna},\ and\ \citenamefont {Wang}}]{Luo2019}%
  \BibitemOpen
  \bibfield  {author} {\bibinfo {author} {\bibfnamefont {L.}~\bibnamefont {Luo}}, \bibinfo {author} {\bibfnamefont {X.}~\bibnamefont {Yang}}, \bibinfo {author} {\bibfnamefont {X.}~\bibnamefont {Liu}}, \bibinfo {author} {\bibfnamefont {Z.}~\bibnamefont {Liu}}, \bibinfo {author} {\bibfnamefont {C.}~\bibnamefont {Vaswani}}, \bibinfo {author} {\bibfnamefont {D.}~\bibnamefont {Cheng}}, \bibinfo {author} {\bibfnamefont {M.}~\bibnamefont {Mootz}}, \bibinfo {author} {\bibfnamefont {X.}~\bibnamefont {Zhao}}, \bibinfo {author} {\bibfnamefont {Y.}~\bibnamefont {Yao}}, \bibinfo {author} {\bibfnamefont {C.-Z.}\ \bibnamefont {Wang}}, \bibinfo {author} {\bibfnamefont {K.-M.}\ \bibnamefont {Ho}}, \bibinfo {author} {\bibfnamefont {I.~E.}\ \bibnamefont {Perakis}}, \bibinfo {author} {\bibfnamefont {M.}~\bibnamefont {Dobrowolska}}, \bibinfo {author} {\bibfnamefont {J.~K.}\ \bibnamefont {Furdyna}},\ and\ \bibinfo {author} {\bibfnamefont {J.}~\bibnamefont {Wang}},\ }\bibfield  {title} {\bibinfo {title} {Ultrafast manipulation of
  topologically enhanced surface transport driven by mid-infrared and terahertz pulses in {Bi$_2$Se$_3$}},\ }\href@noop {} {\bibfield  {journal} {\bibinfo  {journal} {Nat. Commun.}\ }\textbf {\bibinfo {volume} {10}},\ \bibinfo {pages} {607} (\bibinfo {year} {2019})}\BibitemShut {NoStop}%
\bibitem [{\citenamefont {Yang}\ \emph {et~al.}(2020)\citenamefont {Yang}, \citenamefont {Luo}, \citenamefont {Vaswani}, \citenamefont {Zhao}, \citenamefont {Yao}, \citenamefont {Cheng}, \citenamefont {Liu}, \citenamefont {Kim}, \citenamefont {Liu}, \citenamefont {Dobrowolska-Furdyna}, \citenamefont {Furdyna}, \citenamefont {Perakis}, \citenamefont {Wang}, \citenamefont {Ho},\ and\ \citenamefont {Wang}}]{Yang2020}%
  \BibitemOpen
  \bibfield  {author} {\bibinfo {author} {\bibfnamefont {X.}~\bibnamefont {Yang}}, \bibinfo {author} {\bibfnamefont {L.}~\bibnamefont {Luo}}, \bibinfo {author} {\bibfnamefont {C.}~\bibnamefont {Vaswani}}, \bibinfo {author} {\bibfnamefont {X.}~\bibnamefont {Zhao}}, \bibinfo {author} {\bibfnamefont {Y.}~\bibnamefont {Yao}}, \bibinfo {author} {\bibfnamefont {D.}~\bibnamefont {Cheng}}, \bibinfo {author} {\bibfnamefont {Z.}~\bibnamefont {Liu}}, \bibinfo {author} {\bibfnamefont {R.~H.~J.}\ \bibnamefont {Kim}}, \bibinfo {author} {\bibfnamefont {X.}~\bibnamefont {Liu}}, \bibinfo {author} {\bibfnamefont {M.}~\bibnamefont {Dobrowolska-Furdyna}}, \bibinfo {author} {\bibfnamefont {J.~K.}\ \bibnamefont {Furdyna}}, \bibinfo {author} {\bibfnamefont {I.~E.}\ \bibnamefont {Perakis}}, \bibinfo {author} {\bibfnamefont {C.}~\bibnamefont {Wang}}, \bibinfo {author} {\bibfnamefont {K.}~\bibnamefont {Ho}},\ and\ \bibinfo {author} {\bibfnamefont {J.}~\bibnamefont {Wang}},\ }\bibfield  {title} {\bibinfo {title} {Light control of
  surface--bulk coupling by terahertz vibrational coherence in a topological insulator},\ }\href@noop {} {\bibfield  {journal} {\bibinfo  {journal} {npj Quantum Mater.}\ }\textbf {\bibinfo {volume} {5}},\ \bibinfo {pages} {13} (\bibinfo {year} {2020})}\BibitemShut {NoStop}%
\bibitem [{\citenamefont {Cheng}\ \emph {et~al.}(2023{\natexlab{c}})\citenamefont {Cheng}, \citenamefont {Cheng}, \citenamefont {Jiang}, \citenamefont {Xia}, \citenamefont {Song}, \citenamefont {Mootz}, \citenamefont {Luo}, \citenamefont {Perakis}, \citenamefont {Yao}, \citenamefont {Guo},\ and\ \citenamefont {Wang}}]{cheng2023chirality}%
  \BibitemOpen
  \bibfield  {author} {\bibinfo {author} {\bibfnamefont {B.}~\bibnamefont {Cheng}}, \bibinfo {author} {\bibfnamefont {D.}~\bibnamefont {Cheng}}, \bibinfo {author} {\bibfnamefont {T.}~\bibnamefont {Jiang}}, \bibinfo {author} {\bibfnamefont {W.}~\bibnamefont {Xia}}, \bibinfo {author} {\bibfnamefont {B.}~\bibnamefont {Song}}, \bibinfo {author} {\bibfnamefont {M.}~\bibnamefont {Mootz}}, \bibinfo {author} {\bibfnamefont {L.}~\bibnamefont {Luo}}, \bibinfo {author} {\bibfnamefont {I.~E.}\ \bibnamefont {Perakis}}, \bibinfo {author} {\bibfnamefont {Y.}~\bibnamefont {Yao}}, \bibinfo {author} {\bibfnamefont {Y.}~\bibnamefont {Guo}},\ and\ \bibinfo {author} {\bibfnamefont {J.}~\bibnamefont {Wang}},\ }\href@noop {} {\bibinfo {title} {Chirality manipulation of ultrafast phase switchings in a correlated {CDW-Weyl} semimetal}} (\bibinfo {year} {2023}{\natexlab{c}}),\ \Eprint {https://arxiv.org/abs/2308.03895} {arXiv:2308.03895} \BibitemShut {NoStop}%
\bibitem [{\citenamefont {Vaswani}\ \emph {et~al.}(2020{\natexlab{b}})\citenamefont {Vaswani}, \citenamefont {Wang}, \citenamefont {Mudiyanselage}, \citenamefont {Li}, \citenamefont {Lozano}, \citenamefont {Gu}, \citenamefont {Cheng}, \citenamefont {Song}, \citenamefont {Luo}, \citenamefont {Kim}, \citenamefont {Huang}, \citenamefont {Liu}, \citenamefont {Mootz}, \citenamefont {Perakis}, \citenamefont {Yao}, \citenamefont {Ho},\ and\ \citenamefont {Wang}}]{vasw2020}%
  \BibitemOpen
  \bibfield  {author} {\bibinfo {author} {\bibfnamefont {C.}~\bibnamefont {Vaswani}}, \bibinfo {author} {\bibfnamefont {L.-L.}\ \bibnamefont {Wang}}, \bibinfo {author} {\bibfnamefont {D.~H.}\ \bibnamefont {Mudiyanselage}}, \bibinfo {author} {\bibfnamefont {Q.}~\bibnamefont {Li}}, \bibinfo {author} {\bibfnamefont {P.~M.}\ \bibnamefont {Lozano}}, \bibinfo {author} {\bibfnamefont {G.~D.}\ \bibnamefont {Gu}}, \bibinfo {author} {\bibfnamefont {D.}~\bibnamefont {Cheng}}, \bibinfo {author} {\bibfnamefont {B.}~\bibnamefont {Song}}, \bibinfo {author} {\bibfnamefont {L.}~\bibnamefont {Luo}}, \bibinfo {author} {\bibfnamefont {R.~H.~J.}\ \bibnamefont {Kim}}, \bibinfo {author} {\bibfnamefont {C.}~\bibnamefont {Huang}}, \bibinfo {author} {\bibfnamefont {Z.}~\bibnamefont {Liu}}, \bibinfo {author} {\bibfnamefont {M.}~\bibnamefont {Mootz}}, \bibinfo {author} {\bibfnamefont {I.~E.}\ \bibnamefont {Perakis}}, \bibinfo {author} {\bibfnamefont {Y.}~\bibnamefont {Yao}}, \bibinfo {author} {\bibfnamefont {K.~M.}\ \bibnamefont {Ho}},\
  and\ \bibinfo {author} {\bibfnamefont {J.}~\bibnamefont {Wang}},\ }\bibfield  {title} {\bibinfo {title} {Light-driven {Raman} coherence as a nonthermal route to ultrafast topology switching in a {Dirac} semimetal},\ }\href@noop {} {\bibfield  {journal} {\bibinfo  {journal} {Phys. Rev. X}\ }\textbf {\bibinfo {volume} {10}},\ \bibinfo {pages} {021013} (\bibinfo {year} {2020}{\natexlab{b}})}\BibitemShut {NoStop}%
\bibitem [{\citenamefont {Patz}\ \emph {et~al.}(2015)\citenamefont {Patz}, \citenamefont {Li}, \citenamefont {Liu}, \citenamefont {Furdyna}, \citenamefont {Perakis},\ and\ \citenamefont {Wang}}]{Patz2015}%
  \BibitemOpen
  \bibfield  {author} {\bibinfo {author} {\bibfnamefont {A.}~\bibnamefont {Patz}}, \bibinfo {author} {\bibfnamefont {T.}~\bibnamefont {Li}}, \bibinfo {author} {\bibfnamefont {X.}~\bibnamefont {Liu}}, \bibinfo {author} {\bibfnamefont {J.~K.}\ \bibnamefont {Furdyna}}, \bibinfo {author} {\bibfnamefont {I.~E.}\ \bibnamefont {Perakis}},\ and\ \bibinfo {author} {\bibfnamefont {J.}~\bibnamefont {Wang}},\ }\bibfield  {title} {\bibinfo {title} {Ultrafast probes of nonequilibrium hole spin relaxation in the ferromagnetic semiconductor {GaMnAs}},\ }\href@noop {} {\bibfield  {journal} {\bibinfo  {journal} {Phys. Rev. B}\ }\textbf {\bibinfo {volume} {91}},\ \bibinfo {pages} {155108} (\bibinfo {year} {2015})}\BibitemShut {NoStop}%
\bibitem [{\citenamefont {Li}\ \emph {et~al.}(2013)\citenamefont {Li}, \citenamefont {Patz}, \citenamefont {Mouchliadis}, \citenamefont {Yan}, \citenamefont {Lograsso}, \citenamefont {Perakis},\ and\ \citenamefont {Wang}}]{femtomag}%
  \BibitemOpen
  \bibfield  {author} {\bibinfo {author} {\bibfnamefont {T.}~\bibnamefont {Li}}, \bibinfo {author} {\bibfnamefont {A.}~\bibnamefont {Patz}}, \bibinfo {author} {\bibfnamefont {L.}~\bibnamefont {Mouchliadis}}, \bibinfo {author} {\bibfnamefont {J.}~\bibnamefont {Yan}}, \bibinfo {author} {\bibfnamefont {T.~A.}\ \bibnamefont {Lograsso}}, \bibinfo {author} {\bibfnamefont {I.~E.}\ \bibnamefont {Perakis}},\ and\ \bibinfo {author} {\bibfnamefont {J.}~\bibnamefont {Wang}},\ }\bibfield  {title} {\bibinfo {title} {Femtosecond switching of magnetism via strongly correlated spin--charge quantum excitations},\ }\href@noop {} {\bibfield  {journal} {\bibinfo  {journal} {Nature}\ }\textbf {\bibinfo {volume} {496}},\ \bibinfo {pages} {69} (\bibinfo {year} {2013})}\BibitemShut {NoStop}%
\bibitem [{\citenamefont {Song}\ \emph {et~al.}(2023)\citenamefont {Song}, \citenamefont {Yang}, \citenamefont {Sundahl}, \citenamefont {Kang}, \citenamefont {Mootz}, \citenamefont {Yao}, \citenamefont {Perakis}, \citenamefont {Luo}, \citenamefont {Eom},\ and\ \citenamefont {Wang}}]{Song2023}%
  \BibitemOpen
  \bibfield  {author} {\bibinfo {author} {\bibfnamefont {B.~Q.}\ \bibnamefont {Song}}, \bibinfo {author} {\bibfnamefont {X.}~\bibnamefont {Yang}}, \bibinfo {author} {\bibfnamefont {C.}~\bibnamefont {Sundahl}}, \bibinfo {author} {\bibfnamefont {J.-H.}\ \bibnamefont {Kang}}, \bibinfo {author} {\bibfnamefont {M.}~\bibnamefont {Mootz}}, \bibinfo {author} {\bibfnamefont {Y.}~\bibnamefont {Yao}}, \bibinfo {author} {\bibfnamefont {I.~E.}\ \bibnamefont {Perakis}}, \bibinfo {author} {\bibfnamefont {L.}~\bibnamefont {Luo}}, \bibinfo {author} {\bibfnamefont {C.~B.}\ \bibnamefont {Eom}},\ and\ \bibinfo {author} {\bibfnamefont {J.}~\bibnamefont {Wang}},\ }\bibfield  {title} {\bibinfo {title} {Ultrafast martensitic phase transition driven by intense terahertz pulses},\ }\href@noop {} {\bibfield  {journal} {\bibinfo  {journal} {Ultrafast Sci.}\ }\textbf {\bibinfo {volume} {3}},\ \bibinfo {pages} {0007} (\bibinfo {year} {2023})}\BibitemShut {NoStop}%
\end{thebibliography}
%

\end{document}